\documentclass[10pt,aps,pra,twocolumn,superscriptaddress,floatfix,longbibliography]{revtex4-2}
\usepackage{placeins}
\usepackage{mathtools}
\usepackage{physics}
\usepackage{bbm}
\usepackage{bm}
\usepackage{amsbsy}
\usepackage{amsthm}
\usepackage{amssymb}
\usepackage{amsfonts}
\usepackage{amsmath}
\usepackage{soul}
\setstcolor{red}
\usepackage{dsfont} 
\usepackage{graphicx} 
\usepackage{hhline}
\usepackage{epsfig}
\usepackage{epstopdf}
\usepackage{dsfont}
\usepackage{xcolor}
\usepackage{color}
\usepackage{verbatim}
\usepackage{nomencl}
\usepackage[normalem]{ulem}
\usepackage{multirow}
\usepackage{makecell}
\usepackage{url}
\usepackage{tikz}
\usetikzlibrary{arrows.meta, decorations.pathreplacing}
\usepackage[colorlinks=true,linkcolor=blue,citecolor=blue,urlcolor=blue]{hyperref}
\DeclareUnicodeCharacter{2215}{/}
\DeclareUnicodeCharacter{0301}{\'{e}}

\usepackage{placeins}

\begin{document}
\title{System versus charger in performance optimization of quantum batteries}

\author{Rohit Kumar Shukla}
\email[]{rohitkrshukla.rs.phy17@itbhu.ac.in}
\affiliation{Department of Chemistry; Institute of Nanotechnology and Advanced Materials; Center for
Quantum Entanglement Science and Technology, Bar-Ilan University, Ramat-Gan, 5290002, Israel}
\author{Rajiv Kumar}
\affiliation{Department of Physics, Indian Institute of Technology (Banaras Hindu University) Varanasi - 221005, India}
\author{Ujjwal Sen}
\affiliation{Harish-Chandra Research Institute, A CI of Homi Bhabha National Institute, Chhatnag Road, Jhunsi, Prayagraj 211019, India}
\author{Sunil K. Mishra}
\affiliation{Department of Physics, Indian Institute of Technology (Banaras Hindu University) Varanasi - 221005, India}



\begin{abstract}
Quantum batteries provide a platform for investigating energy storage
and extraction in quantum many-body systems. Here, we study a charging
protocol in which battery and chargoid roles are assigned to different
Hamiltonian components of a standalone many-body spin system. By
externally controlling the contribution of the intrinsic battery
Hamiltonian during charging, we reveal a tunable competition between
intrinsic and charging dynamics. We find that suppressing the intrinsic
battery contribution can substantially enhance both the maximum stored
energy and charging power, with the magnitude of the enhancement
determined by the interaction structure and range. We further investigate
the protocol in a Markovian open-system setting that incorporates energy
relaxation and pure dephasing. While environmental effects generally
degrade charging performance, they can instead enhance energy-storage
and energy-extraction dynamics in certain interacting systems. The
enhancement associated with the controlled suppression of the intrinsic
battery dynamics remains robust in the presence of environmental
coupling.
\end{abstract}

\maketitle

\section{Introduction}
\vspace{0.25cm}
A quantum battery is a device designed to exploit quantum mechanical principles to enhance the efficiency of energy storage and retrieval. Unlike traditional batteries, which rely on chemical reactions \cite{liu2014comparative,chaoui2017aging,chen2012state,fleischer2013adaptive,rivera2017soc}, quantum batteries utilize features such as superposition and entanglement to store and release energy more efficiently and with potentially greater capacity \cite{alicki2013entanglement,hovhannisyan2013entanglement,binder2015quantacell,campaioli2017enhancing,le2018spin,ghosh2020enhancement,farina2019charger,PhysRevA.105.022628}. The ability of quantum systems to exist in superposed states allows energy to be distributed in ways not possible classically, while entanglement may provide additional advantages in energy storage, although it alone does not guarantee improved performance \cite{campaioli2018quantum,gyhm2024beneficial}.

Extensive research has focused on demonstrating how quantum effects can enhance both the stored energy and the charging power of these systems. Various protocols—from single-qubit setups to complex many-body systems—have been developed to highlight this quantum advantage \cite{le2018spin,julia2020bounds,rossini2020quantum,rosa2020ultra,ferraro2018high,binder2015quantacell,francica2024quantum}. Models such as the Dicke quantum battery show collective effects that yield modest speedups \cite{ferraro2018high,andolina2019quantum}, while qubit-based schemes exhibit power scaling linearly with the number of cells, indicating strong scalability \cite{binder2015quantacell}. Coherent charging methods outperform classical approaches, providing enhanced power and robustness against noise and decoherence \cite{salvia2023quantum}. Recent findings also highlight the importance of interaction topology, revealing that optimal charging in fermionic systems occurs when the connectivity matches the interaction order \cite{francica2024quantum}. Additionally, charging mechanisms rooted in the Jaynes–Cummings model show that non-Gaussian cavity states can reduce energy fluctuations, achieving near-perfect charging fidelity \cite{rinaldi2024reliable}. These theoretical advances are supported by experimental progress in superconducting circuits and trapped ion platforms \cite{quach2022superabsorption,hu2022optimal,dou2023superconducting,joshi2022experimental}. Techniques such as periodic modulation of spin systems \cite{mondal2022periodically, shukla2026many,Romero2026kicked} and controlled dissipation in open quantum systems have also been explored as means to enhance charging efficiency \cite{ghosh2021fast,liu2024better,santos2021quantum,bai2020floquet,quach2020using,carrega2020dissipative,zhao2021quantum,barra2019dissipative}. Among these, global charging—where all battery units are charged simultaneously—has been found to deliver the strongest quantum advantage, with power scaling quadratically with system size \cite{gyhm2022quantum}.

The investigation of many-body quantum systems remains central to understanding and optimizing quantum battery performance. Foundational studies have explored the XXZ spin chain with both local and non-local interactions \cite{le2018spin}, the Hubbard model, and bosonic and fermionic lattices \cite{konar2022quantum}. Spin-photon hybrid systems exhibit complex dynamical behaviors and rich interplay between components \cite{ferraro2018high,crescente2020ultrafast,dou2022cavity,rodriguez2024optimal,erdman2024reinforcement,zhang2023enhanced,salvia2023quantum,hogan2024quench}, while non-Hermitian models provide insight into energy transport and dissipation in open quantum systems \cite{konar2024quantum}. Comparisons with classical analogues demonstrate superior quantum retention and transfer efficiencies \cite{andolina2019quantum}, influenced by factors such as impurity-induced dimensional effects \cite{PhysRevA.105.022628}. Interestingly, environmental noise, typically viewed as detrimental, can under specific conditions accelerate the charging process \cite{ghosh2021fast}.

A key element underpinning many quantum battery protocols is the utilization of quantum correlations, such as entanglement \cite{hovhannisyan2013entanglement,campaioli2017enhancing,francica2017daemonic} and quantum coherence \cite{gumberidze2019measurement,gumberidze2022pairwise,tirone2025quantum,catalano2024frustrating}, which have been extensively investigated for their role in performance enhancement. Experimental realizations of these principles span platforms including superconducting circuits \cite{hu2022optimal}, semiconductor quantum dots \cite{maillette2023experimental}, and nuclear magnetic resonance setups \cite{joshi2022experimental}. The theoretical progress has been further extended to protocols for work extraction from partially characterized quantum energy sources, broadening the operational framework of quantum batteries \cite{joshi2024maximal}.

In contrast to conventional quantum-battery architectures, where the battery and charger are physically distinct systems, a natural question arises: {\it can the charging process be meaningfully realized within a single interacting many-body system, and how does the intrinsic dynamics of the battery itself influence this process?} In such a setting, the separation between storage and charging is no longer associated with different physical subsystems, but rather with different Hamiltonian components of the same system. This provides a distinct framework in which the energy stored in the battery is determined by one Hamiltonian component, while another, externally controlled component drives the system and modifies its energy with respect to the battery Hamiltonian.

Motivated by this perspective, we investigate a unified spin-based architecture in which the battery and charging roles are assigned to different Hamiltonian components of a single many-body system. To distinguish the controlled charging component from a conventional charger subsystem, we refer to it as the ``chargoid.'' This framework allows us to address a central question: {\it how does the intrinsic battery dynamics compete with the externally controlled charging dynamics, and under what conditions does this competition enhance or suppress energy storage?} To investigate this question, we introduce a tunable {\it countereffect} that externally controls the contribution of the battery Hamiltonian during the charging process. This provides a systematic means of tuning the balance between the intrinsic and charging dynamics and of examining its consequences for energy storage and charging power. Our results reveal that the intrinsic battery dynamics can oppose energy accumulation, while suitable control of its contribution can significantly improve the charging performance. These findings highlight the role of Hamiltonian competition in same-system quantum-battery architectures and provide a framework for exploring charging mechanisms beyond the conventional battery--charger paradigm.

We further examine the robustness of this control mechanism in an open-system setting by incorporating energy relaxation and pure dephasing within a Markovian framework. The influence of environmental coupling depends strongly on the underlying battery--chargoid architecture. While decoherence generally suppresses charging performance, we find that in certain interacting configurations it can instead enhance the energy-storage and energy-extraction dynamics. This qualitative behavior remains consistent both in the presence and absence of the battery countereffect. Moreover, when the intrinsic battery contribution is completely suppressed during the charging process, the energy-storage and energy-extraction performance is consistently enhanced across all considered configurations compared with their counterparts without the countereffect. These results demonstrate that externally controlled Hamiltonian dynamics can provide a robust means of improving energy storage and energy extraction even in the presence of environmental effects, highlighting the combined role of Hamiltonian control and environmental coupling in many-body quantum batteries.

\par
The manuscript is organized as follows: Section~\ref{model} introduces the setup involving the combined battery and chargoid system, and defines the key physical quantities used in the analysis. Section~\ref{result} presents the results for energy storage and charging power in the closed-system setting. Section~\ref{Open_system_dynamics} investigates the effects of environmental interactions on the charging and discharging dynamics within a Markovian open-system framework. Finally, section~\ref{conclusion} summarizes the main findings, and concludes the manuscript.
\vspace{0.25cm}
\section{Set-up}
\label{model}
\vspace{0.25cm}
We consider a quantum battery formed by a finite ensemble of $N$ spin-$1/2$ particles. The ensemble constitutes a single physical system whose energy structure and dynamics are determined by different contributions to its Hamiltonian. Depending on the configuration, the spins may be noninteracting or may exhibit many-body interactions, allowing us to investigate how the interaction structure influences energy storage and charging performance.

The battery and chargoid in our framework are not separate physical subsystems. Instead, they correspond to two distinct Hamiltonian contributions acting on the same Hilbert space,
\begin{equation}
\hat{H}(t)=\hat{H}_B+\hat{H}_C(t),
\label{eq:H_total}
\end{equation}
where $\hat{H}_B$ defines the energy observable of the battery, while $\hat{H}_C(t)$ represents the controllable charging Hamiltonian. Although both terms act on the same set of spins, they correspond to distinct physical contributions and can be controlled independently. Such a battery--chargoid construction, in which the charging dynamics are generated directly within the battery degrees of freedom, has also been considered in recent many-body quantum-battery studies~\cite{Shukla2026Collective}.

The charging protocol begins with the system prepared in an eigenstate of $\hat{H}_B$, which we take to be its ground state. At $t=0$, the charging Hamiltonian is activated, and the system evolves under the total Hamiltonian for a finite charging interval $0\leq t\leq t_{\mathrm{ch}}$. Energy transfer into the battery is enabled by the noncommutativity
\begin{equation}
[\hat{H}_B,\hat{H}_C]\neq0,
\end{equation}
which allows $\hat{H}_C$ to induce coherent transitions between eigenstates of $\hat{H}_B$ with different energies. The instantaneous battery energy is therefore
\begin{equation}
E_B(t)=\langle\hat{H}_B\rangle_t,
\end{equation}
and the energy stored relative to the initial state is
\begin{equation}
\Delta E(t)
=
\langle\hat{H}_B\rangle_t
-
\langle\hat{H}_B\rangle_0.
\label{eq:deltaE}
\end{equation}
The corresponding charging power characterizes the rate at which this energy is accumulated during the charging interval. At $t=t_{\mathrm{ch}}$, the charging Hamiltonian is switched off, and the resulting state defines the charged state of the battery.

The influence of interactions can be examined by varying their distribution between $\hat{H}_B$ and $\hat{H}_C$. Interactions incorporated into $\hat{H}_B$ modify the battery spectrum and the structure of the states in which energy is stored, whereas interactions incorporated into $\hat{H}_C$ modify the dynamical pathways through which energy is transferred. By comparing these complementary realizations within the same physical ensemble, we can distinguish the effects arising from interactions intrinsic to the battery from those generated by the charging dynamics. This provides a systematic framework for investigating how the interaction type and spatial range affect quantum-battery performance.

\subsection*{Battery--Chargoid Configurations}

To elucidate how interactions influence the charging dynamics, we consider three representative battery--chargoid configurations in which the interaction structure is distributed differently between $\hat{H}_B$ and $\hat{H}_C$. This allows us to separately examine the role of interactions in the energy-storage Hamiltonian and in the charging dynamics. We use a compact notation to characterize the interaction type and spatial range of each Hamiltonian. The superscript $0$ denotes a noninteracting component, while $\mathrm{I}$ and $\mathrm{XY}$ denote Ising and anisotropic XY interactions, respectively. The labels $\mathrm{NN}$ and $\mathrm{ATA}$ indicate nearest-neighbor and all-to-all interactions, respectively. This notation is used throughout to distinguish the different battery--chargoid configurations considered below.

\paragraph*{Configuration I: noninteracting battery and interacting chargoid.}
The battery consists of $N$ noninteracting spins, with its Hamiltonian given by
\begin{equation}
\hat H_B^{0}
=
h_z\sum_{j=1}^{N}\hat\sigma_j^z,
\label{eq:HB0}
\end{equation}
where $h_z$ sets the local energy scale and $\hat\sigma_j^\mu$ ($\mu=x,y,z$) denotes the Pauli operator acting on the $j$th spin. The eigenvalues of $\hat H_B^{0}$ range from $E_{\min}=-Nh_z$ to $E_{\max}=Nh_z$. The ground and highest-energy states are nondegenerate, while the intermediate energy levels possess the usual binomial degeneracies. Consequently, the maximum energy that can be stored relative to the ground state is
\begin{equation}
\Delta E_{\max}
=
E_{\max}-E_{\min}
=
2Nh_z.
\label{eq:DEmax_nonint}
\end{equation}

During the charging interval, an interaction Hamiltonian is externally activated, causing the initially noninteracting spins to undergo interacting many-body dynamics. Our charging protocol considers both Ising and anisotropic XY interactions, with each interaction type studied for both nearest-neighbor (NN) and all-to-all (ATA) coupling structures under periodic boundary conditions (PBCs).

For the Ising chargoid, the nearest-neighbor Hamiltonian is
\begin{equation}
\hat H_C^{\mathrm{I,NN}}
=
J\sum_{j=1}^{N}
\hat\sigma_j^x\hat\sigma_{j+1}^x,
\label{eq:HC_INN}
\end{equation}
where $J$ denotes the interaction strength. Periodic boundary conditions are imposed, $\hat\sigma_{N+1}^{x}=\hat\sigma_1^{x}.$ To investigate ATA interactions, we introduce the distance-dependent coupling $J_k=\frac{J}{2^{k-1}},$ and define the corresponding ATA interacting Ising chargoid as
\begin{equation}
\hat H_C^{\mathrm{I,ATA}}
=
\sum_{j=1}^{N}
\sum_{k=1}^{\mathcal K}
J_k\,
\hat\sigma_j^x\hat\sigma_{j+k}^x.
\label{eq:HC_IATA}
\end{equation}
Here, $\mathcal K$ specifies the maximum interaction range. Under periodic boundary conditions,
$\hat\sigma_{N+k}^{\mu}=\hat\sigma_k^{\mu}$, and we choose
\begin{equation}
\mathcal K=
\begin{cases}
(N-1)/2, & N\ \text{odd},\\[2mm]
N/2, & N\ \text{even},
\end{cases}
\end{equation}
so that each spin pair is included only once. The nearest-neighbor model is recovered by setting $\mathcal K=1$.

We also consider an anisotropic XY chargoid. Its nearest-neighbor form is
\begin{equation}
\hat H_C^{\mathrm{XY,NN}}
=
J\sum_{j=1}^{N}
\left[
(1+\Gamma)\hat\sigma_j^x\hat\sigma_{j+1}^x
+
(1-\Gamma)\hat\sigma_j^y\hat\sigma_{j+1}^y
\right],
\label{eq:HC_XYNN}
\end{equation}
where $\Gamma$ is the XY anisotropy parameter. The corresponding ATA interacting form is
\begin{equation}
\hat H_C^{\mathrm{XY,ATA}}
=
\sum_{j=1}^{N}
\sum_{k=1}^{\mathcal K}
J_k
\left[
(1+\Gamma)\hat\sigma_j^x\hat\sigma_{j+k}^x
+
(1-\Gamma)\hat\sigma_j^y\hat\sigma_{j+k}^y
\right].
\label{eq:HC_XYATA}
\end{equation}
Thus, the parameters $\mathrm{I}$ and $\mathrm{XY}$ specify the interaction symmetry, whereas $\mathrm{NN}$ and $\mathrm{ATA}$ specify its spatial range.

This configuration provides a setting in which the battery contains no intrinsic spin--spin interactions, while the charging dynamics is generated by an externally activated interacting chargoid. By comparing the NN and ATA coupling structures, we investigate the influence of interaction range on energy accumulation and charging power, while the comparison between Ising and XY chargoids allows us to assess the role of the interaction structure in the charging dynamics.

\paragraph*{Configuration II: interacting battery and noninteracting chargoid.}

We next reverse the roles of the two components and consider an interacting battery driven by a noninteracting chargoid. The battery may be described by either a nearest-neighbor Ising or XY Hamiltonian,
\begin{equation}
\hat H_B^{\mathrm{I,NN}}
=
J\sum_{j=1}^{N}
\hat\sigma_j^x\hat\sigma_{j+1}^x,
\label{eq:HB_INN}
\end{equation}
or
\begin{equation}
\hat H_B^{\mathrm{XY,NN}}
=
J\sum_{j=1}^{N}
\left[
(1+\Gamma)\hat\sigma_j^x\hat\sigma_{j+1}^x
+
(1-\Gamma)\hat\sigma_j^y\hat\sigma_{j+1}^y
\right].
\label{eq:HB_XYNN}
\end{equation}
The chargoid is taken to be noninteracting,
\begin{equation}
\hat H_C^{0}
=
h_z\sum_{j=1}^{N}\hat\sigma_j^z.
\label{eq:HC0}
\end{equation}
This reversed arrangement allows us to examine whether the effect of the counterterm persists when the intrinsic interaction is associated with the battery rather than the chargoid.

\paragraph*{Configuration III: interacting battery and interacting chargoid.}

Finally, we consider the case in which both the battery and chargoid possess nontrivial many-body interactions. As a representative configuration, we take an Ising battery with a nearest-neighbor interaction,
\begin{equation}
\hat H_B^{\mathrm{I,NN}}
=
J\sum_{j=1}^{N}
\hat\sigma_j^x\hat\sigma_{j+1}^x,
\label{eq:HB_I_configIII}
\end{equation}
and an anisotropic XY chargoid,
\begin{equation}
\hat H_C^{\mathrm{XY,NN}}
=
J\sum_{j=1}^{N}
\left[
(1+\Gamma)\hat\sigma_j^x\hat\sigma_{j+1}^x
+
(1-\Gamma)\hat\sigma_j^y\hat\sigma_{j+1}^y
\right].
\label{eq:HC_XY_configIII}
\end{equation}
We also investigate the reversed arrangement,
\begin{equation}
\hat H_B=
\hat H_B^{\mathrm{XY,NN}},
\qquad
\hat H_C=
\hat H_C^{\mathrm{I,NN}}.
\label{eq:config_III_reverse}
\end{equation}
These two realizations allow us to examine the effect of exchanging the roles of the Ising and XY interactions between the battery and chargoid. In this way, both components contribute nontrivially to the charging dynamics, providing a more general setting in which to study the countereffect.

\paragraph*{Countereffect of battery.} 
In the self-contained battery--chargoid framework, the battery Hamiltonian $\hat{H}_B$ remains an active part of the system dynamics while the chargoid is applied. Although $\hat{H}_C$ drives the system toward higher-energy states of $\hat{H}_B$, the simultaneous evolution generated by $\hat{H}_B$ can interfere with this charging dynamics. We refer to this intrinsic contribution of the battery Hamiltonian to the charging dynamics as the \emph{countereffect of the battery}. It represents the dynamical competition between the charging process generated by $\hat{H}_C$ and the intrinsic evolution generated by $\hat{H}_B$. 

To control this effect, we introduce an externally tunable counterterm proportional to the battery Hamiltonian i.e. $-\lambda\hat{H}_B$. Specifically, during the charging interval, we modify the total Hamiltonian according to
\begin{equation}
    \hat{H}_{\rm ch}(\lambda)
    =
    (1-\lambda)\hat{H}_B+\hat{H}_C,
    \qquad \lambda\in[0,1].
\end{equation}
Here, $\lambda$ controls the degree of compensation: $\lambda=0$ corresponds to the unmodified charging dynamics, while increasing $\lambda$ progressively suppresses the dynamical contribution of $\hat{H}_B$. At $\lambda=1$, the contribution of the intrinsic battery Hamiltonian is completely cancelled during charging, such that $\hat{H}_{\rm ch}=\hat{H}_C$. Importantly, $\hat{H}_B$ itself remains unchanged and continues to serve as the reference Hamiltonian for quantifying the energy stored in the battery.

The complete charging protocol can consequently be expressed as
\begin{equation}
    \hat{H}(t)=
    \begin{cases}
        \hat{H}_B, & t<0,\\[4pt]
        (1-\lambda)\hat{H}_B+\hat{H}_C,
        & 0\leq t\leq \tau_{\rm ch},\\[4pt]
        \hat{H}_B, & t>\tau_{\rm ch},
    \end{cases}
\end{equation}
where $\tau_{\rm ch}$ denotes the duration of the charging interval. During the charging interval, $\hat{H}_C$ drives the system away from its initial state, while the controlled contribution $(1-\lambda)\hat{H}_B$ determines the extent to which the intrinsic battery dynamics participate in the charging process. This protocol therefore provides a direct means of tuning the interplay between the intrinsic and chargoid-induced dynamics while keeping the definition of the battery energy fixed throughout the process.

\par

\par
\paragraph*{Energy Storage and Charging Power.} To analyze the counter effect of the battery during the charging process, we calculate two key metrics: storage energy and power. The energy stored in the battery during the charging process is defined as:
\begin{equation}
\Delta E = \mathrm{Tr}[\hat{\rho}(t) \hat{H}_B] - \mathrm{Tr}[\hat{\rho}(0) \hat{H}_B],
\end{equation}
where $\hat{\rho}(0) = |\psi_0\rangle \langle \psi_0|$, and $|\psi_0\rangle$ is the ground state of the battery Hamiltonian. The time evolution of the system's state is given as: $\ket{\psi(t)} = \hat{U}(t) \ket{\psi_0}$, where $\hat{U}(t) = e^{-i\hat{H}t}$ is the time-evolution operator.
\par
The charging power quantifies the rate at which energy is transferred into the system during the charging process and is given by:
\begin{equation}
P = \frac{\Delta E}{T}.
\end{equation}
Here, $P$ represents the average power over the entire charging duration, and $T$ denotes the total charging time. This formulation provides a measure of how efficiently energy is accumulated, offering insight into the performance and effectiveness of the energy storage process.

\paragraph*{Experimental realization.}
The charging protocol considered above can, in principle, be implemented
using programmable superconducting-qubit architectures, which provide
independent control over local qubit energies, two-qubit interactions, and
externally tunable Hamiltonian parameters. In particular, tunable
superconducting-qubit platforms allow the strength and effective sign of
qubit--qubit interactions to be controlled experimentally
\cite{Stehlik2021,Mitchell2021}, while microwave resonators can mediate
coherent interactions between spatially separated qubits
\cite{Majer2007}. These capabilities provide the essential ingredients
required to realize the different battery--chargoid cases considered in
this work.

For Configuration I, the noninteracting battery is described by  $ \hat H_B^{0}=h_z\sum_{j=1}^{N}\hat{\sigma}_j^z,$  where $h_z$ sets the local energy scale of each battery spin. In a
superconducting-qubit implementation, the corresponding qubit transition
frequencies can be controlled through external flux biases or microwave
fields \cite{Koch2007}. The chargoid can then be implemented by activating
programmable two-qubit interactions during the charging interval. As a representative example, consider the ATA Ising chargoid, $\hat H_C^{\mathrm I,\mathrm{ATA}}
    =
    \sum_{j=1}^{N}
    \sum_{k=1}^{\mathcal K}
   J_k
    \hat{\sigma}_j^x\hat{\sigma}_{j+k}^x ,$ where  $J_k=\frac{J}{2^{k-1}}$ denotes the coupling between spins separated by $k$ sites. Such
distance-dependent couplings can, in principle, be engineered using
tunable couplers or parametrically controlled interactions
\cite{Stehlik2021,Mitchell2021}. For the finite system sizes considered
here, only a finite number of coupling strengths is required. For example,
for $N=8$, the relevant couplings are
$J$, $J/2$, $J/4$, and $J/8$. Their implementation would require individual
calibration of the corresponding coupling channels and therefore
represents one of the principal experimental challenges of realizing the
ATA chargoid.

The same platform can also realize the NN Ising and XY chargoids by
restricting the interaction to neighboring qubits. Thus, the transition
between the ATA and NN limits corresponds experimentally to changing the
set of active coupling channels rather than introducing a fundamentally
different physical system. Similarly, the anisotropy parameter $\Gamma$
in the XY chargoid can be controlled by independently tuning the effective
$x$- and $y$-type interaction strengths.

The charging protocol involves three physically distinct contributions:
the intrinsic battery Hamiltonian, the chargoid Hamiltonian, and an
externally controlled countereffect. The charging Hamiltonian can
therefore be written as
\begin{equation}
    \hat H_{\mathrm{ch}}
    =
    \underbrace{\hat H_B}_{\text{intrinsic battery}}
    +
    \underbrace{\hat H_C}_{\text{chargoid}}
    +
    \underbrace{\hat H_{\mathrm{counter}}}_{\text{external control}},
\end{equation}
with  $\hat H_{\mathrm{counter}}
    =
    -\lambda \hat H_B .$
Equivalently, $\hat H_{\mathrm{ch}}
    =
    (1-\lambda)\hat H_B+\hat H_C.$ The three-term representation is retained here to emphasize the distinct
physical roles of the two intrinsic Hamiltonians and the externally
controlled counterterm.

For the noninteracting battery considered in Configuration I, the counterterm is
given explicitly by
$ \hat H_{\mathrm{counter}}
    =
    -\lambda h_z
    \sum_{j=1}^{N}\hat{\sigma}_j^z .$
This term can be implemented by dynamically tuning the local qubit energy
splittings through external flux or microwave control \cite{Koch2007}.
Thus, $\lambda$ serves as a tunable control parameter that determines the
strength of the externally applied countereffect. Importantly, this
external control does not redefine the battery Hamiltonian. The operator
$\hat H_B$ remains the fixed reference Hamiltonian with respect to which
the stored energy and ergotropy are evaluated throughout the protocol.

Configuration II and III can be realized within the same experimental framework.
For Configuration II, the interacting battery--noninteracting chargoid combination
is obtained by activating the appropriate NN Ising or XY couplings as part
of $\hat H_B$ while using the local $z$-field as the chargoid Hamiltonian. For Configuration III, both the battery and chargoid are interacting, with their respective NN coupling structures and interaction types specified independently. Hence, the three cases do not require fundamentally different hardware; they correspond to different choices of
which programmable Hamiltonian terms are assigned to the battery and
chargoid roles.

\section{Closed-System Dynamics}
\label{result}
\vspace{0.25cm}

We first examine the charging dynamics in the absence of environmental
effects, with the evolution governed solely by the controlled
Hamiltonian of the battery--chargoid system. Our focus is on how the
intrinsic battery dynamics influences energy transfer when the chargoid
Hamiltonian is externally activated. Using the three battery--chargoid
configurations introduced above, we systematically investigate how their
different interaction structures affect energy storage, energy extraction,
and charging performance. We begin with the noninteracting
battery--interacting chargoid configuration.

\subsection{Configuration I: Non-interacting battery and interacting chargoid}
\label{nonint_battery}

We consider the noninteracting battery--interacting chargoid configuration introduced above. The battery is described by $\hat H_B^{0}$ [Eq.~\ref{eq:HB0}], while the chargoid is taken to be either an Ising or an anisotropic XY system with nearest-neighbor (NN) or all-to-all (ATA) interactions. Starting from the ground state of the battery Hamiltonian, we examine how the externally controlled countereffect, characterized by $\lambda$, modifies the charging dynamics. For each chargoid structure, we investigate the evolution over $\lambda\in[0,1]$, focusing on the resulting stored energy and charging power.

\begin{figure*}
\includegraphics[width=0.32\linewidth,height=0.28\linewidth]{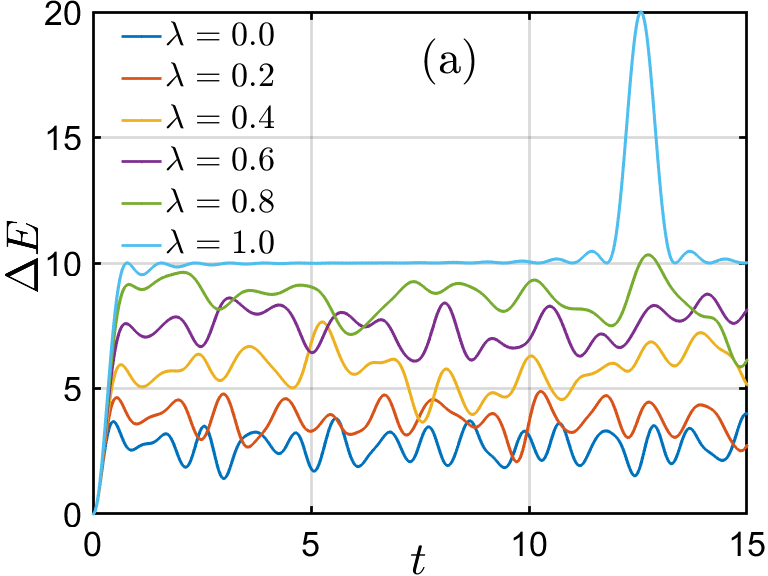}
\includegraphics[width=0.32\linewidth,height=0.28\linewidth]{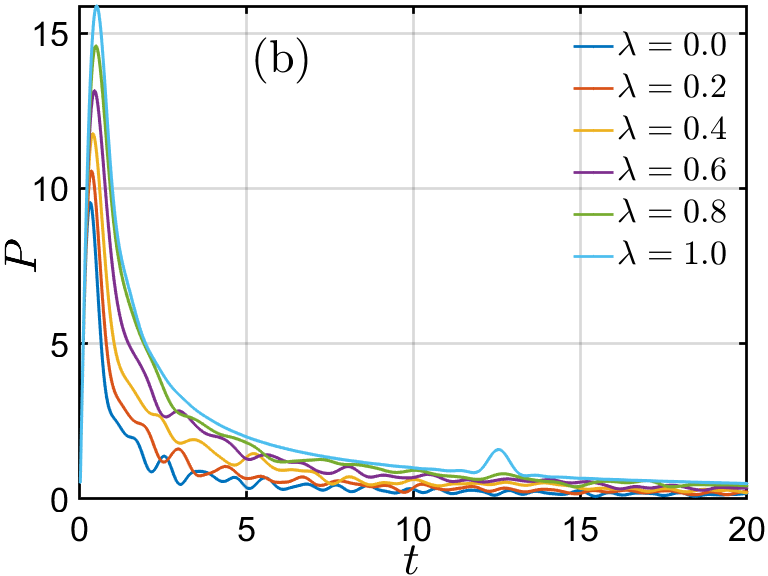}\\
\includegraphics[width=0.32\linewidth,height=0.28\linewidth]{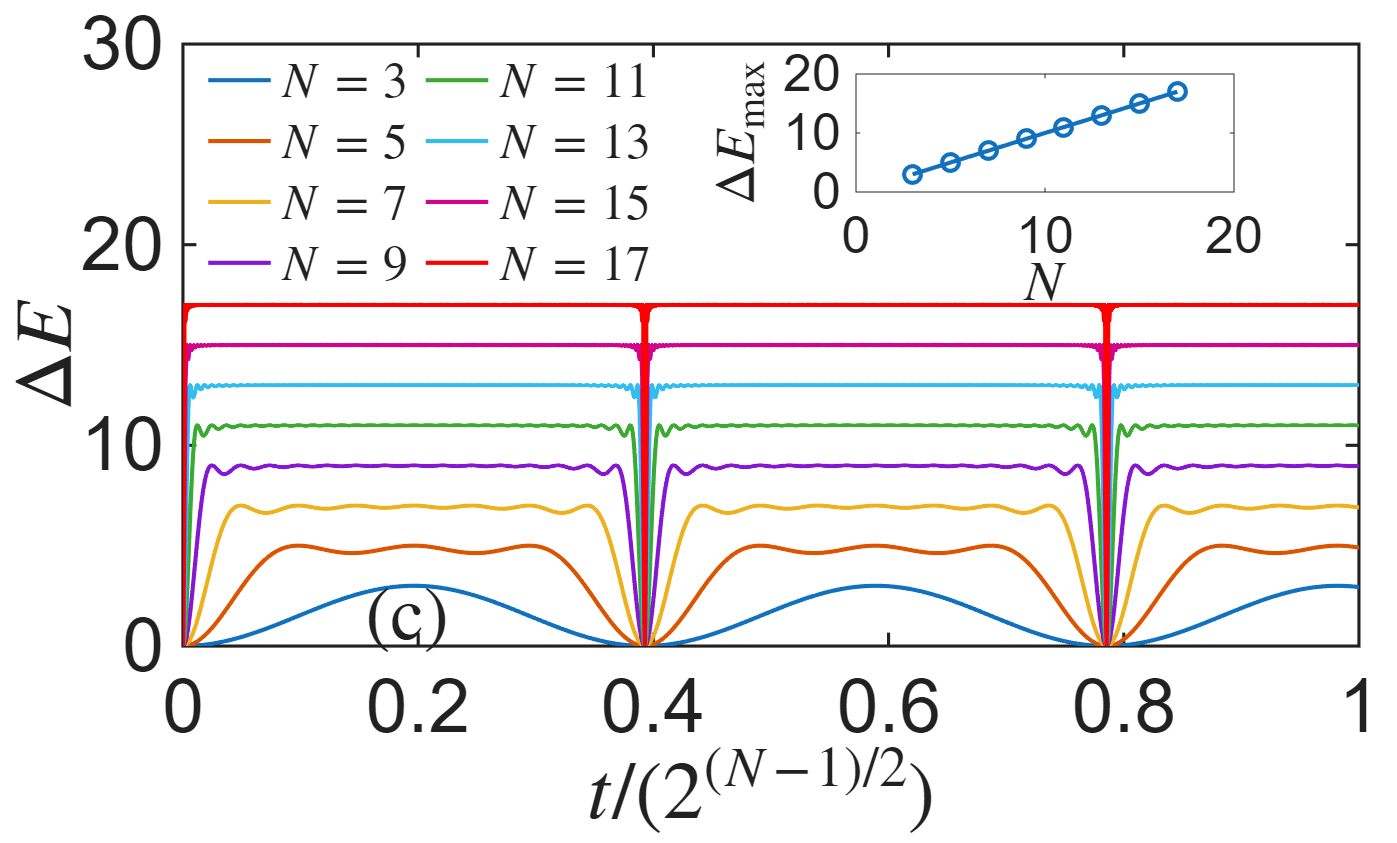}
\includegraphics[width=0.32\linewidth,height=0.28\linewidth]{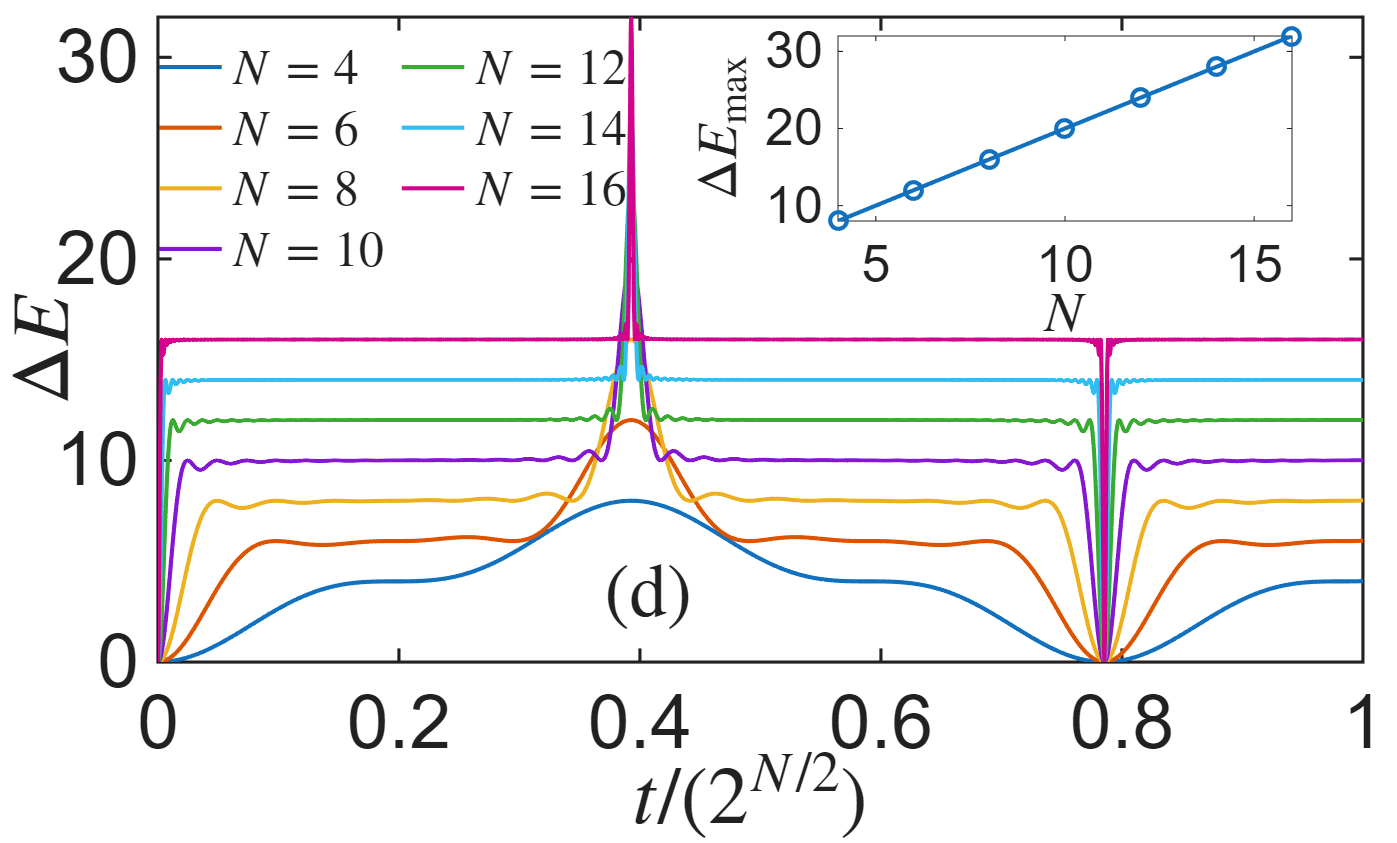}
\includegraphics[width=0.32\linewidth,height=0.28\linewidth]{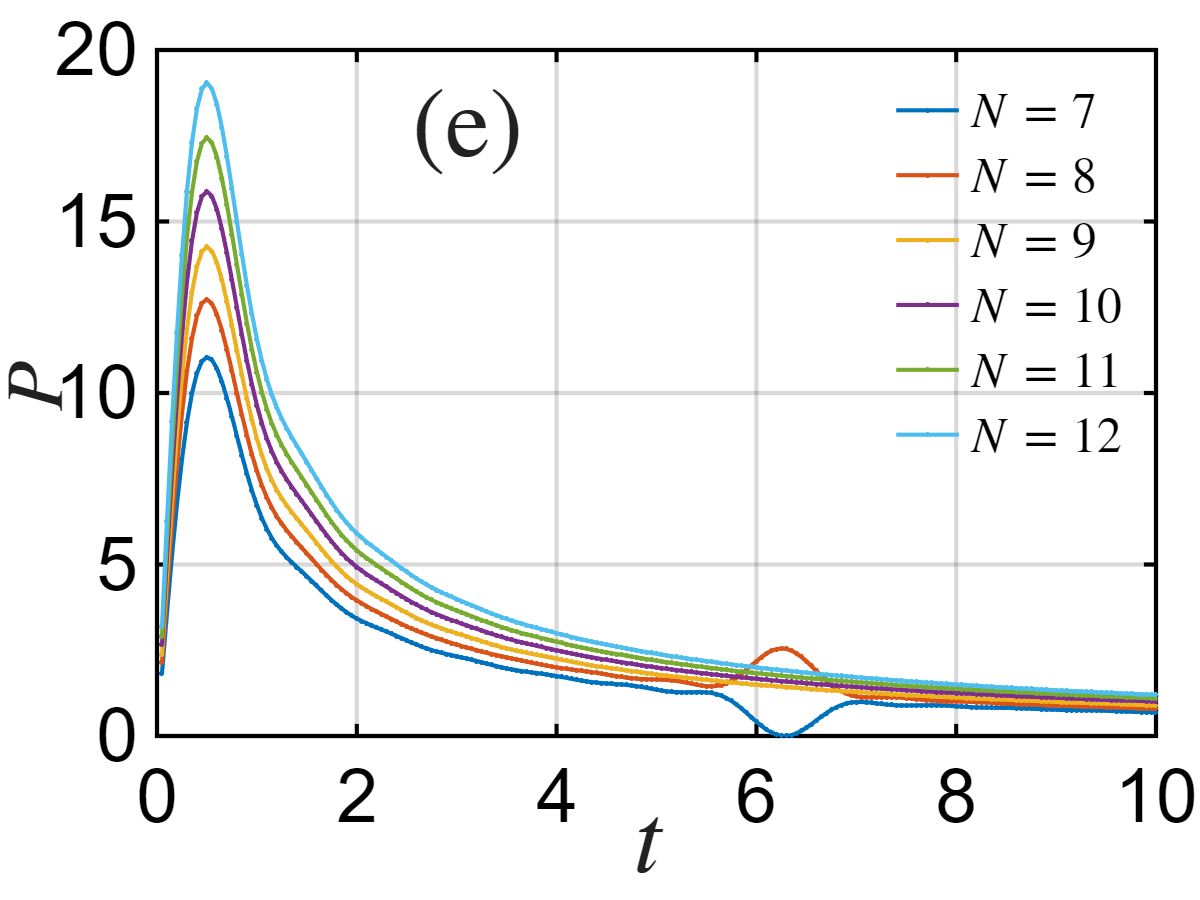}
     \caption{Non-interacting battery charged by an ATA interacting Ising spin chargoid. (a) Stored energy $\Delta E$ and (b) charging power $P$ as functions of time $t$ for different values of the countereffect strength $\lambda$, with system size fixed at $N = 10$. Panels (c) and (d) show the time evolution of the stored energy $\Delta E$ as a function of the rescaled time $t/2^{(N-1)/2}$ for odd $N$ and $t/2^{N/2}$ for even $N$, respectively, at fixed $\lambda=1$. The insets show the corresponding maximum stored energy $\Delta E_{\max}$ as a function of the system size $N$. (e) Charging power $P$ as a function of time $t$ for different system sizes $N$. All simulations are performed with interaction strength $J = 1$, external field $h_z = 1$, and periodic boundary conditions.}
\label{Ising_long_fig_v1}
\end{figure*}

We begin with the ATA Ising chargoid and examine the time evolution of the energy stored in the noninteracting battery. For $0\leq\lambda<1$, the stored energy initially increases and subsequently undergoes persistent oscillations around a characteristic value, with both the amplitude and frequency depending on $\lambda$ [Fig.~\ref{Ising_long_fig_v1}(a)]. A qualitatively different behavior is observed for $\lambda=1$. In this case, the stored energy initially increases toward a nearly saturated value and subsequently exhibits a sharp increase at a characteristic time, at which the battery reaches its maximum possible energy gain,
$\Delta E_{\rm max}=2Nh_z$. This corresponds to a complete inversion of the battery, from the ground state to the highest-energy state of $\hat H_B^{0}$. Thus, complete suppression of the intrinsic battery contribution produces a pronounced enhancement of energy storage for the ATA Ising chargoid, as shown in Fig.~\ref{Ising_long_fig_v1}(a).
\par
The storage power exhibits a similar behavior for all values of \(\lambda\), increases, reaches its maximum value at a specific time, and then starts to decrease. As the value of $\lambda$ increases, the maximum storage power also increases. [Fig.~\ref{Ising_long_fig_v1}(b)]. 

\par

Furthermore, we investigate the system-size dependence of the stored energy at
$\lambda=1$. The ATA Ising chargoid exhibits a pronounced odd--even dependence
in the charging dynamics. Before discussing the maximum stored energy, we note
that the characteristic timescale of the charging dynamics increases rapidly
with the system size. This increase originates from the exponentially
decreasing interaction strength with distance, which introduces progressively
slower dynamical timescales as $N$ increases. To compare different system
sizes on a common timescale, we therefore plot the stored energy using the
rescaled time $t/2^{(N-1)/2}$ for odd $N$ and $t/2^{N/2}$ for even $N$, as shown
in Fig.~\ref{Ising_long_fig_v1}(c) and Fig.~\ref{Ising_long_fig_v1}(d),
respectively. With this rescaling, the charging dynamics for different system
sizes exhibits a common periodic structure within each parity sector,
demonstrating that the apparent increase in the oscillation period is governed
by a systematic system-size-dependent timescale. In particular, increasing
$N$ by two spins approximately doubles the characteristic recurrence time
within a given parity sector.

The maximum stored energy also displays a clear parity dependence. For odd
$N$, the maximum stored energy reaches $\Delta E_{\max}=Nh_z$, corresponding
to half of the full energy range of the battery. In contrast, even $N$ allows
complete population inversion and hence reaches the upper bound
$\Delta E_{\max}=2Nh_z$. This difference originates from the interaction
structure of the ATA chargoid under periodic boundary conditions: for even
$N$, the presence of a unique antipodal interaction introduces an unsquared
contribution to the spin polarization, which can change sign and thereby
allows complete inversion. For odd $N$, the interaction distances occur in
pairs, leading to squared contributions that restrict the polarization and
prevent complete inversion. The systematic increase of the maximum stored
energy with $N$ is shown in the insets of Figs.~\ref{Ising_long_fig_v1}(c)
and \ref{Ising_long_fig_v1}(d) for odd and even system sizes, respectively.
The analytical origin of the parity dependence, the system-size scaling of
the characteristic timescale, and the corresponding large-$N$ behavior are
derived in Appendix~\ref{sec:odd_even_analytical}.

In the thermodynamic limit, the recurrence time diverges exponentially with
system size. Consequently, the periodic finite-size behavior, including the
complete-inversion maximum found for finite even $N$, does not remain a
finite-time feature as $N\rightarrow\infty$. Instead, the stored-energy
density approaches a common thermodynamic-limit behavior and tends to $h_z$
at long times. Thus, the $2Nh_z$ maximum observed for finite even systems is a
finite-size effect associated with the exponentially long time required to
reach complete inversion. Further details of the thermodynamic-limit behavior
are given in Appendix~\ref{sec:odd_even_analytical}.

\par
The storage power of the battery is analyzed as a function of the system size with a fixed value of \(\lambda = 1\). It is observed that while the storage power depends on the system size, it does not exhibit the odd-even effect. For all system sizes, the storage power increases over time, reaches a maximum value, and then begins to decline. The maximum storage energy increases with the growth of the system size. [Fig.~\ref{Ising_long_fig_v1}(e)].

\subsubsection*{Optimum storage energy and charging power}

We next compare the charging performance for the different interacting
chargoid structures considered in this configuration. In particular, we
analyze the maximum stored energy and the corresponding charging power as
functions of the countereffect parameter $\lambda$ and the system size.
Both Ising and anisotropic XY chargoids are considered with nearest-neighbor
(NN) and all-to-all (ATA) interaction structures, as defined in
Eqs.~(\ref{eq:HC_INN})--(\ref{eq:HC_XYATA}). This comparison enables us to
examine how the interaction type and interaction range of the chargoid
influence the energy-storage and charging performance of the noninteracting
battery.

\begin{figure}
\includegraphics[width=0.48\linewidth,height=0.4\linewidth]{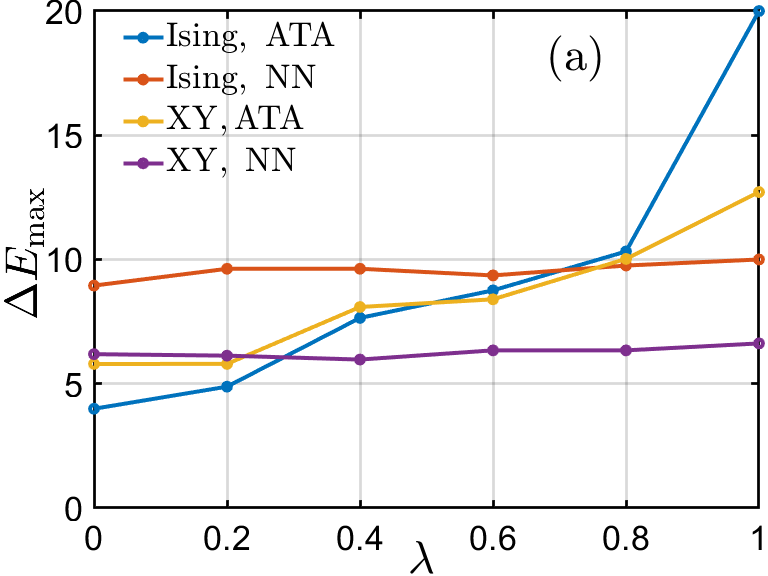}
\includegraphics[width=0.49\linewidth,height=0.4\linewidth]{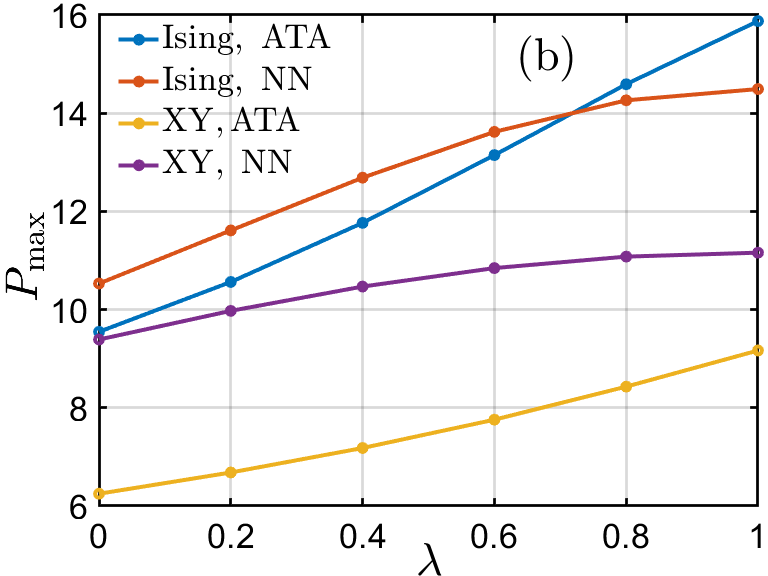}
\includegraphics[width=0.48\linewidth,height=0.4\linewidth]{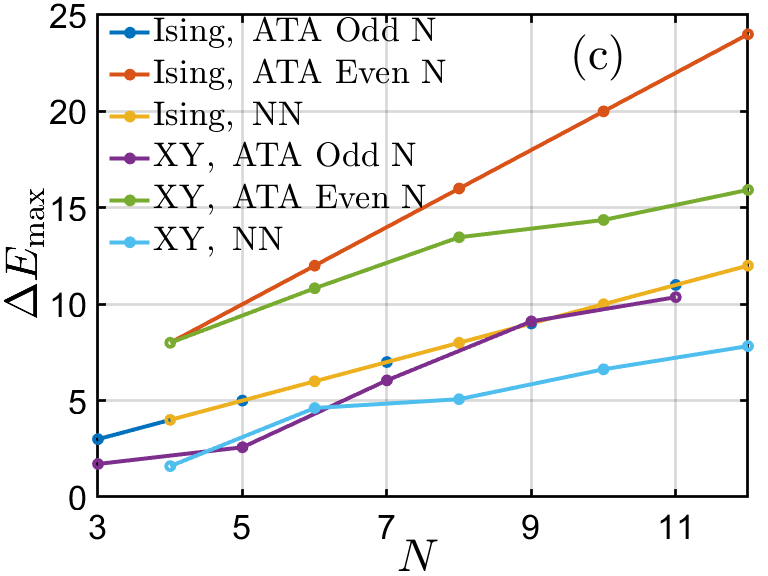}
\includegraphics[width=0.49\linewidth,height=0.4\linewidth]{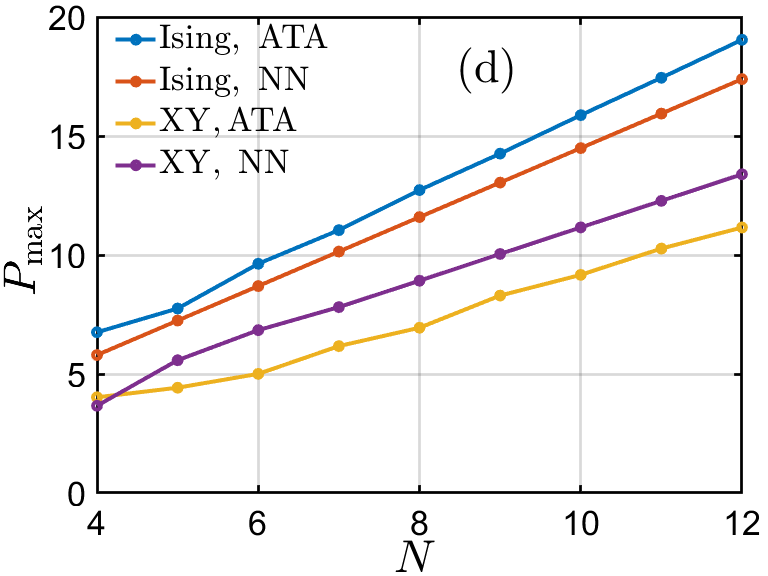}
\caption{(a) Maximum stored energy and (b) maximum power as functions of the battery’s countereffect strength $\lambda$ for a non-interacting battery charged by four types of chargoids: Ising ATA, Ising NN, XY ATA, and XY NN, with system size $N = 10$. (c) Maximum stored energy and (d) maximum power as functions of system size $N$ for the same four chargoid types, with the countereffect strength fixed at $\lambda = 1$. The parameters are $J = 1$, $h_z = 1$, and $\Gamma = 0.5$.}
\label{Max_E_P}
\end{figure}

The maximum stored energy increases monotonically with the countereffect
strength for all four chargoid structures and reaches its highest value at
$\lambda=1$, as shown in Fig.~\ref{Max_E_P}(a). A clear dependence on the
interaction range is observed for both Ising and anisotropic XY chargoids.
For the Ising ATA chargoid, the battery reaches the full energy range,
$\Delta E_{\max}=2h_zN$, whereas the corresponding NN chargoid is limited to
$\Delta E_{\max}=h_zN$. The XY ATA chargoid also provides an enhancement over
the NN case, with $\Delta E_{\max}>h_zN$, although it does not reach the
upper bound $2h_zN$. In contrast, the XY NN chargoid remains below
$\Delta E_{\max}=h_zN$. Thus, extending the interaction range from NN to ATA
generally enhances the achievable energy storage, with the enhancement being
particularly pronounced for the Ising chargoid. The common optimum at
$\lambda=1$ indicates that fully suppressing the intrinsic battery
contribution provides the most favorable condition for energy accumulation
among the cases considered.
\par
The corresponding maximum charging power is presented in Fig.~\ref{Max_E_P}(b). Similar to the stored energy, the maximum charging power increases with $\lambda$ and attains its largest value at $\lambda=1$ for all considered chargoid structures. The charging power also depends strongly on the interaction structure. The Ising chargoids yield higher maximum charging power than the corresponding XY chargoids for both ATA and NN interactions, indicating a faster energy-transfer dynamics for the Ising case. Moreover, the ATA interactions generally provide a higher charging power than their NN counterparts, with the enhancement being particularly evident for the Ising chargoid. Thus, the interaction type and range of the chargoid influence not only the maximum amount of energy that can be stored but also the rate at which this energy is accumulated.
\par
Furthermore, we examine the system-size dependence of the maximum stored
energy and charging power at the optimal countereffect strength,
$\lambda=1$. For the ATA chargoids, the stored energy exhibits a pronounced
parity dependence. In the Ising ATA case, even system sizes can reach the
upper bound $\Delta E_{\max}=2Nh_z$, corresponding to complete inversion of
the battery, whereas for odd $N$ the maximum is limited to
$\Delta E_{\max}=Nh_z$. This parity-dependent behavior persists for larger
finite systems. However, the complete inversion observed for even $N$ is a
finite-size effect: the recurrence time required to reach
$\Delta E_{\max}=2Nh_z$ increases exponentially with $N$ and diverges in the
thermodynamic limit. Consequently, at any fixed finite time, the
stored-energy density approaches the smooth thermodynamic-limit value
$h_z$, rather than the finite-size maximum $2h_z$. The analytical origin of
this odd--even effect and its system-size dependence are discussed in
Appendix~\ref{sec:odd_even_analytical}.
A similar parity dependence is observed for the XY ATA chargoid, although the
enhancement is weaker: the maximum stored energy remains below $Nh_z$ for odd
$N$ and exceeds $Nh_z$ for even $N$ [Fig.~\ref{Max_E_P}(c)].

The NN chargoids exhibit a simple system-size dependence. Since only
nearest-neighbor interactions are present, the charging dynamics is governed
by a single interaction scale $J$, independent of $N$. For the Ising NN
chargoid, the stored energy is $\Delta E_N^{\mathrm{NN}}(t)
=
Nh_z\sin^2(2Jt),$ as derived in Appendix~\ref{NN_ana}. Thus, the maximum stored energy is
$\Delta E_{\max}^{\mathrm{NN}}=Nh_z$, independent of the parity of $N$,
while the charging period remains fixed at
$T_{\mathrm{NN}}=\pi/(2J)$. Increasing the system size therefore increases
the total stored energy extensively without changing the characteristic
charging timescale. In the thermodynamic limit, the stored-energy density
remains finite and reaches $\lim_{N\rightarrow\infty}
\frac{\Delta E_{\max}^{\mathrm{NN}}}{N}
=h_z,$
while the charging dynamics retains the same period $T_{\mathrm{NN}}$.
The corresponding system-size dependence of the maximum stored energy is
shown in Fig.~\ref{Max_E_P}(c), with the detailed discussion given in
Appendix~\ref{NN_ana}.

The maximum charging power likewise increases linearly with the system size for all considered chargoid configurations [Fig.~\ref{Max_E_P}(d)]. Among them, the Ising ATA chargoid exhibits the largest charging power, reflecting the combined enhancement arising from the interaction structure and all-to-all couplings. The XY ATA configuration, in contrast, yields the smallest power, indicating a comparatively slower energy-transfer process. Together, these results demonstrate that both the parity of the system size and the interaction structure of the chargoid can substantially influence the charging performance.

\label{int_battery}
 \begin{figure}
\includegraphics[width=0.48\linewidth,height=0.4\linewidth]{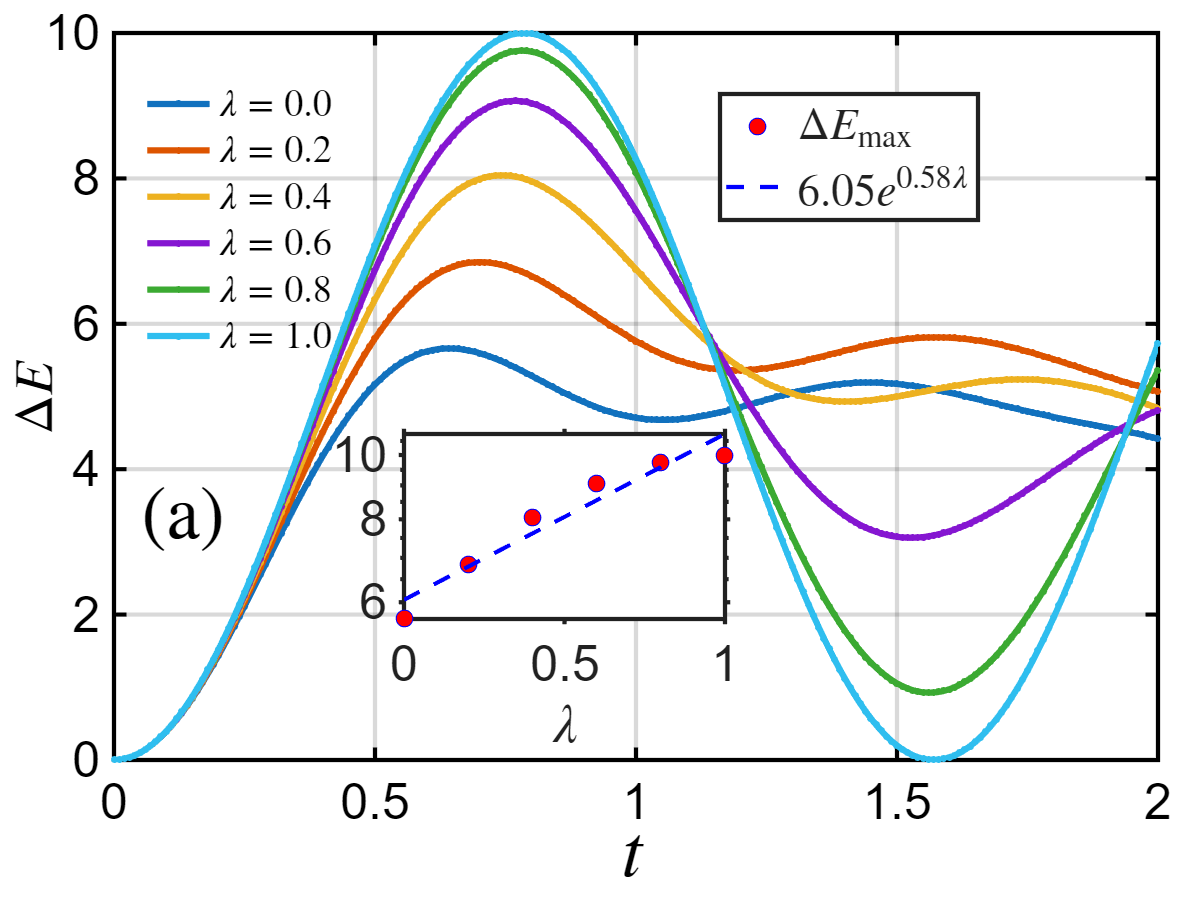}
\includegraphics[width=0.49\linewidth,height=0.4\linewidth]{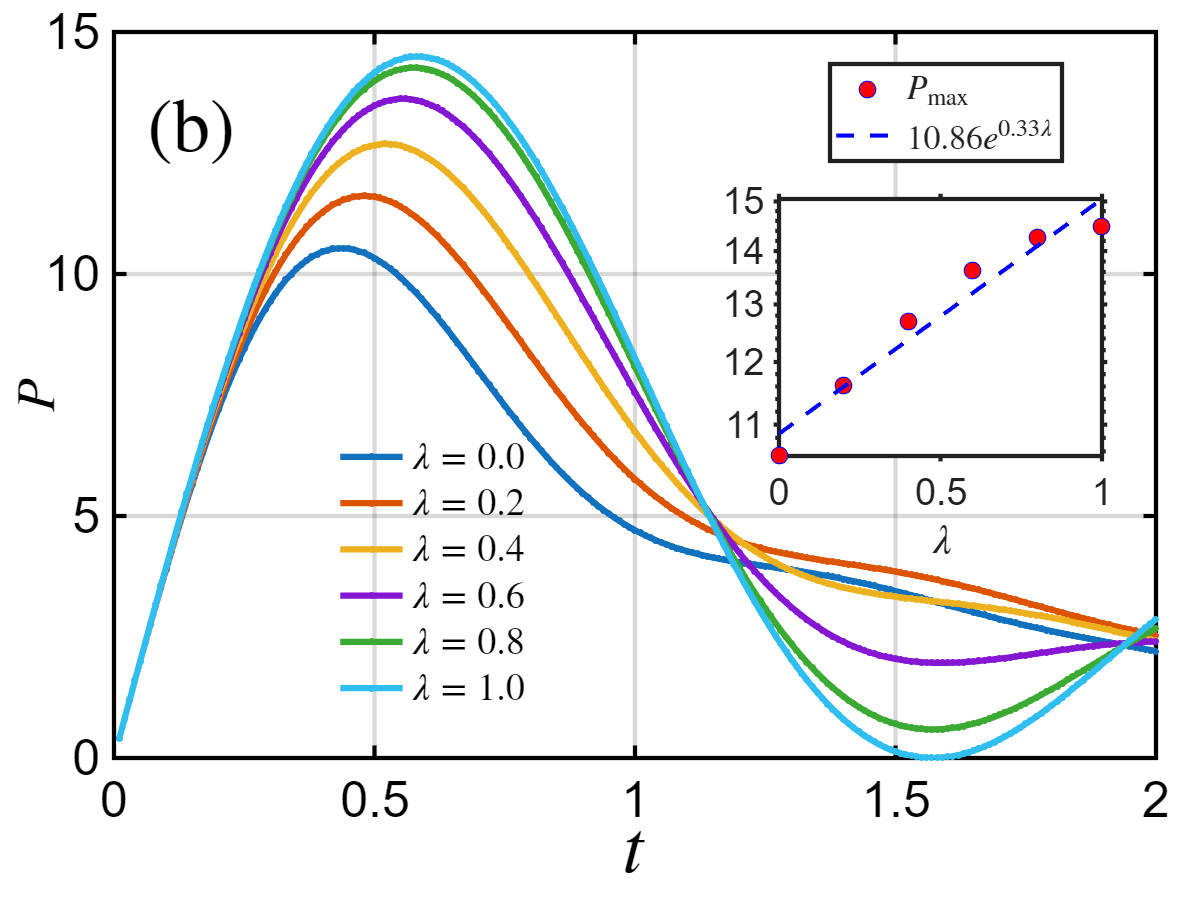}
\includegraphics[width=0.48\linewidth,height=0.4\linewidth]{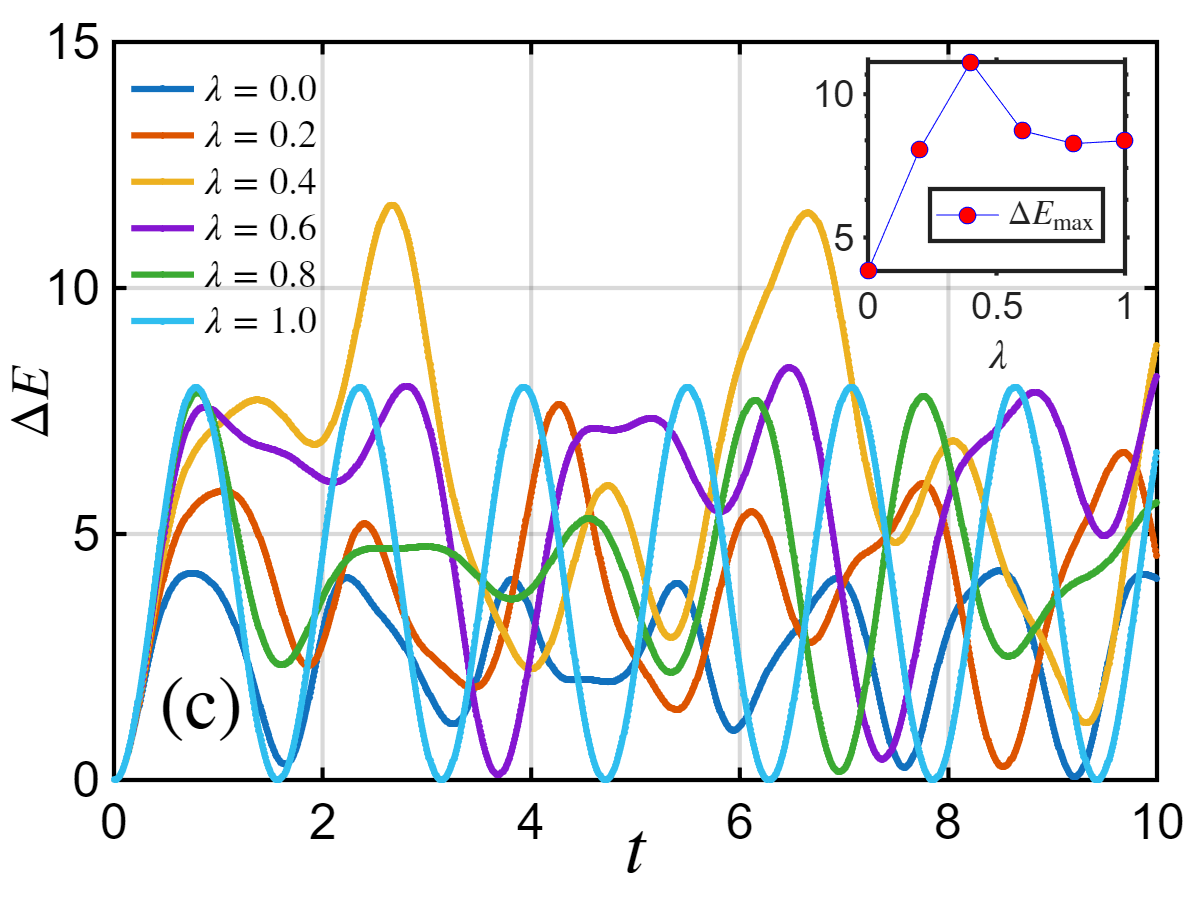}
\includegraphics[width=0.49\linewidth,height=0.4\linewidth]{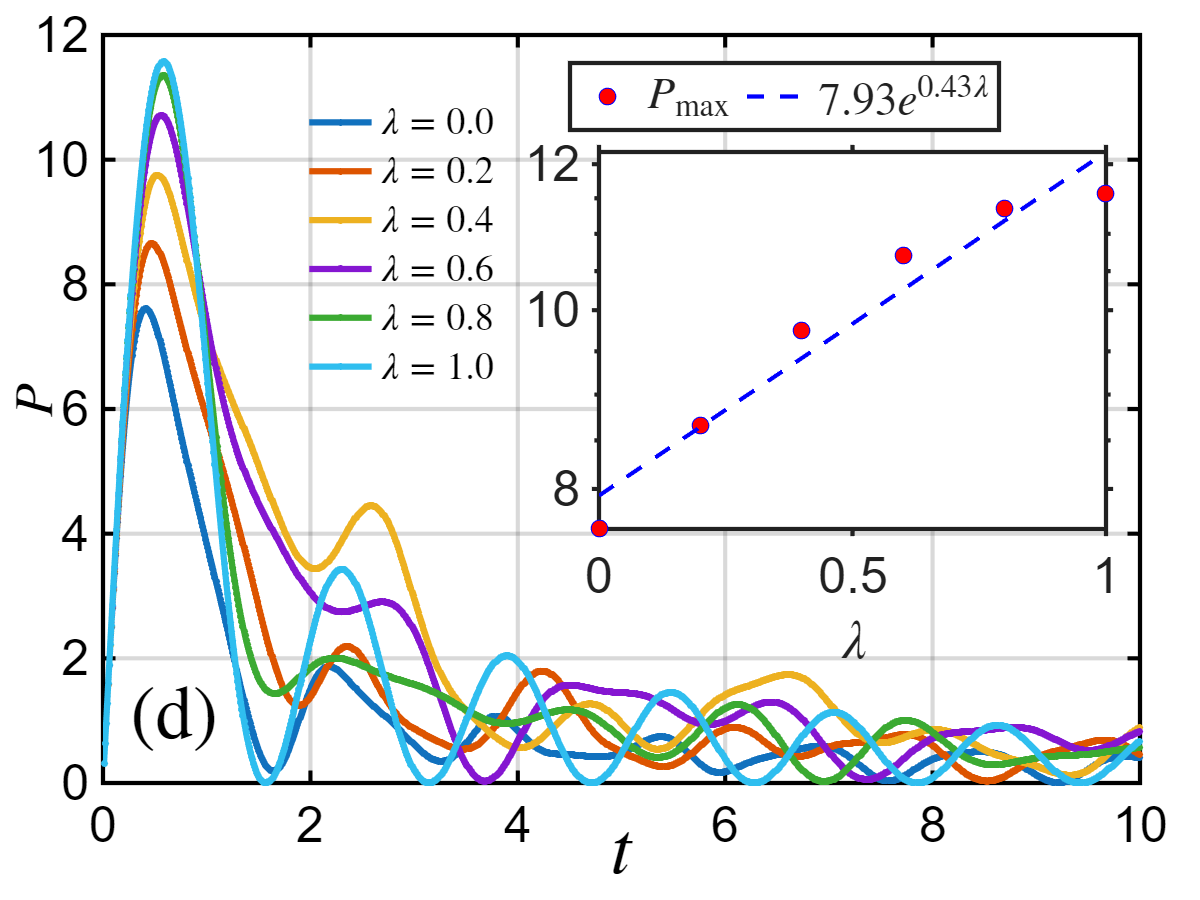}
\caption{\textbf{Interacting battery and non-interacting chargoid.}
(a) Stored energy $\Delta E$ and (b) charging power $P$ as functions of time $t$ for an interacting battery described by a NN Ising Hamiltonian and charged by an external field corresponding to a non-interacting chargoid. (c) Stored energy $\Delta E$ and (d) charging power $P$ for an interacting battery described by a NN XY Hamiltonian and charged by the same non-interacting chargoid. The insets in the energy plots show the maximum stored energy as a function of the counter-effect parameter $\lambda$, while those in the power plots show the corresponding maximum charging power. The red data points represent the numerical results, and the blue dashed lines indicate the corresponding fits. The system parameters are $N=10$, $J=1$, $h_z=1$, and $\Gamma=0.5$, with periodic boundary conditions.}
\label{Ising_B}
\end{figure}

\subsection{Configuration II: Interacting battery and non-interacting chargoid}
\label{int_battery_nonint_chargoid}

We next consider the reversed setup, in which the battery is interacting
while the chargoid is non-interacting. The battery is taken to be either
an NN Ising or an NN anisotropic XY system, described by
$\hat H_B^{\mathrm{I,NN}}$ [Eq.~\ref{eq:HB_INN}] and $\hat H_B^{\mathrm{XY,NN}}$ [Eq.~\ref{eq:HB_XYNN}],
respectively, while the chargoid is described by the non-interacting
Hamiltonian $\hat H_C^{0}$ [Eq.~\ref{eq:HC0}]. Starting from the ground state of the
corresponding battery Hamiltonian, we investigate the charging dynamics
for $\lambda\in[0,1]$ and analyze the stored energy and charging power.
In this configuration, the countereffect acts by reducing the
contribution of the intrinsic battery interaction to the charging
dynamics. We first examine the NN Ising battery and subsequently
consider the NN XY battery, allowing us to assess how the internal
interaction structure of the battery affects the response to the
countereffect.

\par

For the NN Ising battery, the stored energy and charging power are
calculated for different values of $\lambda$. As shown in
Fig.~\ref{Ising_B}(a), increasing the countereffect strength leads to a
systematic enhancement of the maximum stored energy. The corresponding
dependence of $\Delta E_{\rm max}$ on $\lambda$ is shown in the inset,
where the numerical results are well described by $\Delta E_{\rm max}\approx 6.05 e^{0.58\lambda}.$ A similar enhancement is observed for the charging power. Figure
~\ref{Ising_B}(b) shows that the maximum charging power increases
monotonically with $\lambda$, with the dependence approximately
described by $ P_{\rm max}\approx 10.8 e^{0.33\lambda},$ as shown in the inset. Thus, for the NN Ising battery, progressively
suppressing the intrinsic battery interaction enhances both the maximum
stored energy and the maximum charging power.

We next turn to the NN XY battery driven by the same non-interacting
chargoid. In contrast to the Ising case, the stored energy exhibits a
nonmonotonic dependence on the countereffect strength. As shown in
Fig.~\ref{Ising_B}(c), increasing $\lambda$ initially enhances
$\Delta E_{\rm max}$, but beyond approximately $\lambda\simeq0.4$ the
maximum stored energy decreases. Moreover, the maximum is attained at a
later time as the charging dynamics evolves. The resulting dependence
of $\Delta E_{\rm max}$ on $\lambda$, shown in the inset, therefore
displays an initial enhancement followed by a suppression at larger
values of $\lambda$.

This behavior demonstrates that the effect of the countereffect depends
on the internal structure of the battery Hamiltonian. In particular,
suppressing the intrinsic interaction is not universally beneficial for
energy storage: in the XY case, an intermediate countereffect strength
provides a more favorable balance between the intrinsic battery
dynamics and the charging field. The delayed occurrence of the maximum
stored energy further indicates that the relevant charging timescale is
also modified by the countereffect.

The charging power, however, behaves differently from the stored energy.
As shown in Fig.~\ref{Ising_B}(d), the maximum power increases
monotonically with $\lambda$. The dependence is approximately
exponential and is well described by $    P_{\rm max}\approx 7.93 e^{0.43\lambda},$ as shown in the inset. Hence, although the stored energy of the XY
battery exhibits a nonmonotonic response, its maximum charging power
continues to increase with the strength of the countereffect.

\begin{figure}
    \centering
\includegraphics[width=0.49\linewidth,height=0.5\linewidth]{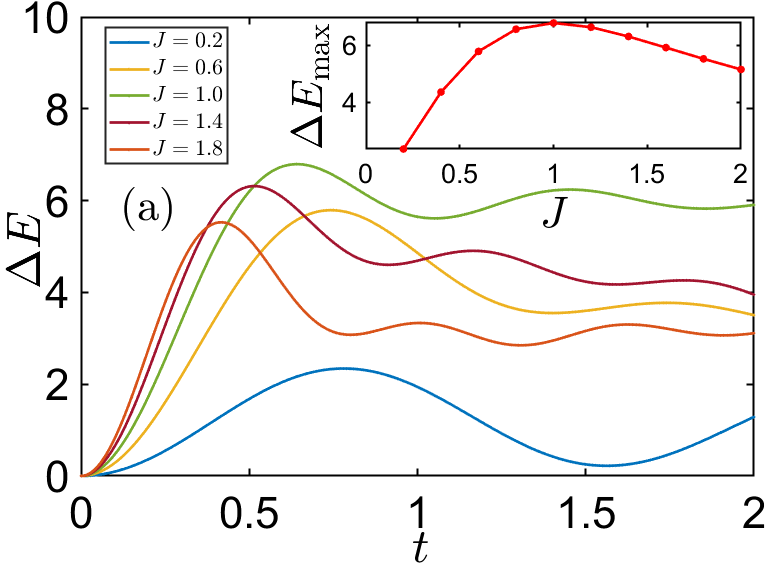}
\includegraphics[width=0.49\linewidth,height=0.5\linewidth]{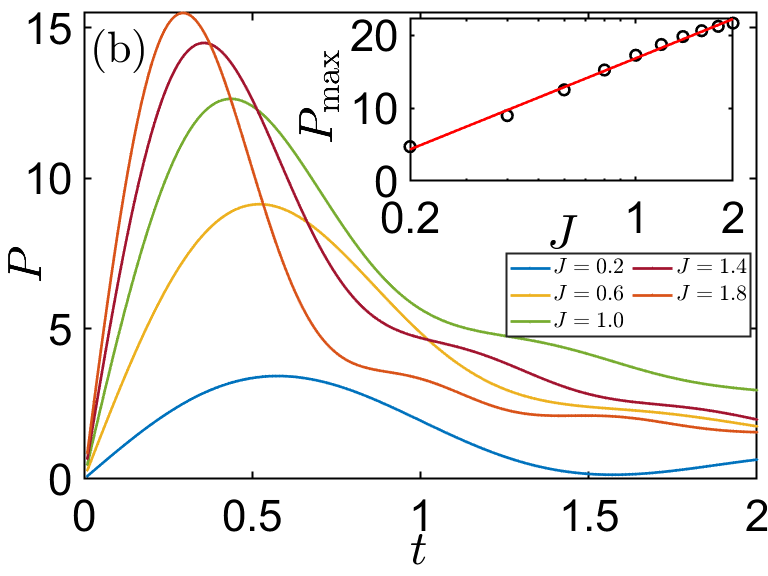}
     \caption{Storage energy $\Delta E$ and power $P$ of the NN Ising spin system as a quantum battery, with non-interacting spins as the chargoid, shown as a function of time for different interaction strengths $J$ (fixing $\lambda=0$). The inset shows the maximum storage energy and power as $J$ is varied. The maximum power scales approximately logarithmically with interaction strengh , following $P_{\rm max} \approx 17.91 \log_{10}(J) + 16.86$.}
 \label{J_Insing_B}
\end{figure}

\paragraph*{Dependence on the Interaction Strength.} To further isolate the role of the intrinsic battery interaction, we
next examine the dependence of the charging performance on the
interaction strength $J$ in the absence of the countereffect. For this analysis, the battery is taken to be 
an NN Ising system, described by
$\hat H_B^{\mathrm{I,NN}}$ [Eq.~\ref{eq:HB_INN}], while the chargoid is described by the non-interacting
Hamiltonian $\hat H_C^{0}$ [Eq.~\ref{eq:HC0}]. Figure
~\ref{J_Insing_B}(a) shows the time-dependent stored energy for
different values of $J$. The stored energy initially increases,
reaches a maximum, and subsequently decreases. Increasing $J$ also
reduces the time required to reach the maximum stored energy. The
corresponding maximum value, shown in the inset, increases with $J$
up to $J=1$ and decreases for stronger interactions. Thus, the optimal
storage is obtained near $ J=h_z=1,$ where the intrinsic interaction scale of the battery matches the local
energy scale of the non-interacting chargoid.

The charging power exhibits a similar time-dependent behavior. As
shown in Fig.~\ref{J_Insing_B}(b), the power increases initially,
reaches a maximum, and then decreases, while the time at which the
maximum occurs becomes shorter as $J$ increases. The maximum charging
power follows an approximately logarithmic dependence on the
interaction strength, given by $ P_{\rm max}\approx 17.91\log_{10}(J)+16.86,$ as shown in the inset of Fig.~\ref{J_Insing_B}(b). These results
highlight that the intrinsic interaction strength of the battery itself
provides an additional control parameter for both the amount of stored
energy and the rate at which it is accumulated. Related enhancement of
quantum-battery performance due to interactions has also been reported
in Refs.~\cite{ghosh2020enhancement,PhysRevA.105.022628}.

\subsection{Configuration III: Interacting battery and interacting chargoid}
\label{int_battery_int_chargoid}

We finally consider the case in which both the battery and the chargoid
possess intrinsic interactions. Two complementary realizations are
investigated. In the first, the battery is described by the NN Ising
Hamiltonian $\hat H_B^{\mathrm{I,NN}}$ [Eq.~\ref{eq:HB_I_configIII}], while the chargoid is described
by the NN anisotropic XY Hamiltonian $\hat H_C^{\mathrm{XY,NN}}$ [Eq.~\ref{eq:HC_XY_configIII}]. In the
second, the roles of the two Hamiltonians are interchanged, with the
battery described by $\hat H_B^{\mathrm{XY,NN}}$  and the chargoid by
$\hat H_C^{\mathrm{I,NN}}$ [Eq.~\ref{eq:config_III_reverse}]. These two cases allow us to examine whether
the effect of the countereffect depends on which interaction structure
is associated with the battery and which with the chargoid.

We first consider the Ising-battery--XY-chargoid case,
corresponding to $\hat H_B^{\mathrm{I,NN}}$ and
$\hat H_C^{\mathrm{XY,NN}}$. The system is initially prepared in the
ground state of the battery Hamiltonian. During the charging interval,
the XY chargoid is activated while the intrinsic battery contribution
is modified by the countereffect according to $\hat H_B^{\mathrm{I,NN}}
    \longrightarrow
    (1-\lambda)\hat H_B^{\mathrm{I,NN}},$ where $\lambda\in[0,1]$ quantifies the strength of the countereffect.
Thus, $\lambda=0$ corresponds to the unmodified charging dynamics,
whereas increasing $\lambda$ progressively suppresses the contribution
of the intrinsic battery interaction to the charging Hamiltonian.

\begin{figure}
\includegraphics[width=0.48\linewidth,height=0.4\linewidth]{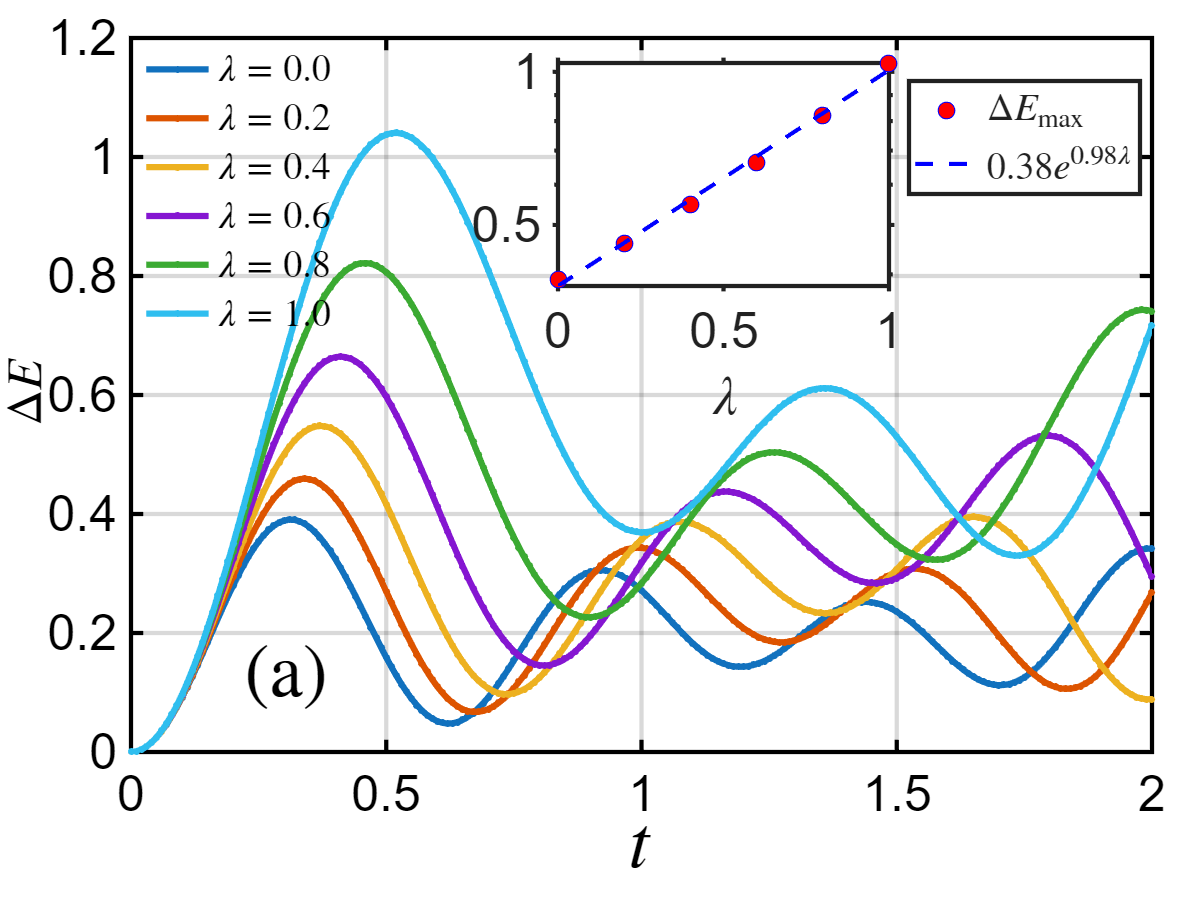}
\includegraphics[width=0.49\linewidth,height=0.4\linewidth]{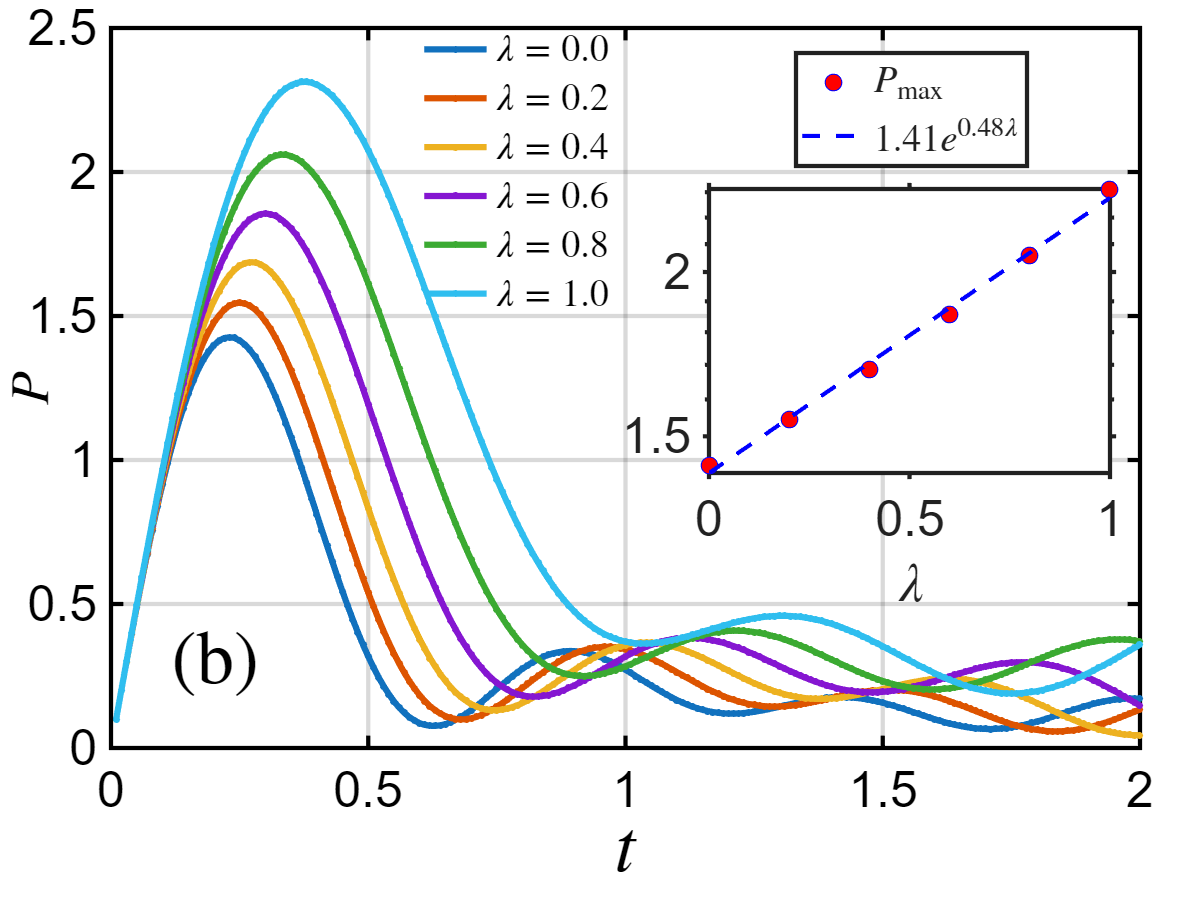}
\includegraphics[width=0.48\linewidth,height=0.4\linewidth]{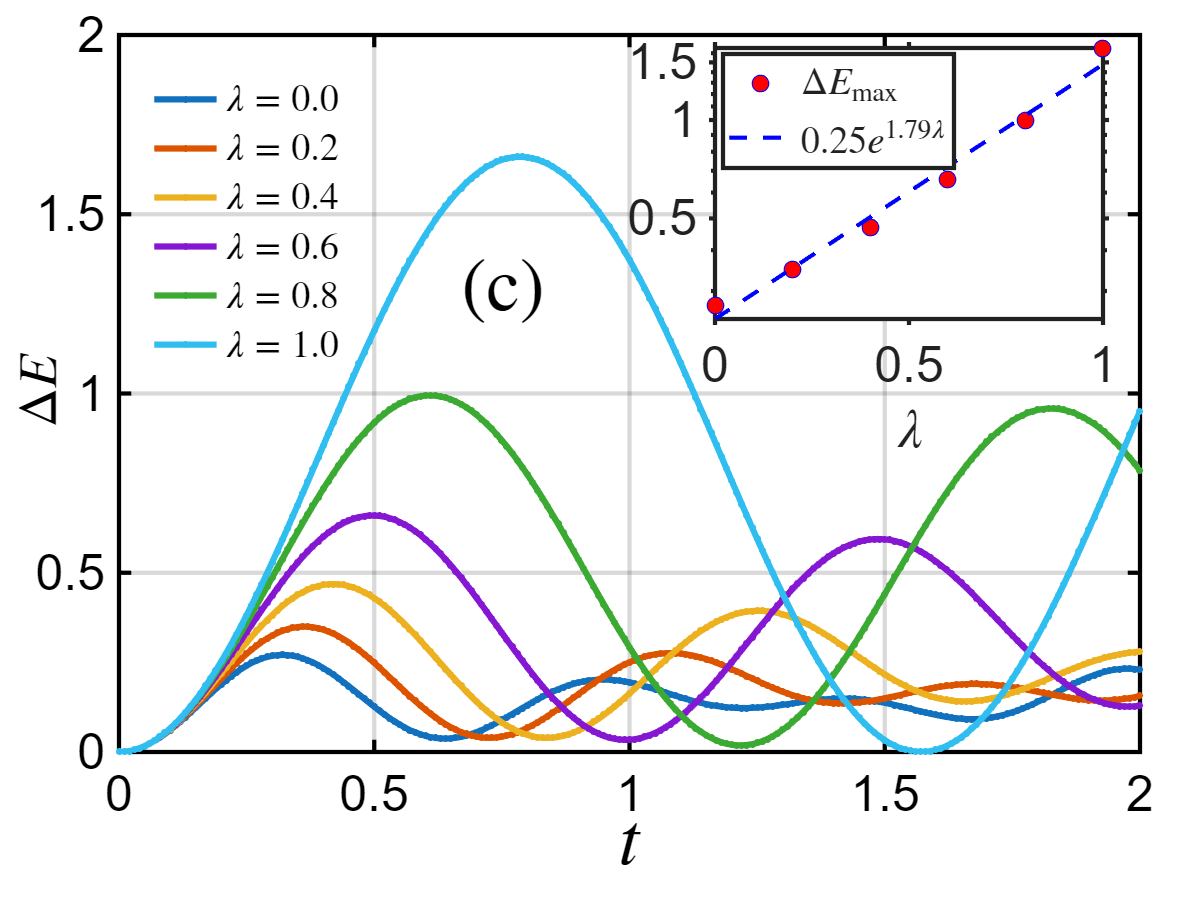}
\includegraphics[width=0.49\linewidth,height=0.4\linewidth]{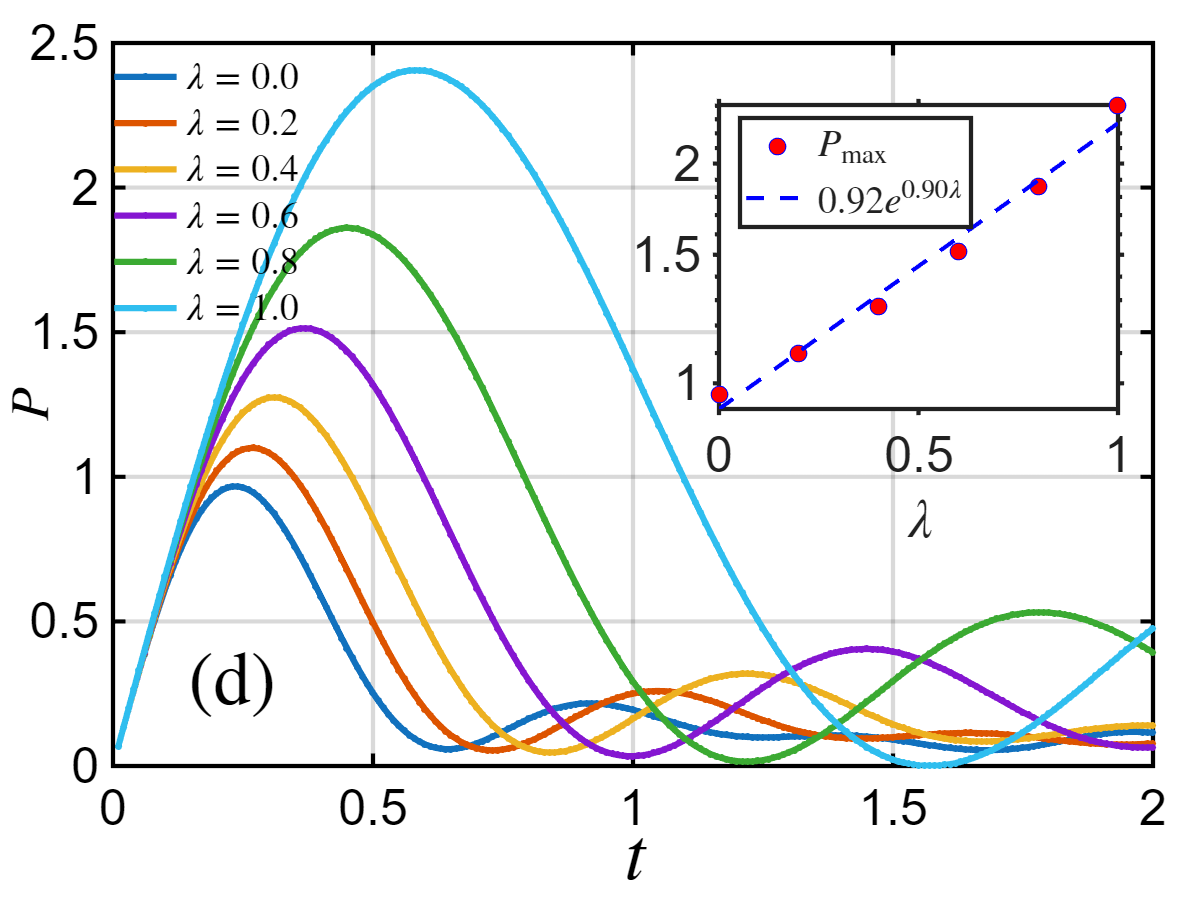}
\caption{\textbf{Interacting battery and interacting chargoid.}
(a) Stored energy $\Delta E$ and (b) charging power $P$ as functions of time $t$ for a battery governed by a NN Ising spin Hamiltonian, with charging dynamics generated by a NN XY interaction. (c) Stored energy $\Delta E$ and (d) charging power $P$ for a battery governed by a NN XY spin Hamiltonian, with charging dynamics generated by a NN Ising interaction.  The insets in the energy panels show the maximum stored energy as a function of the counter-effect parameter $\lambda$, while those in the power panels display the corresponding maximum charging power. The red data points denote numerical results, whereas the blue dashed lines represent the corresponding fits. The calculations are performed for $N=10$, $J=1$, and $\Gamma=0.5$, with periodic boundary conditions.}
\label{Int_B_XX_C_XY}
\end{figure}

We first examine the dependence of the charging dynamics on $\lambda$.
As shown in Fig.~\ref{Int_B_XX_C_XY}(a), the maximum stored energy
increases monotonically with the countereffect strength. The dependence
of the maximum stored energy on $\lambda$ is well described by the exponential form $
    \Delta E_{\rm max}
    \approx 0.38 e^{0.98\lambda},$ as shown in the inset of Fig.~\ref{Int_B_XX_C_XY}(a). A similar enhancement is observed for the charging power. Figure~\ref{Int_B_XX_C_XY}(b) shows that the maximum charging power increases systematically with $\lambda$, with the dependence approximately described by $  P_{\rm max}  \approx 1.41 e^{0.48\lambda}.$  Thus, for an interacting Ising battery driven by an interacting XY chargoid, suppressing the intrinsic battery interaction during charging enhances both the maximum stored energy and the maximum charging power.

We next consider the reverse case, in which the battery is described by
the NN XY Hamiltonian $\hat H_B^{\mathrm{XY,NN}}$ and the chargoid by the
NN Ising Hamiltonian $\hat H_C^{\mathrm{I,NN}}$. The same countereffect
protocol is applied, with the intrinsic battery contribution modified as $  \hat H_B^{\mathrm{XY,NN}} \longrightarrow  (1-\lambda)\hat H_B^{\mathrm{XY,NN}}.$ The resulting charging dynamics again exhibits a systematic enhancement with increasing $\lambda$. As shown in Fig.~\ref{Int_B_XX_C_XY}(c), the maximum stored energy increases monotonically with the countereffect strength. The dependence is well captured by $  \Delta E_{\rm max}   \approx 0.29 e^{1.79\lambda},$  as shown in the inset of Fig.~\ref{Int_B_XX_C_XY}(c). The maximum charging power follows the same qualitative behavior. As shown in Fig.~\ref{Int_B_XX_C_XY}(d), $P_{\rm max}$ increases with $\lambda$ and
is approximately described by $
    P_{\rm max}
    \approx 0.92 e^{0.90\lambda}.$

The results obtained for the two realizations demonstrate that the
countereffect enhances both energy storage and charging power even when
the battery and chargoid are themselves interacting. Moreover, the
enhancement persists when the roles of the Ising and XY interactions are
interchanged. This indicates that the observed benefit of the
countereffect is not restricted to a particular assignment of the
interaction structures to the battery and chargoid. Instead, it reflects
the general influence of suppressing the intrinsic battery contribution
during the charging dynamics. Taken together with the results of
Configurations I and II, these findings show that the countereffect
provides a robust mechanism for improving charging performance across
different combinations of interacting and non-interacting battery and
chargoid Hamiltonians.
 \

\section{Open-System Dynamics}
\label{Open_system_dynamics}
To assess the robustness of the battery dynamics under realistic conditions, we extend the analysis beyond the idealized closed-system setting by considering the coupling of the battery to an external environment. Such coupling introduces decoherence and dissipation, which can modify the charging process and influence the subsequent energy-extraction dynamics. We therefore examine the effects of environmental interactions on the stored energy, charging power, and discharging process. In addition, we characterize the quality of the stored energy through the ergotropy, which quantifies the maximum amount of energy that can be extracted as useful work by unitary operations. The open-system dynamics is described within the Markovian framework by the master equation \cite{gorini1976completely,lindblad1976generators}
\begin{equation}
\frac{d\rho}{dt}
=
-i[H_{\rm ch},\rho]
+
\mathcal{D}_{\mathrm{env}}(\rho),
\label{eq:master}
\end{equation}
where $\rho$ denotes the density matrix of the battery, initially prepared in the ground state of the battery Hamiltonian $H_B$, $H_{\rm ch}$ is the Hamiltonian governing the system dynamics, and $\mathcal{D}_{\mathrm{env}}$ describes the dissipative contribution arising from the coupling to the environment.

To model the dominant environmental effects, we consider two representative decoherence channels for spin systems: energy relaxation and pure dephasing. The former describes energy exchange between the battery and its surroundings, while the latter suppresses quantum coherences without directly changing the populations. Within the Markovian approximation, these processes are described by the Lindblad dissipator
\begin{equation}
\mathcal{D}_{\mathrm{env}}(\rho)
=
\gamma_{1}\sum_{j=1}^{N}\mathcal{D}[\sigma_j^-]\rho
+
\gamma_{\phi}\sum_{j=1}^{N}\mathcal{D}[\sigma_j^{z}]\rho,
\label{eq:dissipator}
\end{equation}
where $\gamma_{1}$ and $\gamma_{\phi}$ denote the relaxation and pure-dephasing rates, respectively. For simplicity, we consider equal rates throughout the analysis, $\gamma_{1}=\gamma_{\phi}=\gamma,$ where $\gamma$ sets the overall strength of the system--environment coupling. Here,
$\sigma_j^- = |\downarrow\rangle_j\langle\uparrow|$
is the spin-lowering operator acting on the $j$th spin, while $\sigma_j^z$ is the corresponding Pauli-$z$ operator. The Lindblad superoperator is defined as
\begin{equation}
\mathcal{D}[L]\rho
=
L\rho L^{\dagger}
-\frac{1}{2}
\left\{
L^{\dagger}L,\rho
\right\},
\end{equation}
where $\{A,B\}=AB+BA$ denotes the anticommutator. The relaxation channel allows energy to dissipate from the battery into the environment, whereas pure dephasing reduces quantum coherences without directly exchanging energy. We assume that both processes act locally and independently on each spin, corresponding to uncorrelated Markovian environmental noise.

The coherent and dissipative contributions can be combined into the Liouvillian superoperator
\begin{equation}
\mathcal{L}\rho
=
-i[H_{\rm ch},\rho]
+
\mathcal{D}_{\mathrm{env}}(\rho),
\label{eq:Liouvillian}
\end{equation}
so that the master equation takes the compact form
\begin{equation}
\frac{d\rho}{dt}
=
\mathcal{L}\rho.
\label{eq:master_liouvillian}
\end{equation}
For time-independent system parameters and dissipative rates, the corresponding formal solution is
\begin{equation}
\rho(t)
=
e^{\mathcal{L}t}\rho(0),
\label{eq:formal_solution}
\end{equation}
where $\rho(0)$ is the initial density matrix (ground state of the battery Hamiltonian) and $e^{\mathcal{L}t}$ denotes the dynamical propagator generated by the Liouvillian. This framework allows us to investigate how environmental decoherence and dissipation modify the charging and energy-extraction dynamics of the battery, as well as the usefulness of the stored energy.

\paragraph*{Storage Energy and Charging Power.}

In the presence of environmental coupling, the battery state evolves non-unitarily, and the energy stored during the charging process is determined by the density matrix obtained from the open-system dynamics. We quantify the stored energy relative to the initial ground-state energy $E_0$ as
\begin{equation}
\Delta E(t)
=
\mathrm{Tr}\left[\rho(t)H_B\right]-E_0,
\label{eq:stored_energy_open}
\end{equation}
where $H_B$ is the battery Hamiltonian and $\rho(t)$ is the time-dependent density matrix obtained from Eq.~(\ref{eq:formal_solution}). The corresponding average charging power at the charging time $T$ is defined as
\begin{equation}
P
=
\frac{\Delta E(T)}{T}.
\label{eq:charging_power_open}
\end{equation}
Thus, $\Delta E(T)$ measures the net energy accumulated in the battery during the charging process, while $P$ characterizes the average rate of energy storage.

Although the definitions of stored energy and charging power retain the same form as in the closed-system case, their dynamics are now governed by the Liouvillian evolution and therefore incorporate both coherent charging and environmental effects. In particular, relaxation can remove energy from the battery, while dephasing can modify the quantum coherences responsible for the charging dynamics. The resulting behavior of $\Delta E$ and $P$ therefore provides a direct measure of the robustness of the charging protocol against environmental noise.

Experimentally, environmental effects in superconducting-qubit platforms can be characterized by the relaxation time $T_1$ and the coherence time $T_2$, which quantify energy relaxation and loss of quantum coherence, respectively. The latter generally receives contributions from both energy relaxation and pure dephasing, such that
\begin{equation}
\frac{1}{T_2}
=
\frac{1}{2T_1}
+
\frac{1}{T_\phi},
\end{equation}
where $T_\phi$ is the pure-dephasing time \cite{Burnett2019,lu2021characterizing}. Both $T_1$ and $T_2$, and consequently $T_\phi$, can be experimentally determined through standard relaxation and coherence measurements of superconducting qubits \cite{Burnett2019,lu2021characterizing}.

To incorporate these experimentally relevant environmental processes into our open-system description, we consider two types of Markovian dissipation. Energy relaxation is modeled by the jump operator $\hat{\sigma}_j^-$, while pure dephasing is described by $\hat{\sigma}_j^z$. The corresponding dissipative rates can be related to the experimentally measured relaxation and dephasing times. Thus, the parameters entering our Lindblad description have a direct physical interpretation in terms of experimentally accessible decoherence processes.

\paragraph*{Ergotropy.}

The energy stored in a quantum battery does not necessarily correspond entirely to energy that can be extracted as useful work. This distinction becomes particularly relevant when the battery state is mixed, as environmental interactions can generate passive contributions to the energy that cannot be extracted by unitary operations. To quantify the extractable part of the stored energy, we consider the ergotropy of the battery state.

For a battery described by the Hamiltonian $H_B$ and density matrix $\rho(t)$, the ergotropy is defined as the maximum amount of energy that can be extracted by a unitary transformation ~\cite{allahverdyan2004maximal},
\begin{equation}
\mathcal{W}(t)
=
\max_{U}
\left\{
\mathrm{Tr}\left[H_B\rho(t)\right]
-
\mathrm{Tr}\left[H_B U\rho(t)U^\dagger\right]
\right\},
\label{eq:ergotropy_definition}
\end{equation}
where $U$ is an arbitrary unitary operator acting on the battery. The minimum energy attainable under unitary transformations is associated with the passive state of $\rho(t)$ ~\cite{pusz1978passive}, denoted by $\rho_{\mathrm{p}}(t)$. A passive state is a state from which no further energy can be extracted through unitary operations. Consequently,
\begin{equation}
\mathcal{W}(t)
=
\mathrm{Tr}\left[H_B\rho(t)\right]
-
\mathrm{Tr}\left[H_B\rho_{\mathrm{p}}(t)\right].
\label{eq:ergotropy_passive}
\end{equation}

To construct the passive state, we consider the spectral decomposition of the density matrix,
\begin{equation}
\rho(t)
=
\sum_{n=1}^{d}
r_n(t)
\ket{r_n(t)}
\bra{r_n(t)},
\end{equation}
where $d=2^N$ is the Hilbert-space dimension and the eigenvalues are ordered as $r_1(t)\geq r_2(t)\geq\cdots\geq r_d(t).$
The battery Hamiltonian is expressed in its energy eigenbasis as
\begin{equation}
H_B
=
\sum_{n=1}^{d}
E_n
\ket{E_n}\bra{E_n},
\end{equation}
with the energy eigenvalues arranged in ascending order, $E_1\leq E_2\leq\cdots\leq E_d.$ The passive state is obtained by assigning the largest eigenvalue of $\rho(t)$ to the lowest energy eigenstate of $H_B$, the second-largest eigenvalue to the second-lowest energy eigenstate, and so forth. It therefore takes the form
\begin{equation}
\rho_{\mathrm{p}}(t)
=
\sum_{n=1}^{d}
r_n(t)
\ket{E_n}\bra{E_n}.
\label{eq:passive_state}
\end{equation}
The corresponding passive-state energy is
\begin{equation}
E_{\mathrm{p}}(t)
=
\mathrm{Tr}\left[H_B\rho_{\mathrm{p}}(t)\right]
=
\sum_{n=1}^{d}r_n(t)E_n.
\label{eq:passive_energy}
\end{equation}
The ergotropy can consequently be evaluated as
\begin{equation}
\mathcal{W}(t)
=
\mathrm{Tr}\left[H_B\rho(t)\right]
-
E_{\mathrm{p}}(t).
\label{eq:ergotropy_final}
\end{equation}

Unlike the open-system case, the closed-system dynamics is unitary. Since the battery is initially prepared in the ground state of $H_B$, the state remains pure during the evolution, and the energy of the corresponding passive state remains fixed at the ground-state energy $E_0$. Thus, $ E_{\mathrm{p}}(t)=E_0.$ The ergotropy, defined as the maximum energy extractable by unitary operations, is therefore $\mathcal{W}(t)
= E_B(t)-E_{\mathrm{p}}(t)= E_B(t)-E_0=\Delta E(t).$ Hence, in the closed-system limit, the energy stored above the ground state is fully extractable, and the stored energy coincides with the ergotropy. In contrast, for finite environmental coupling, the evolution is non-unitary and the battery state can become mixed. The passive-state energy is then generally different from $E_0$, leading to $\mathcal{W}(t)
=\Delta E(t)-\left[E_{\mathrm{p}}(t)-E_0\right],$ so that the stored energy and its extractable part need not coincide.


To assess the robustness of the charging protocol against environmental
effects, we investigate the stored energy, charging power, and ergotropy for
different strengths of energy relaxation and pure dephasing. We take equal
rates for the two decoherence channels, $\gamma_1=\gamma_{\phi}=\gamma$, and
consider $\gamma=0,10^{-3},10^{-2},$ and $10^{-1}$, where $\gamma=0$
corresponds to the closed-system limit. Throughout this analysis, we focus
on the fully suppressed countereffect limit, $\lambda=1$, in which the
intrinsic battery contribution is completely compensated during charging. For comparison,
the $\lambda=0$ limit, where the intrinsic battery contribution remains
fully active, is discussed in Appendix~\ref{open_System_dynamics_l0}. This
comparison allows us to assess how the intrinsic battery dynamics modifies
the charging performance in the presence of environmental coupling. The
environmental effects are examined for the three battery--chargoid
configurations introduced above, thereby allowing us to determine how the
interaction structure of the battery and chargoid influences the response
to relaxation and dephasing.

\begin{figure*}
    \centering
\includegraphics[width=0.30\linewidth,height=0.25\linewidth]{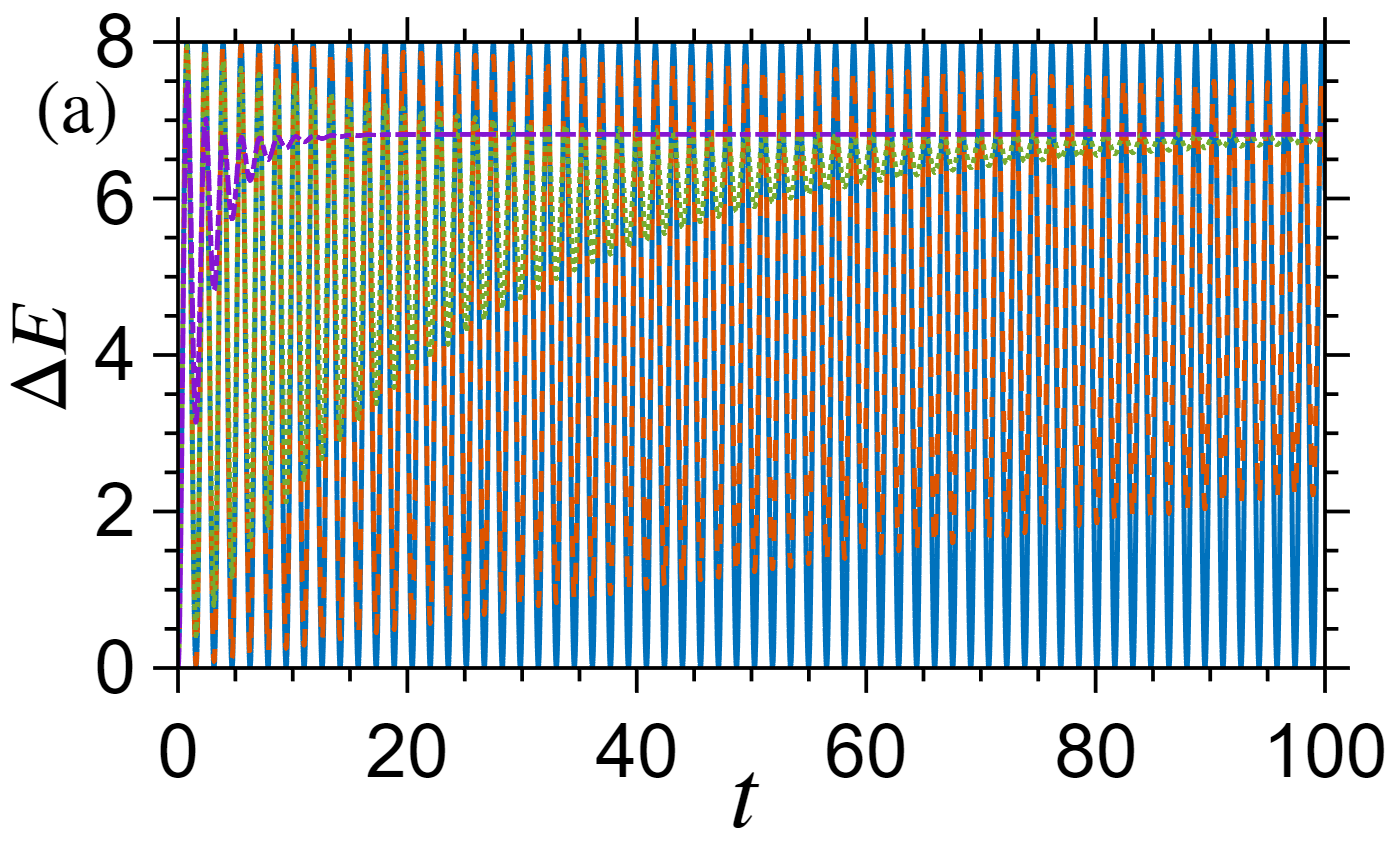}
\includegraphics[width=0.30\linewidth,height=0.25\linewidth]{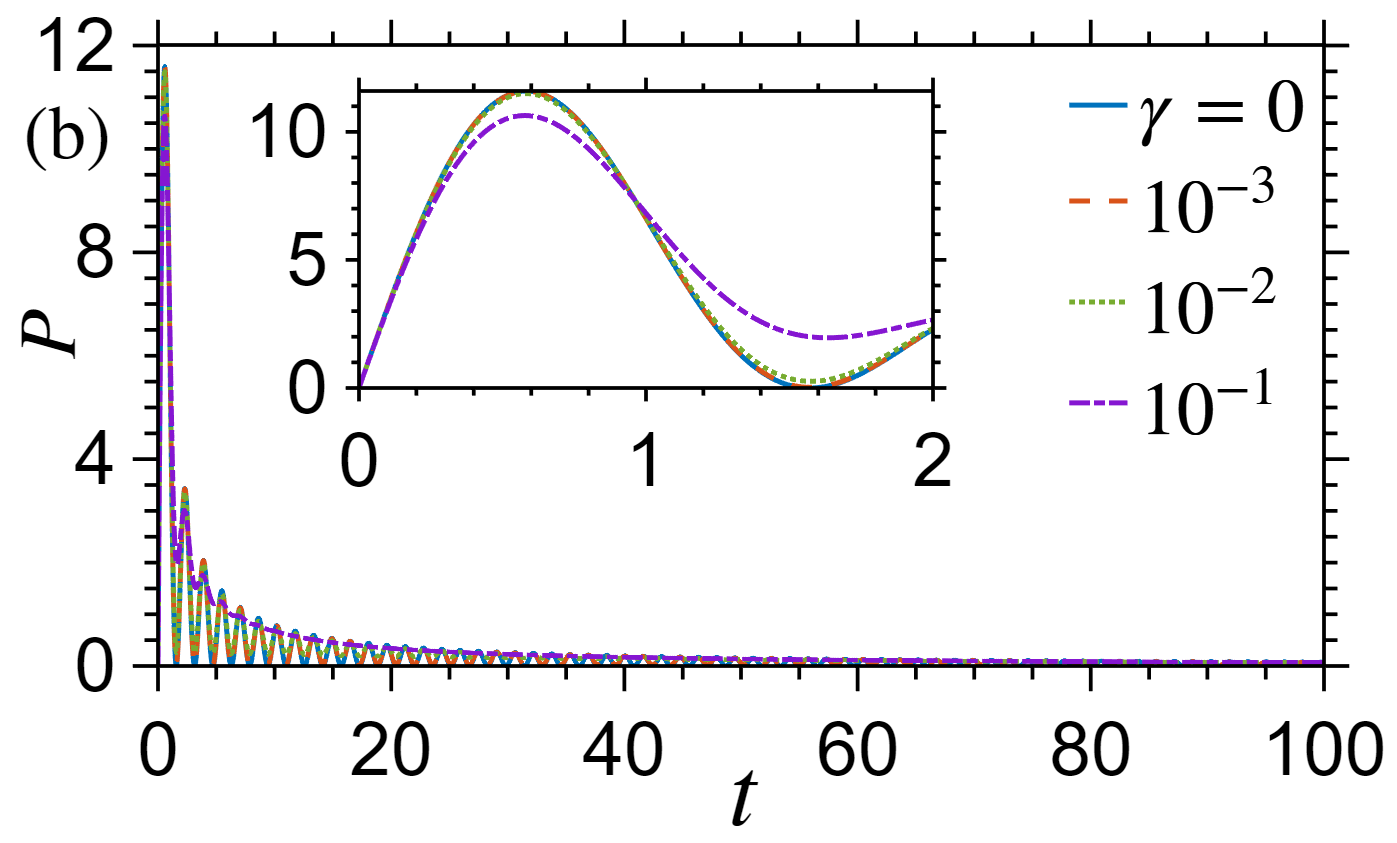}
\includegraphics[width=0.30\linewidth,height=0.25\linewidth]{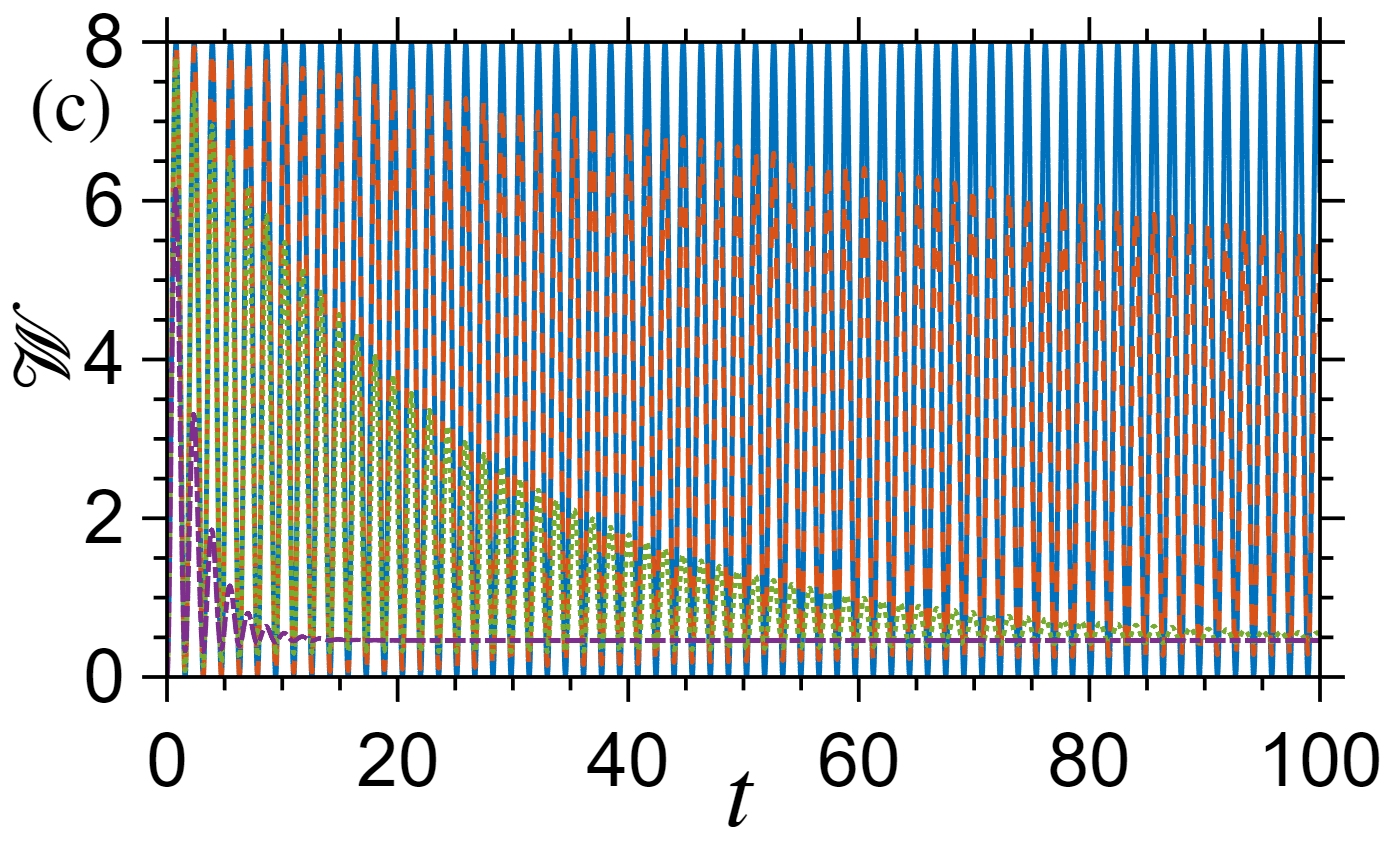}
\includegraphics[width=0.30\linewidth,height=0.25\linewidth]{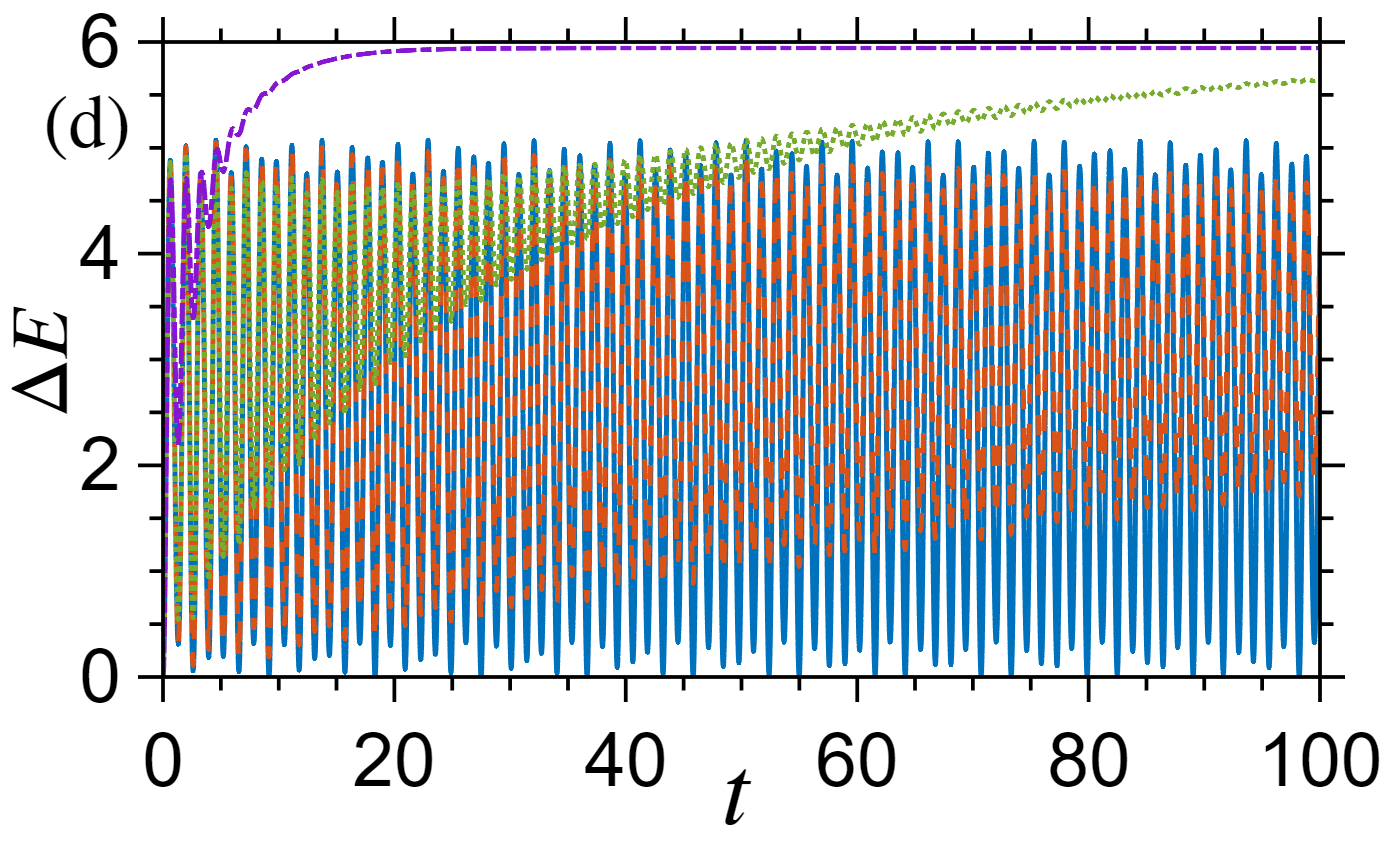}
\includegraphics[width=0.30\linewidth,height=0.25\linewidth]{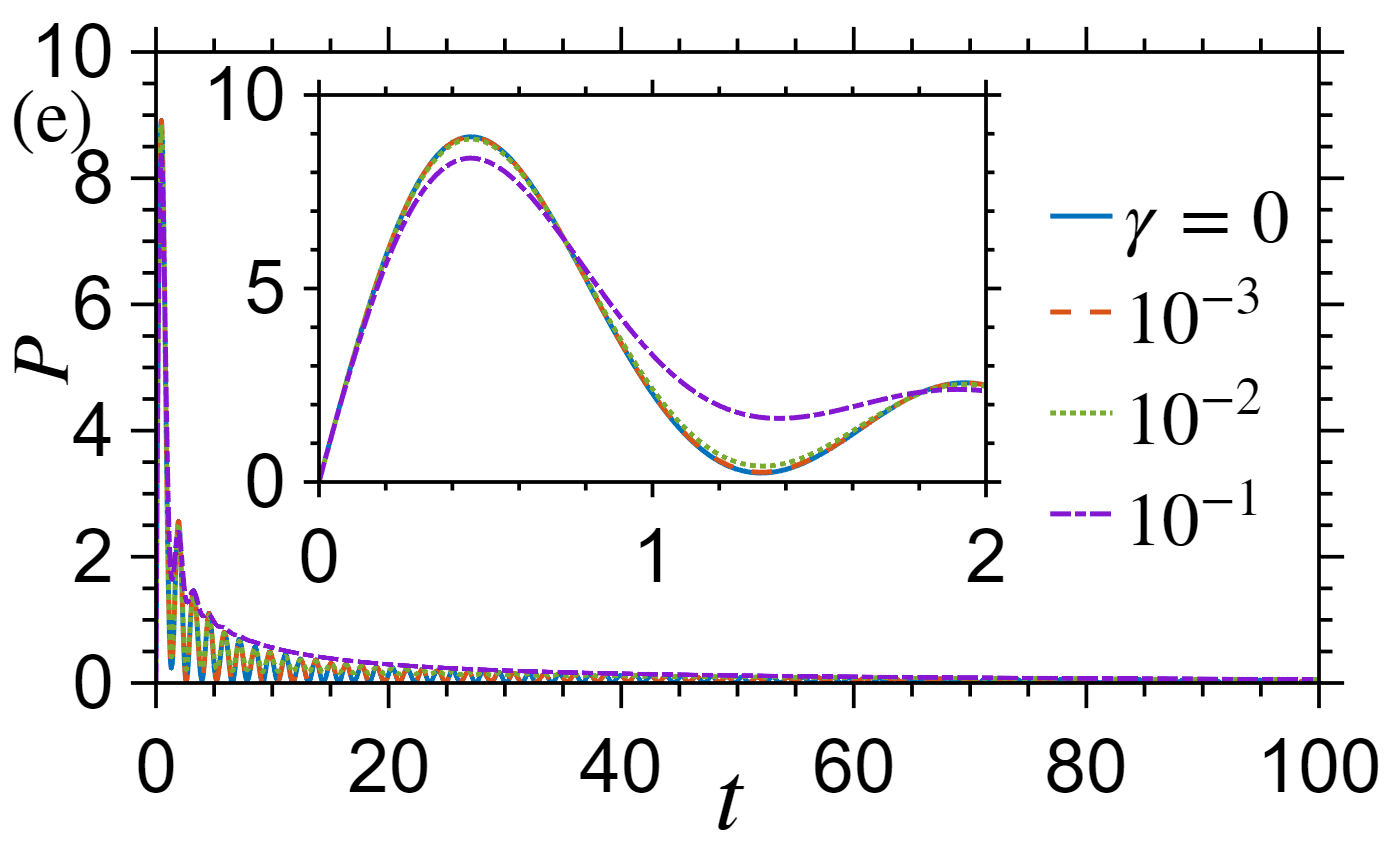}
\includegraphics[width=0.30\linewidth,height=0.25\linewidth]{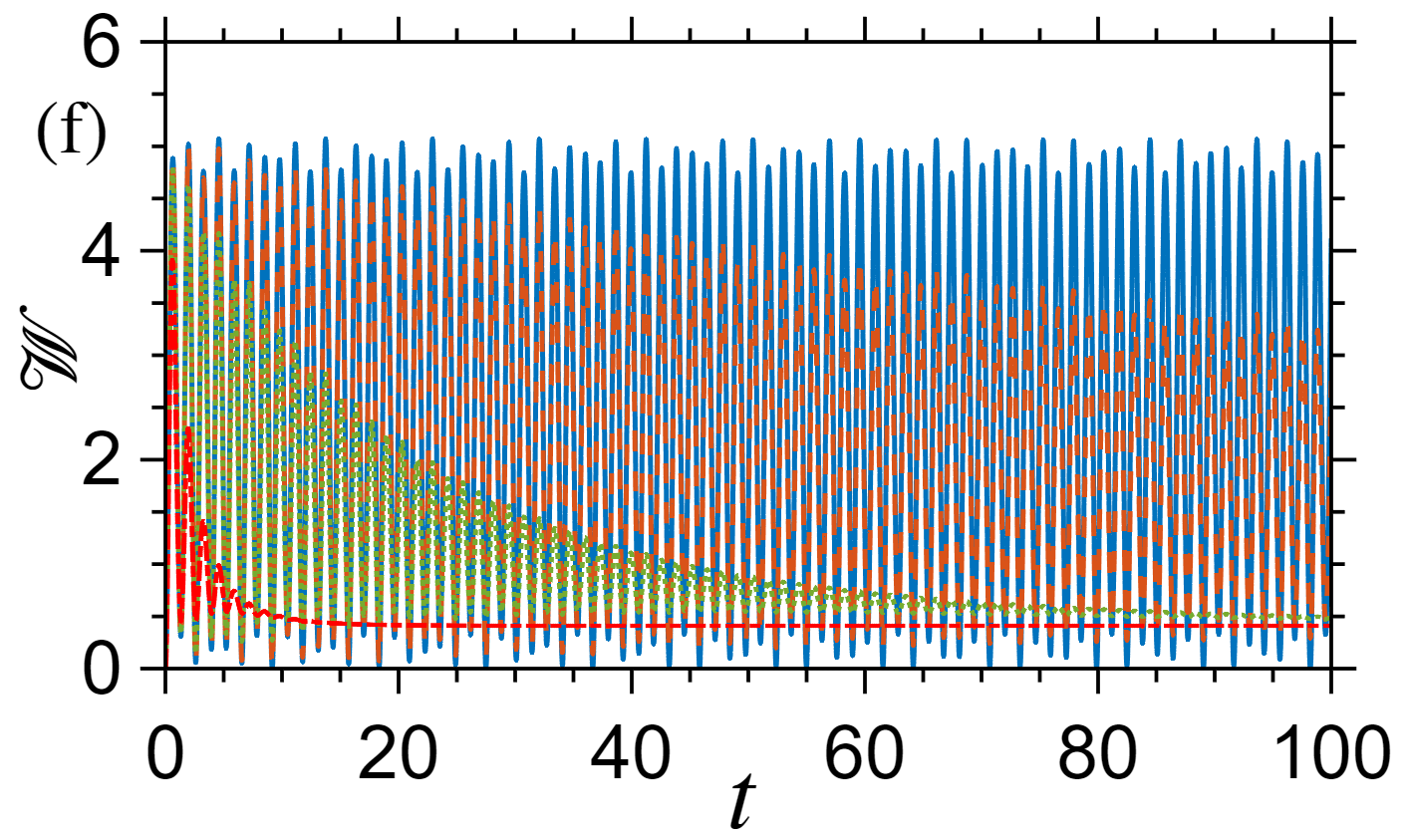} \caption{\textbf{Noninteracting batteries charged by interacting chargoids.} Charging dynamics of a noninteracting battery driven by interacting chargoids: (a--c) an NN Ising chargoid and (d--f) an NN XY chargoid. Panels (a,d), (b,e), and (c,f) show the stored energy, charging power, and ergotropy, respectively. The inset in the power panels highlights the short-time charging behavior. The system size is fixed at $N=8$ with periodic boundary conditions. The parameters are $J=1$, $\Gamma=0.5$, and $h_z=1$, and the battery countereffect parameter is set to $\lambda=1$.}
\label{open_nonint_B_int_C}
\end{figure*}

\subsection{Configuration I: Noninteracting battery with interacting chargoid.}

We first consider Configuration I, comprising a noninteracting battery
$\hat{H}_B^{0}$ [Eq.~\ref{eq:HB0}] and an externally activated
nearest-neighbor (NN) interacting chargoid. We consider both the Ising
$\hat{H}_C^{\rm I,NN}$ [Eq.~\ref{eq:HC_INN}] and anisotropic XY
$\hat{H}_C^{\rm XY,NN}$ [Eq.~\ref{eq:HC_XYNN}] chargoids with the
countereffect set to $\lambda=1$. We begin with the NN Ising chargoid.
In the closed-system limit, $\gamma=0$, the stored energy exhibits
pronounced periodic oscillations, reflecting the coherent many-body
dynamics generated by the interacting chargoid, as shown in
Fig.~\ref{open_nonint_B_int_C}(a).

When environmental effects are included, $\gamma>0$, these coherent
oscillations are progressively modified. At weak decoherence, the
charging dynamics retains oscillatory features at early times, although
the dynamics becomes increasingly irregular. At longer times, the stored
energy approaches a stationary value rather than continuing the
persistent oscillations observed in the closed-system case. Increasing
the decoherence strength further shortens the transient regime and
drives the system more rapidly toward this stationary behavior.
Moreover, the stationary stored energy decreases systematically with
increasing $\gamma$ and remains below the maximum energy attained in the
closed-system dynamics. Thus, environmental decoherence suppresses the
coherent energy accumulation and reduces the amount of energy that can
ultimately be stored in the battery.
The corresponding charging power is shown in Fig.~\ref{open_nonint_B_int_C}(b). The maximum charging power decreases systematically with increasing decoherence strength, indicating that environmental interactions suppress the rapid coherent energy-transfer processes responsible for fast charging. Thus, decoherence reduces not only the attainable stored energy but also the rate at which energy is accumulated.

We further examine the extractable energy through the ergotropy, shown in Fig.~\ref{open_nonint_B_int_C}(c). In the closed-system limit, the battery state remains pure throughout the unitary evolution. Since the initial state is the ground state of $H_B$, the passive-state energy remains equal to the ground-state energy, and hence the ergotropy coincides with the stored energy, $\mathcal{W}(t)=\Delta E(t)$. Accordingly, the ergotropy displays the same periodic behavior as the stored energy for $\gamma=0$. For finite decoherence, the periodic oscillations are progressively suppressed, giving way to a transient regime of irregular oscillations followed by saturation at a finite value. The duration of this transient regime decreases as the decoherence strength increases. These features demonstrate that relaxation and dephasing progressively destroy the coherent dynamics underlying the reversible charging and discharging cycles, leading to reduced energy storage and charging performance.

A qualitatively different response is observed for the stored energy with the XY chargoid. As shown in Fig.~\ref{open_nonint_B_int_C}(d), the maximum stored energy increases with increasing decoherence strength, whereas the corresponding charging power decreases [Fig.~\ref{open_nonint_B_int_C}(e)]. This behavior indicates a trade-off between the amount of energy accumulated and the rate at which it is stored. The enhancement of the stored energy can be attributed to the distinct many-body dynamics generated by the XY interaction. In the coherent limit, interference among different excitation-transfer pathways can restrict the population of highly excited battery states. The presence of environmental coupling modifies these coherent pathways and redistributes the populations, allowing the battery to access higher-energy configurations. At the same time, the loss of coherence suppresses the rapid energy-transfer processes, leading to a reduction in the charging power.

Interestingly, the increase in stored energy is accompanied by a decrease in ergotropy, as shown in Fig.~\ref{open_nonint_B_int_C}(f). Thus, the additional energy accumulated in the presence of environmental coupling is not fully available for extraction as useful work. This demonstrates that enhanced energy storage does not necessarily imply improved battery performance, since the quality of the stored energy, as quantified by the ergotropy, can simultaneously deteriorate.

\begin{figure*}
    \centering
\includegraphics[width=0.30\linewidth,height=0.25\linewidth]{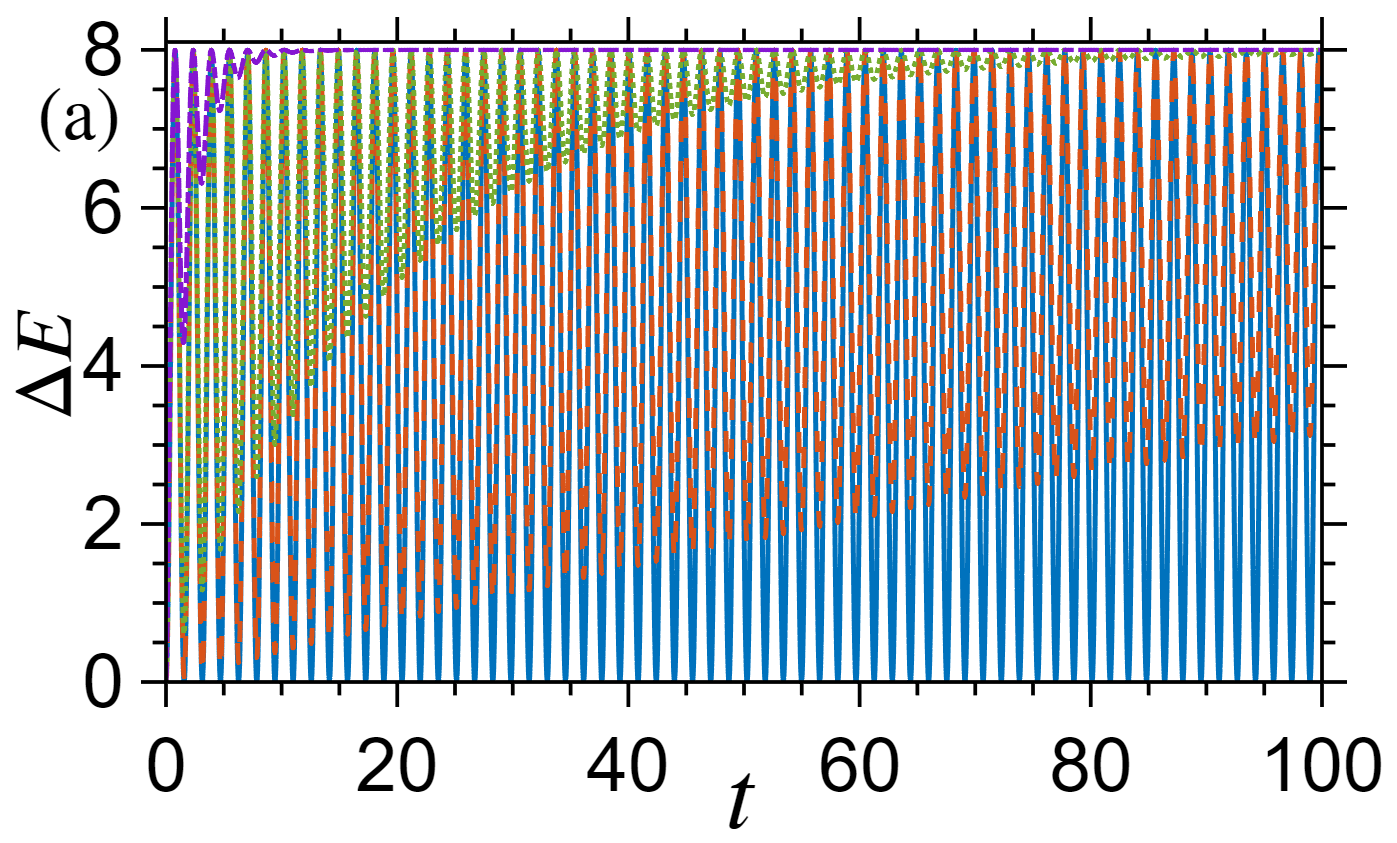}
\includegraphics[width=0.30\linewidth,height=0.25\linewidth]{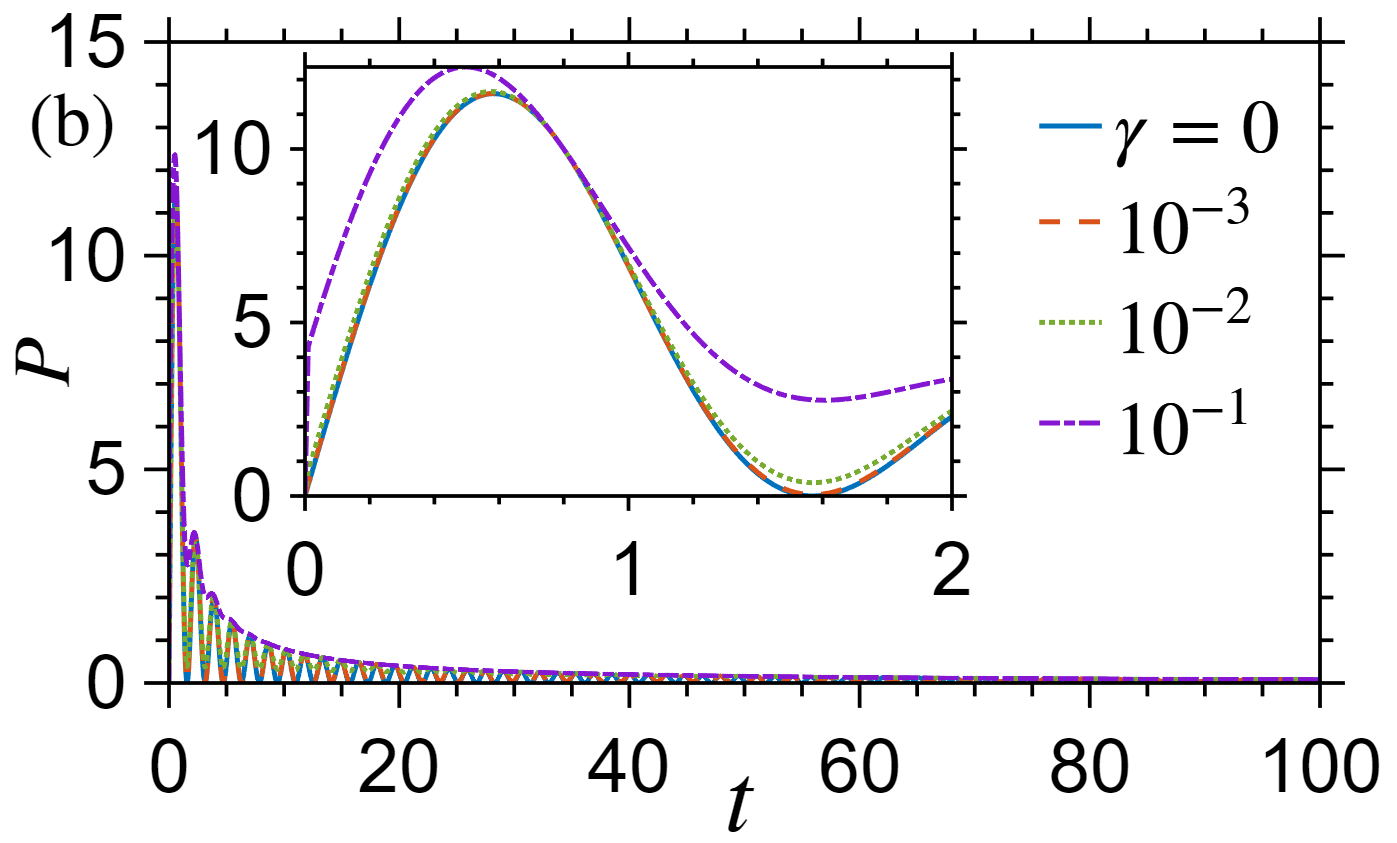}
\includegraphics[width=0.30\linewidth,height=0.25\linewidth]{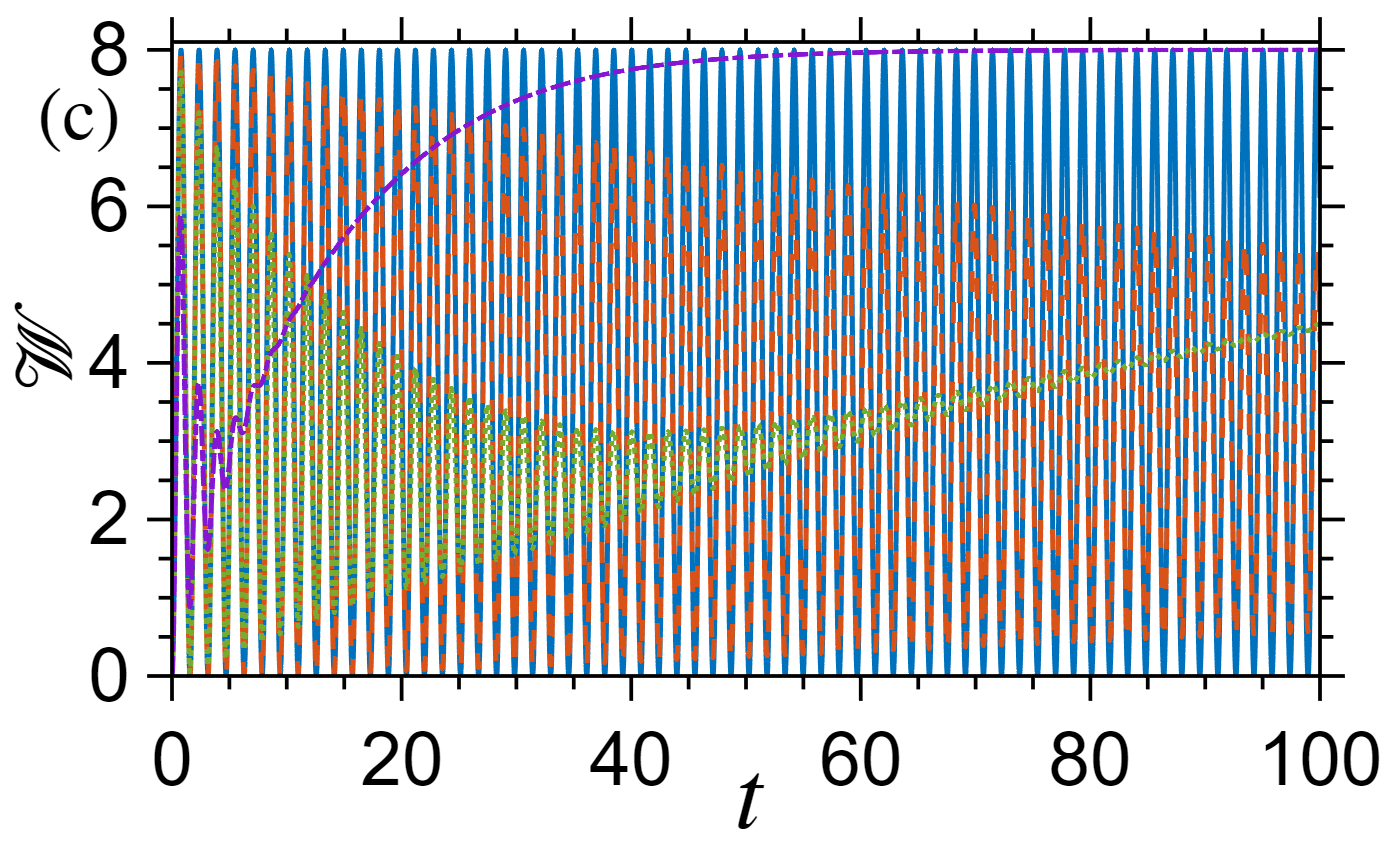}
\includegraphics[width=0.30\linewidth,height=0.25\linewidth]{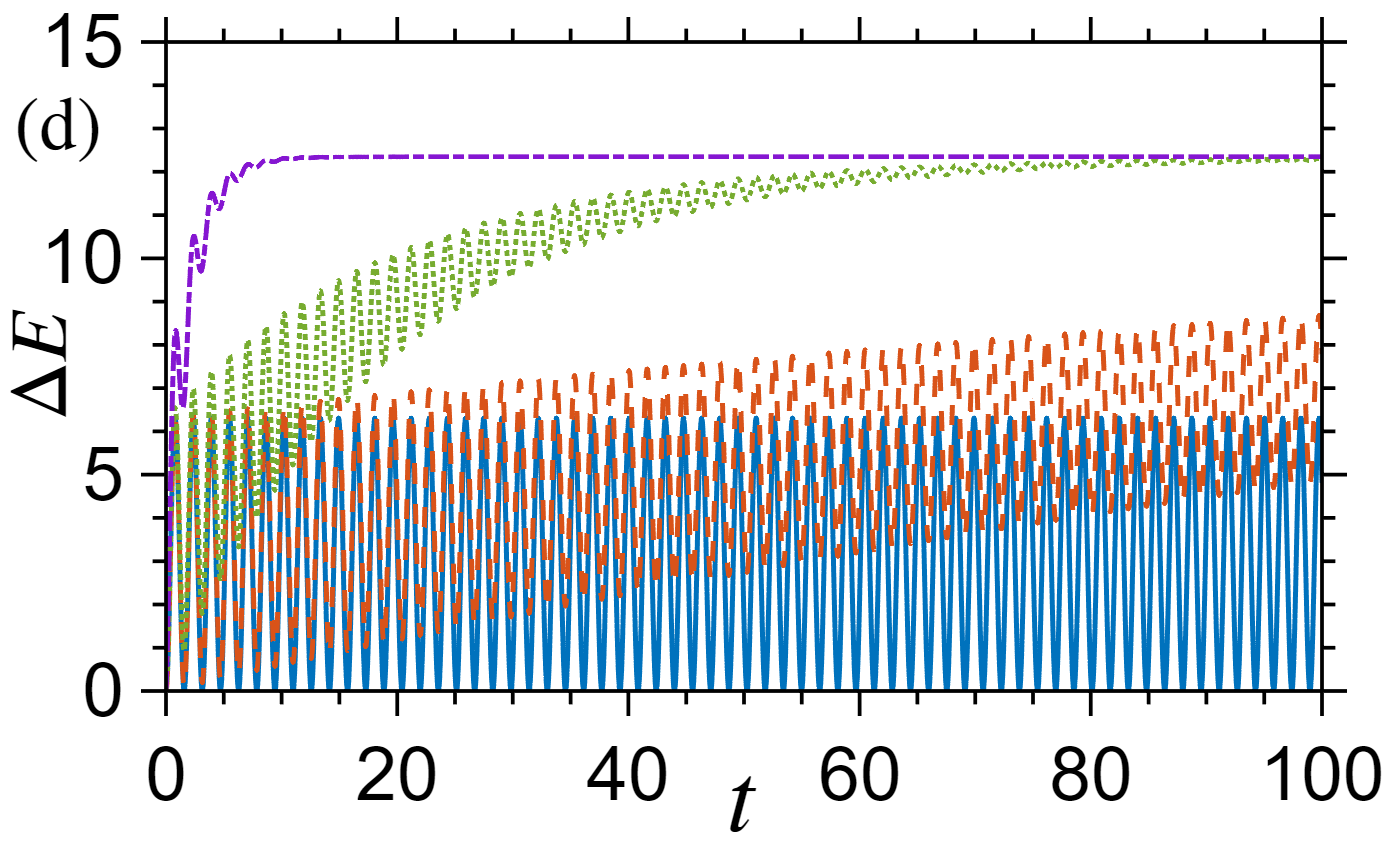}
\includegraphics[width=0.30\linewidth,height=0.25\linewidth]{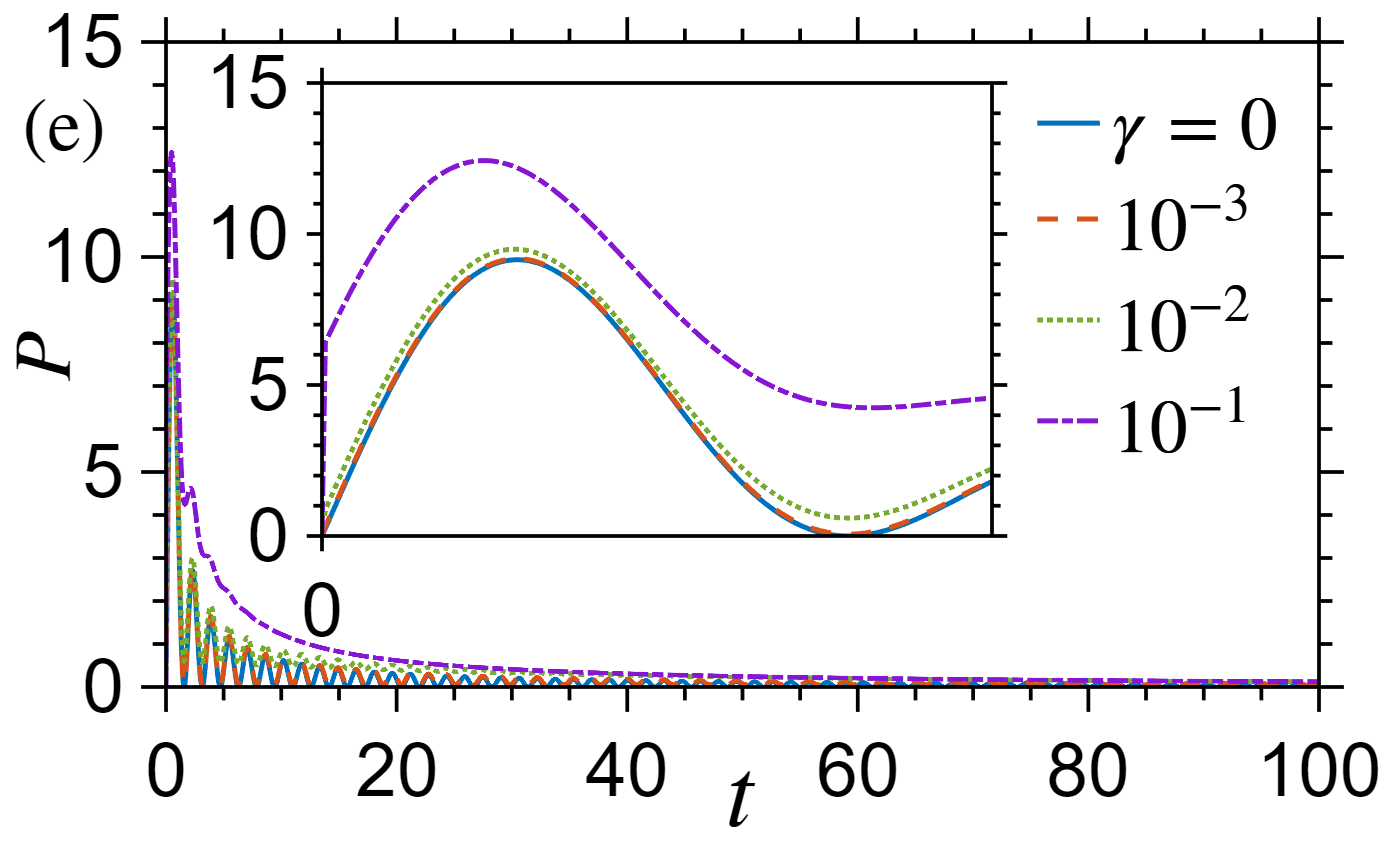}
\includegraphics[width=0.30\linewidth,height=0.25\linewidth]{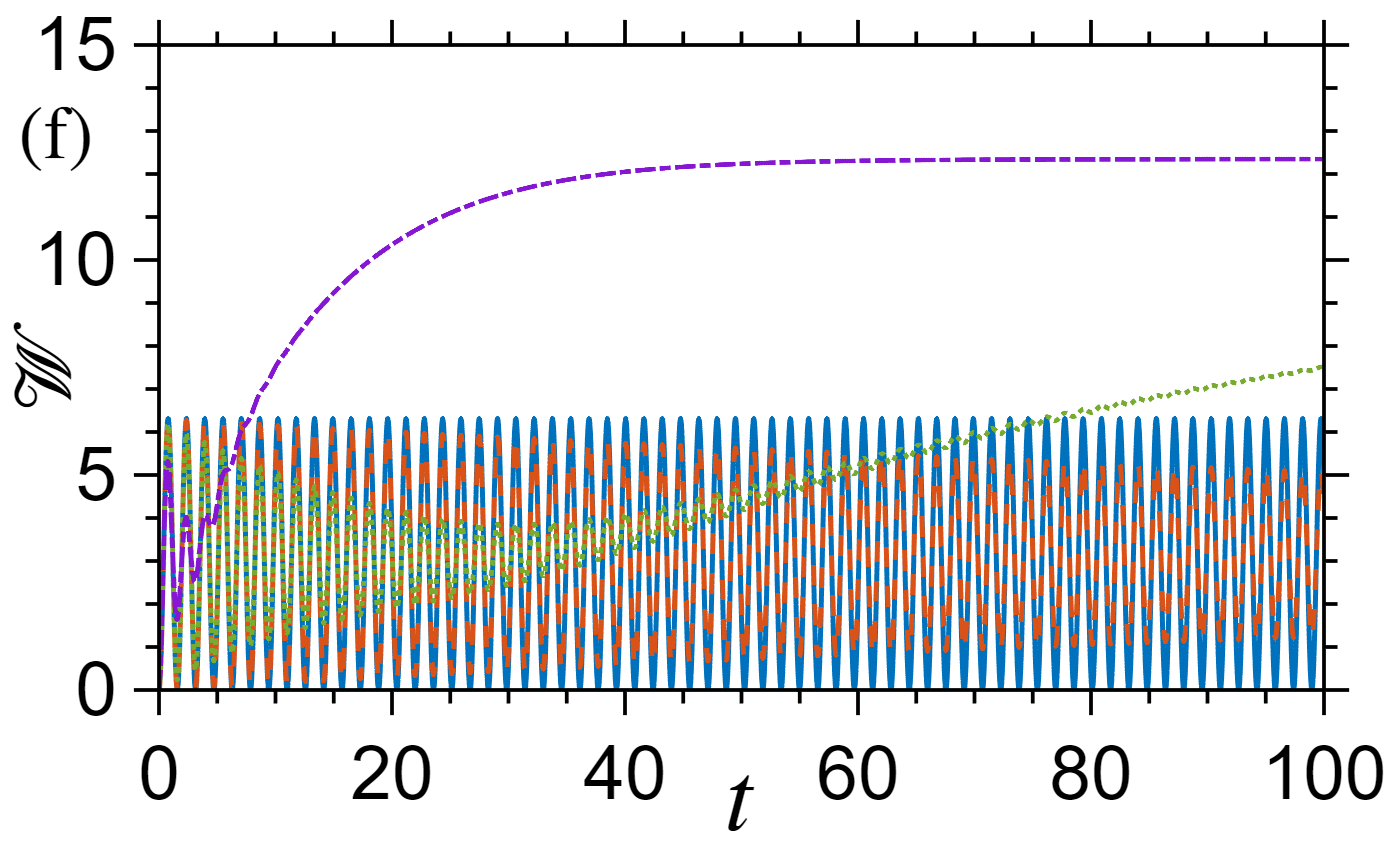}
\caption{\textbf{Interacting batteries charged by noninteracting chargoids.} Charging dynamics of interacting batteries driven by a noninteracting chargoid: (a--c) an NN Ising interacting battery and (d--f) an NN XY interacting battery. Panels (a,d), (b,e), and (c,f) show the stored energy, charging power, and ergotropy, respectively, for the corresponding battery--chargoid configurations. The inset in the power panels highlights the short-time charging behavior. The system size is fixed at $N=8$ with periodic boundary conditions. The parameters are $J=1$, $\Gamma=0.5$, and $h_z=1$, and the battery countereffect parameter is set to $\lambda=1$.}
 \label{open_int_B_nonint_C}
\end{figure*}

\subsection{Configuration II: Interacting battery with noninteracting chargoid.}

We next consider Configuration II, in which the battery possesses intrinsic spin--spin interactions, while the chargoid is noninteracting. Specifically, the battery is described by either the nearest-neighbor (NN) Ising Hamiltonian $\hat H_B^{\mathrm{I,NN}}$ [Eq.~\ref{eq:HB_INN}] or the NN anisotropic XY Hamiltonian $\hat H_B^{\mathrm{XY,NN}}$ [Eq.~\ref{eq:HB_XYNN}], whereas the chargoid is represented by the noninteracting Hamiltonian $\hat H_C^{0}$ [Eq.~\ref{eq:HC0}], corresponding to a uniform field along the $z$ direction. We set the countereffect parameter to $\lambda=1$, such that the intrinsic battery contribution is fully compensated during the charging interval.

For the NN Ising battery, the closed-system dynamics
($\gamma=0$) exhibits pronounced oscillations in both the stored energy
and the ergotropy, as shown in Fig.~\ref{open_int_B_nonint_C}(a,c).
These oscillations originate from the coherent exchange of energy
between the interacting battery degrees of freedom and the external
charging field. Upon introducing environmental effects
($\gamma>0$), the coherent oscillations are progressively damped and
the dynamics approaches a stationary regime. Interestingly, the maximum
stored energy remains nearly unchanged over the range of decoherence
strengths considered, indicating that the maximum charging capacity is
relatively robust against environmental coupling. In contrast, the
maximum charging power increases with $\gamma$, as shown in
Fig.~\ref{open_int_B_nonint_C}(b). The damping of the coherent
oscillations suppresses energy backflow and allows the battery to reach
its maximum stored energy on a shorter timescale, thereby increasing the
maximum charging power.

A similar trend is observed for the ergotropy. As shown in
Fig.~\ref{open_int_B_nonint_C}(c), increasing the decoherence strength
progressively suppresses the oscillations of the ergotropy and drives
the dynamics toward a stationary value. The maximum ergotropy remains
comparable to that obtained in the closed-system limit, indicating that
the environmental coupling does not substantially reduce the maximum
extractable energy in this configuration. Thus, for the NN Ising battery
with a noninteracting chargoid, decoherence strongly modifies the
temporal charging dynamics while leaving the maximum stored energy and
ergotropy comparatively robust. At the same time, the suppression of
coherent energy backflow can lead to an enhancement of the maximum
charging power.

The XY-interacting battery exhibits a qualitatively different response to environmental coupling. As shown in Fig.~\ref{open_int_B_nonint_C}(d--f), the maximum stored energy, charging power, and ergotropy all increase with increasing decoherence strength. The enhancement of stored energy reflects the modification of the many-body population dynamics induced by the dissipative processes. In particular, environmental coupling can suppress interference effects that limit energy accumulation in the coherent dynamics, while the damping of coherent oscillations reduces energy backflow and allows the battery to reach higher-energy states. Importantly, the simultaneous increase in ergotropy shows that the additional stored energy is not merely passive but remains available for extraction as useful work. Hence, for the XY-interacting battery, environmental coupling can simultaneously enhance the amount of stored energy, the charging rate, and its extractable component.

\begin{figure*}
    \centering
\includegraphics[width=0.30\linewidth,height=0.25\linewidth]{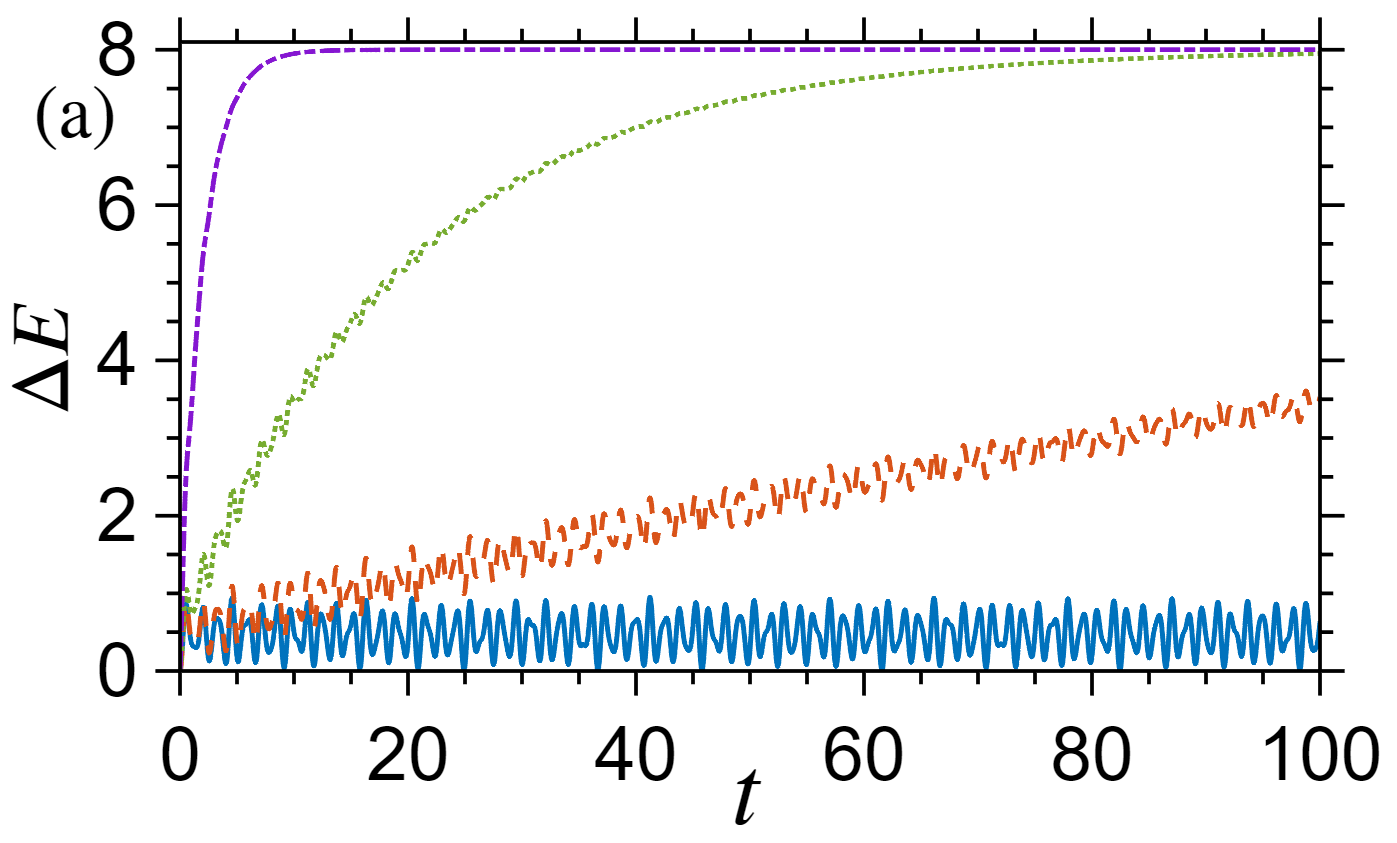}
\includegraphics[width=0.30\linewidth,height=0.25\linewidth]{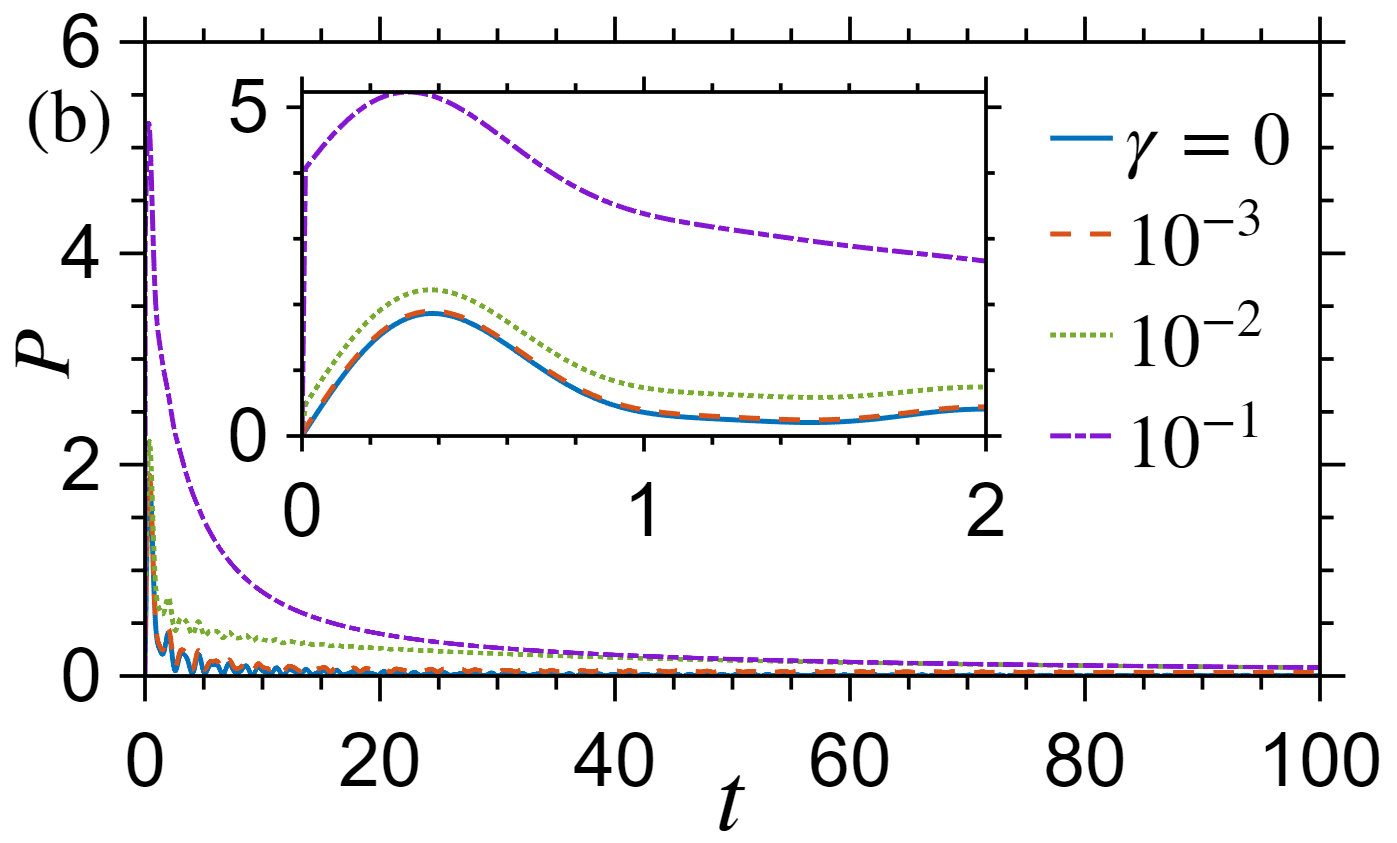}
\includegraphics[width=0.30\linewidth,height=0.25\linewidth]{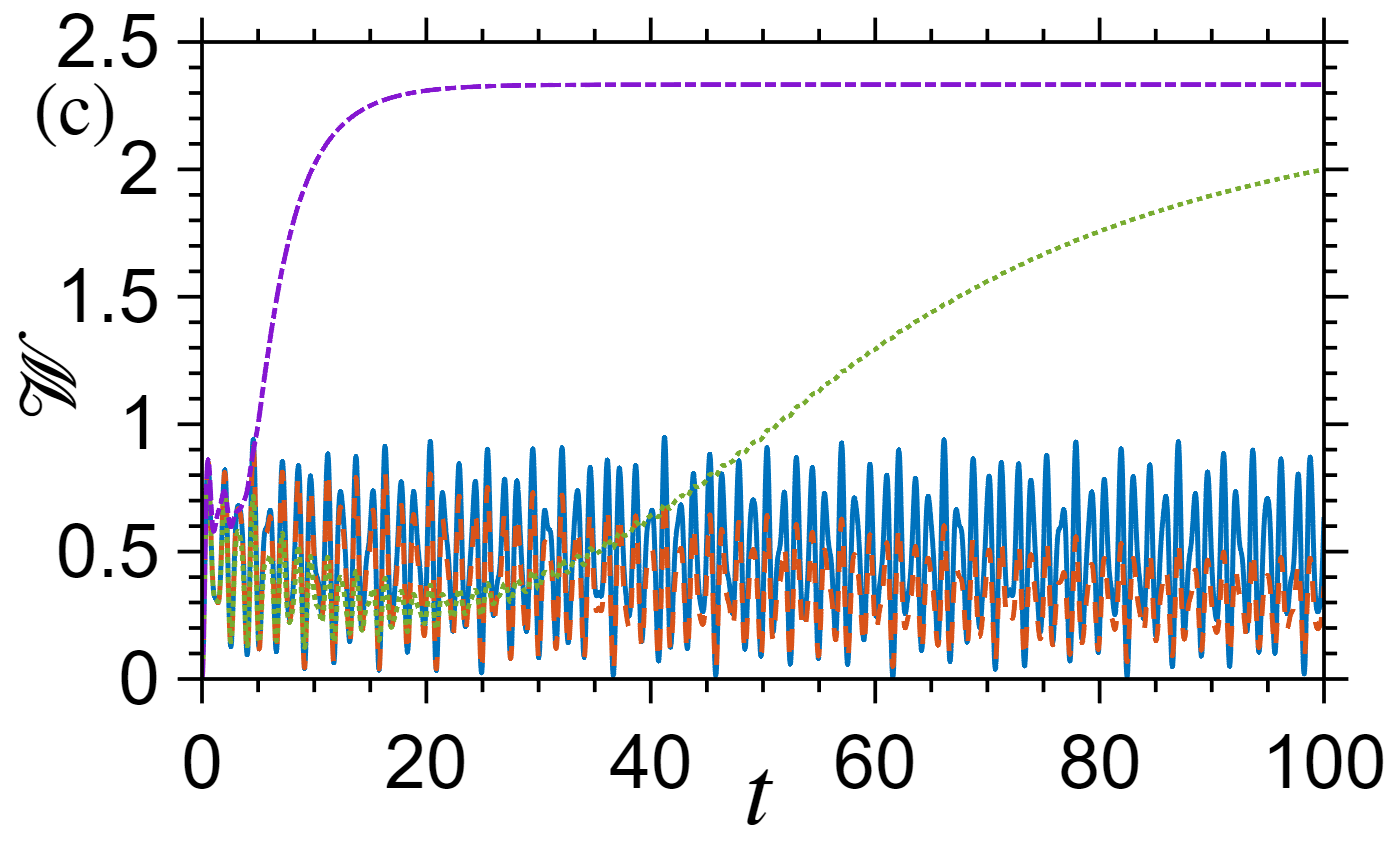}
\includegraphics[width=0.30\linewidth,height=0.25\linewidth]{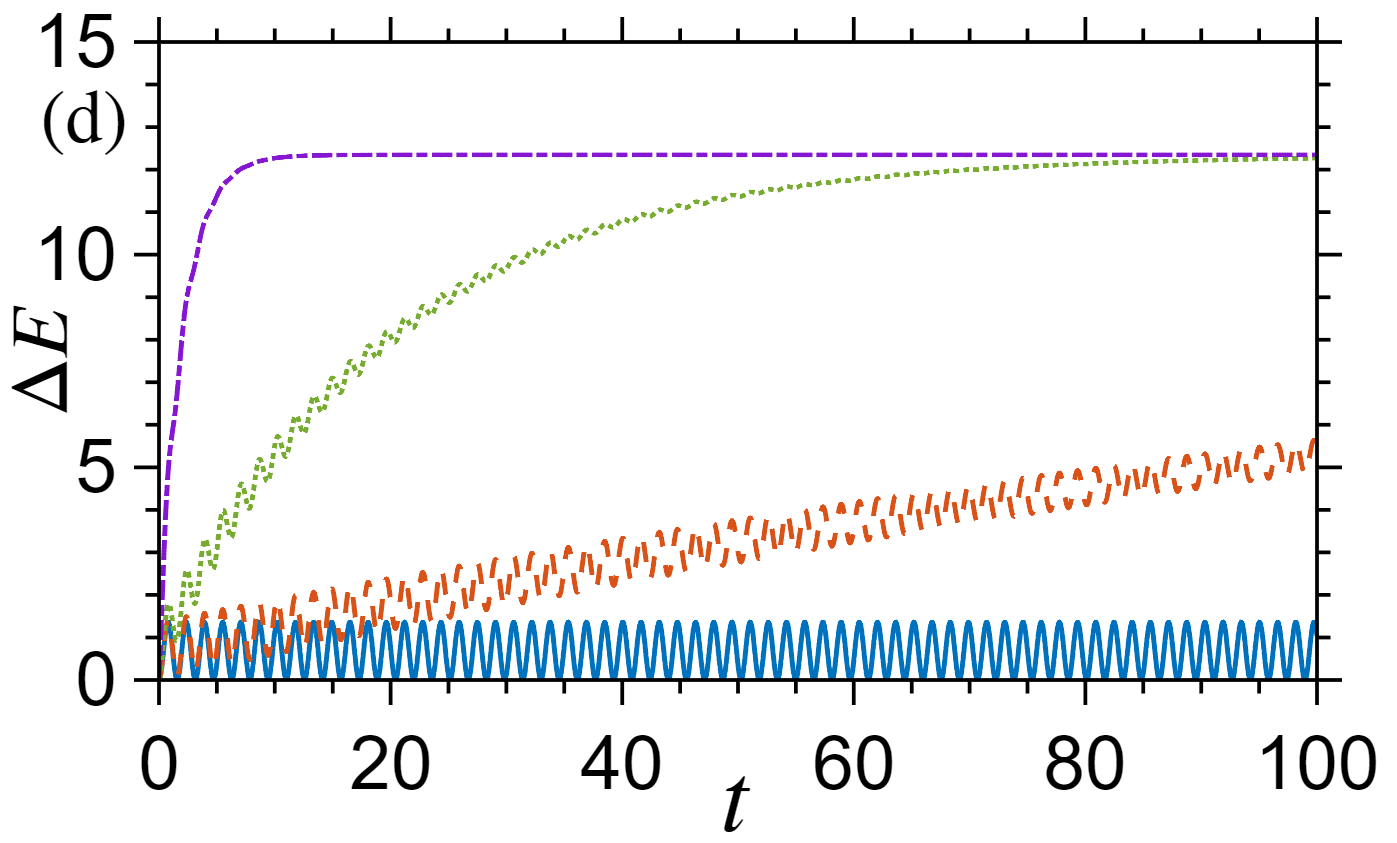}
\includegraphics[width=0.30\linewidth,height=0.25\linewidth]{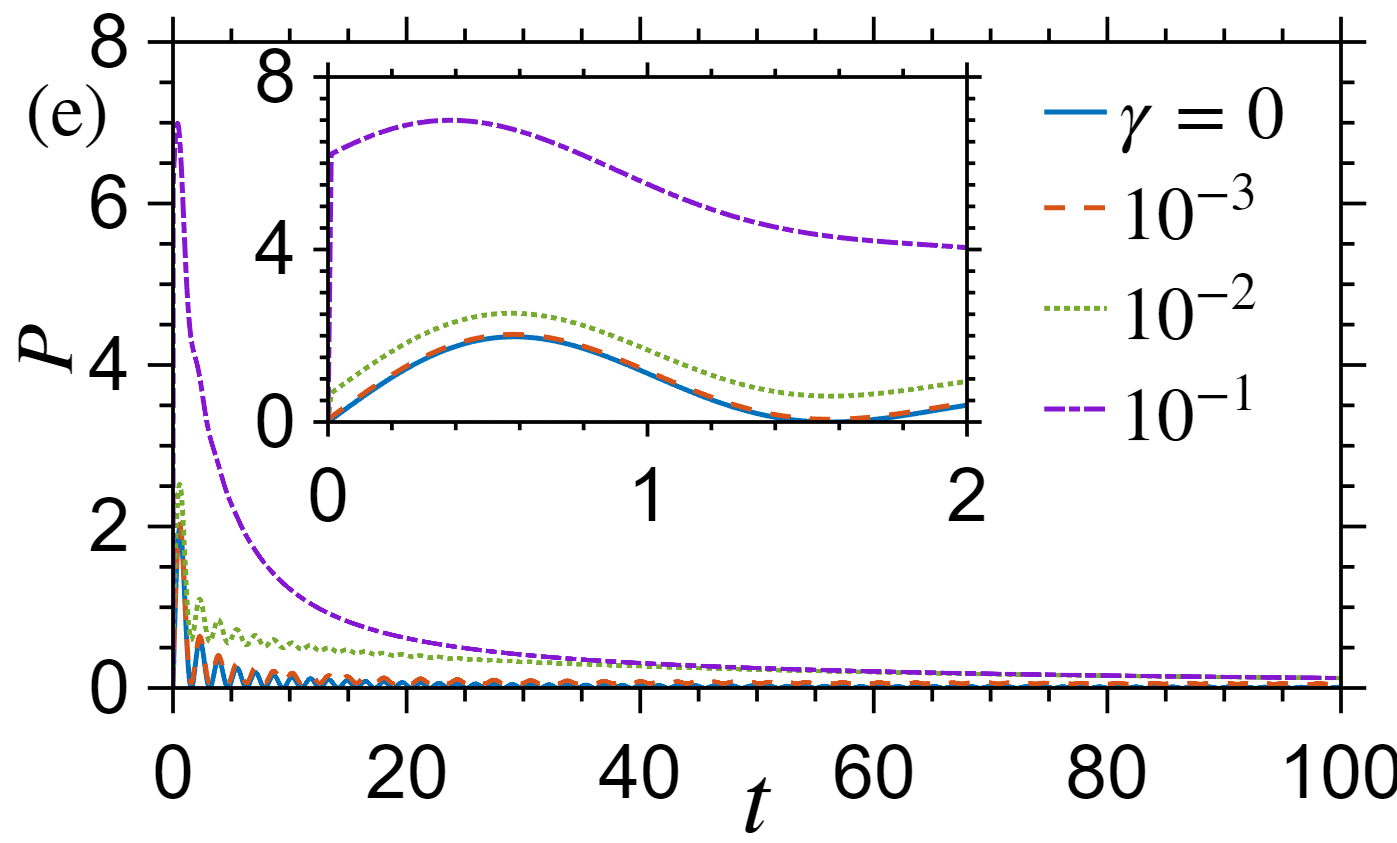}
\includegraphics[width=0.30\linewidth,height=0.25\linewidth]{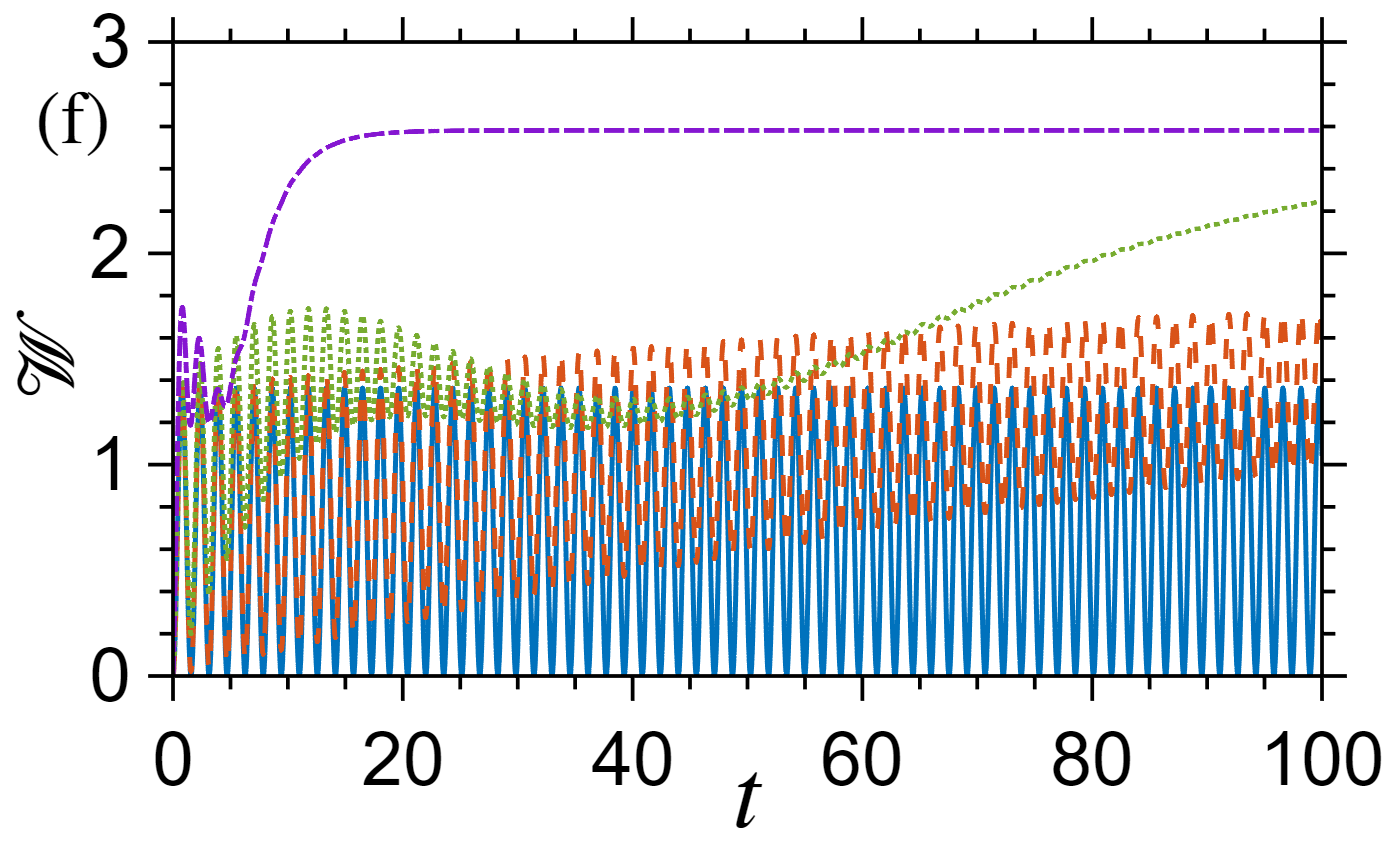}
   \caption{\textbf{Interacting batteries charged by interacting chargoids.} Charging dynamics of interacting batteries driven by interacting chargoids: (a--c) an NN Ising interacting battery driven by an NN XY interacting chargoid and (d--f) an NN XY interacting battery driven by an NN Ising interacting chargoid. Panels (a,d), (b,e), and (c,f) show the stored energy, charging power, and ergotropy, respectively, for the corresponding battery--chargoid configurations. The inset in the power panels highlights the short-time charging behavior. The system size is fixed at $N=8$ with periodic boundary conditions. The parameters are $J=1$, $\Gamma=0.5$, and $h_z=1$, and the battery countereffect parameter is set to $\lambda=1$.}

\label{open_int_B_int_C}
\end{figure*}

\subsection{Configuration III: Interacting battery with interacting chargoid.}

We finally consider the case in which both the battery and chargoid possess intrinsic interactions. Two complementary realizations are examined: an NN Ising-interacting battery with an NN anisotropic-XY chargoid, and an NN anisotropic-XY battery with an NN Ising chargoid. In this open-system setting, we investigate how energy relaxation and pure dephasing modify the charging dynamics for these two interaction structures. We focus on the fully compensated limit $\lambda=1$, for which the intrinsic battery contribution is suppressed during charging. The corresponding stored energy, charging power, and ergotropy are shown in Fig.~\ref{open_int_B_int_C}(a--f).

For both interaction arrangements, the stored energy, charging power, and ergotropy increase with increasing decoherence strength, while the pronounced oscillations of the closed-system dynamics are progressively suppressed. The enhancement of stored energy indicates that environmental coupling modifies the many-body dynamics in a manner that favors energy accumulation. In particular, the damping of coherent oscillations reduces the backflow of energy during the charging process, allowing the battery to retain a larger fraction of the transferred energy. At the same time, this suppression of coherent back-and-forth dynamics enables the battery to reach higher-energy states more rapidly, resulting in an enhancement of the charging power.

Importantly, the simultaneous increase in ergotropy demonstrates that the additional energy accumulated in the presence of decoherence is not predominantly passive. Instead, a substantial fraction remains available for extraction as useful work. Thus, for both interaction arrangements, environmental coupling produces a constructive effect, enhancing not only the total stored energy but also its charging rate and extractable component.

For completeness, we also examine the environmental response in the complementary limit ($\lambda=0$), where the intrinsic battery contribution remains fully active during the charging dynamics. As the decoherence strength ($\gamma$) is increased, the stored energy, charging power, and ergotropy exhibit the same overall qualitative trends observed under the fully compensated condition ($\lambda=1$), although their magnitudes are generally reduced. Thus, environmental relaxation and dephasing modify the charging performance in a qualitatively similar manner even when the intrinsic battery dynamics is retained. The detailed results for ($\lambda=0$) are presented in Appendix~\ref{open_System_dynamics_l0}.

Comparing the two limits, we find that the primary effect of the countereffect is to control the magnitude of the charging performance, rather than to qualitatively alter its response to environmental decoherence. In particular, the fully compensated case ($\lambda=1$) generally allows larger stored energy, charging power, and ergotropy than the corresponding ($\lambda=0$) case. This comparison shows that the enhancement produced by the countereffect persists in the presence of environmental coupling, while the qualitative influence of relaxation and dephasing remains largely unchanged by whether the intrinsic battery contribution is fully active or compensated.


\section{Conclusion}
\label{conclusion}
\vspace{0.25cm}

We have investigated the charging dynamics of a many-body quantum battery within a unified battery--chargoid framework, where the battery and chargoid represent distinct Hamiltonian roles of the same many-body system. This framework provides a systematic way to examine how the intrinsic battery dynamics influences charging when the charging Hamiltonian is externally activated. By introducing the countereffect parameter $\lambda$, we obtain direct control over the contribution of the battery Hamiltonian during the charging process and investigate its impact on stored energy and charging power for a range of battery--chargoid configurations.

Our results show that the intrinsic battery dynamics can oppose the energy-transfer process induced by the chargoid. Consequently, compensating for this contribution generally improves charging performance, with the optimal behavior typically achieved in the fully compensated limit, $\lambda=1$. Across the configurations considered, suppressing the counteracting battery dynamics can substantially enhance both the maximum stored energy and the charging power. This demonstrates that improving quantum-battery performance need not rely solely on strengthening the charging interaction; controlling the intrinsic battery dynamics provides an additional and independent route for optimizing energy storage.

The interaction structure of the chargoid provides another important control parameter. In particular, all-to-all interactions can significantly enhance the achievable stored energy compared with nearest-neighbor interactions. For the all-to-all Ising chargoid, complete inversion of the battery, corresponding to $\Delta E_{\max}=2Nh_z$, can be achieved for even finite system sizes. This enhancement is accompanied by a pronounced odd--even effect originating from the parity-dependent structure of the all-to-all interactions under periodic boundary conditions. Importantly, however, this complete-inversion behavior is a finite-size effect. The corresponding recurrence time grows exponentially with system size and diverges in the thermodynamic limit, while the stored-energy density approaches a smooth limiting behavior. Thus, the enhanced energy storage observed for finite systems should be clearly distinguished from the long-time and thermodynamic-limit behavior.

We further considered configurations in which the battery, the chargoid, or both possess intrinsic interactions. These results demonstrate that charging performance is governed by the combined structure of the two Hamiltonian roles rather than by the chargoid interaction alone. Changing the interaction character assigned to the battery and chargoid can modify both the amount of energy stored and the rate at which it is transferred. This highlights the importance of treating the intrinsic battery dynamics as an active component of the charging protocol rather than as a fixed background contribution.

To assess the robustness of these findings in realistic settings, we also extended the analysis to open-system dynamics by incorporating energy relaxation and pure dephasing. Environmental coupling does not produce a universal suppression of charging performance; instead, its effect depends on the underlying battery--chargoid architecture. In some configurations, decoherence suppresses coherent charging oscillations and consequently reduces the stored energy and charging power. In others, it can suppress coherent energy backflow and thereby enhance the stored energy, charging power, and ergotropy simultaneously. The latter result is particularly noteworthy, as it shows that an increase in stored energy need not correspond predominantly to passive energy: the ergotropy can increase as well, indicating that a substantial fraction of the stored energy remains extractable as useful work. The qualitative influence of environmental coupling remains similar when the intrinsic battery contribution is retained, although the fully compensated case $\lambda=1$ generally yields superior charging performance.

An additional feature appears in noninteracting battery and ATA chargoid configuration, where the maximum charging power occurs at $\lambda>1$, beyond the regime in which the countereffect parameter admits a direct physical interpretation. Although this regime is not directly relevant for practical implementation, the emergence of such behavior points to nontrivial dynamical features of the underlying Hamiltonian structure and may motivate further theoretical investigation (for detail discussion, see Appendix~\ref{Appendix_lambd_g1}).

\vspace{0.25cm}
\section*{ Data availability}
\vspace{0.25cm}
The data supporting the findings of this study are openly available at \cite{github_data_QB}.

\appendix
\section{Odd--Even Effects in a Noninteracting Battery with an All-to-All Interacting Ising Chargoid}
\label{sec:odd_even_analytical}

In the main text, we showed that the ATA interaction structure of the chargoid gives rise to a pronounced odd--even effect when the battery is noninteracting. Here, we provide an analytical treatment of this parity-dependent behavior for this configuration. In particular, by setting $\lambda=1$, the intrinsic battery contribution is completely compensated, so that the charging dynamics is governed solely by the ATA chargoid Hamiltonian. This limit allows us to isolate the role of the long-range interaction structure and analytically investigate the origin and system-size dependence of the observed odd--even effect.

We consider a quantum battery initially prepared in the fully polarized ground state,
\begin{equation}
\ket{\psi_0}
= \ket{\downarrow\downarrow\cdots\downarrow},
\end{equation}
with the noninteracting battery Hamiltonian
\begin{equation}
\hat{H}_B^0
=h_z\sum_{j=1}^{N}\hat{\sigma}_j^z.
\end{equation}
The charging protocol is implemented by switching on an ATA interacting Ising chargoid of the form
\begin{equation}
\hat{H}_C^{\rm I, ATA}
= J\sum_{j=1}^{N}
\sum_{k=1}^{\mathcal{K}}
\frac{1}{2^{k-1}}
\hat{\sigma}_j^x\hat{\sigma}_{j+k}^x,
\label{eq:ATA_charger}
\end{equation}
with periodic boundary conditions. Here, the interaction strength decreases geometrically with the distance, $J_k=\frac{J}{2^{k-1}},$
and the maximum interaction range is $\mathcal{K}=(N-1)/2$ for odd $N$ and $\mathcal{K}=N/2$ for even $N$. The Hamiltonian governing the charging dynamics is therefore
\begin{equation}
\hat{H}_{\rm ch}
=\hat{H}_C+(1-\lambda)\hat{H}_B,
\end{equation}
where the countereffect parameter $\lambda$ controls the contribution of the intrinsic battery Hamiltonian during charging. In particular, for the fully compensated case $\lambda=1$, the battery contribution is completely suppressed and the dynamics is generated solely by the chargoid,
\begin{equation}
\left.\hat{H}_{\rm ch}\right|_{\lambda=1}
=\hat{H}_C.
\end{equation}
This limit allows us to isolate the role of the chargoid interaction structure and analytically characterize the odd--even effect observed in the main text. In the following, we analyze the stored energy and its temporal periodicity for even and odd system sizes separately, and further examine the behavior of the stored-energy density in the thermodynamic limit.

\subsection{Stored enegy}
The initial energy of the battery is $E_0
=
\langle\psi_0|\hat{H}_B|\psi_0\rangle
=
-Nh_z.$ The stored energy is defined as $\Delta E_N(t)
= E_N(t)-E_0,$ where $E_N(t) = \langle\psi_0| e^{i\hat{H}_Ct} \hat{H}_B e^{-i\hat{H}_Ct} |\psi_0\rangle.$

A crucial property of the ATA charger is that all pairwise interaction terms commute. Since all terms contain only Pauli operators of the same $x$ component,
\begin{equation}
\left[
\hat{\sigma}_i^x\hat{\sigma}_j^x,
\hat{\sigma}_l^x\hat{\sigma}_m^x
\right]
=0
\end{equation}
for arbitrary sites $i,j,l,m$. Consequently, the time-evolution operator can be factorized as
\begin{equation}
\hat{U}(t)
=
e^{-i\hat{H}_Ct}
=
\prod_{j,k}
\exp\left[
-iJ_k t
\hat{\sigma}_j^x\hat{\sigma}_{j+k}^x
\right].
\label{eq:factorized_U}
\end{equation}
For a single interaction between sites $j$ and $l$,
\begin{equation}
\hat{U}_{jl}(t)
=
\exp\left(
-iJ_k t\hat{\sigma}_j^x\hat{\sigma}_l^x
\right),
\end{equation}
the operator $\hat{\sigma}_j^z$ transforms according to
\begin{equation}
\hat{U}_{jl}^{\dagger}(t)
\hat{\sigma}_j^z
\hat{U}_{jl}(t)
=
\hat{\sigma}_j^z\cos(2J_k t)
+
\hat{\sigma}_j^y\hat{\sigma}_l^x\sin(2J_k t).
\label{eq:single_rotation}
\end{equation}
Because the initial state is an eigenstate of every $\hat{\sigma}_j^z$ operator, $\langle\psi_0|
\hat{\sigma}_j^y\hat{\sigma}_l^x
|\psi_0\rangle
=0,$ and hence every interaction connected to site $j$ contributes a cosine factor to the $z$-polarization.

\paragraph*{Odd system size.} For odd $N$, the maximum interaction range is $\mathcal{K}=\frac{N-1}{2}$. At every distance $k=1,\ldots,\frac{N-1}{2},$ each spin has two equivalent partners on the periodic ring. Consequently, each interaction range contributes two identical cosine factors, yielding
\begin{equation}
\left\langle\hat{\sigma}_j^z(t)\right\rangle_{\mathrm{odd}}
=
-
\prod_{k=1}^{(N-1)/2}
\cos^2(2J_k t).
\label{eq:sigma_z_odd}
\end{equation}
Using $J_k=\frac{J}{2^{k-1}}$, this becomes
\begin{equation}
\left\langle\hat{\sigma}_j^z(t)\right\rangle_{\mathrm{odd}}
=
-
\prod_{k=1}^{(N-1)/2}
\cos^2
\left(
\frac{2Jt}{2^{k-1}}
\right).
\label{eq:sigma_z_odd_explicit}
\end{equation}
The corresponding battery energy is
\begin{equation}
E_N^{\mathrm{odd}}(t)
=
-Nh_z
\prod_{k=1}^{(N-1)/2}
\cos^2
\left(
\frac{2Jt}{2^{k-1}}
\right).
\label{eq:energy_odd}
\end{equation}
Since $E_0=-Nh_z$, the stored energy is
\begin{equation}
\Delta E_N^{\mathrm{odd}}(t)
=
Nh_z
\left[
1-
\prod_{k=1}^{(N-1)/2}
\cos^2
\left(
\frac{2Jt}{2^{k-1}}
\right)
\right].
\label{eq:stored_energy_odd}
\end{equation}
Because $0\leq
\cos^2(2J_k t)
\leq1,$ we have $0\leq
\Delta E_N^{\mathrm{odd}}(t)
\leq Nh_z.$ Therefore, the maximum dynamically accessible stored energy for odd $N$ is
\begin{equation}
\Delta E_{\max}^{\mathrm{odd}}
=
Nh_z.
\label{eq:max_stored_odd}
\end{equation}
Thus, although the maximum possible energy difference between the ground and fully excited battery states is $2Nh_z$, the odd-$N$ ATA charging dynamics accesses only half of this energy range.

\paragraph*{Even system size:} For even $N$, the maximum interaction range is $\mathcal{K}=\frac{N}{2}.$ The interaction at $k=\frac{N}{2}$ is special because it connects each spin to its unique diametrically opposite spin. Since the Hamiltonian in Eq.~\eqref{eq:ATA_charger} contains a sum over all $j$, this antipodal bond is counted twice. Hence, its effective coupling is $2J_{N/2}$. The resulting polarization is
\begin{equation}
\left\langle\hat{\sigma}_j^z(t)\right\rangle_{\mathrm{even}}
=
-
\left[
\prod_{k=1}^{N/2-1}
\cos^2(2J_k t)
\right]
\cos(4J_{N/2}t).
\label{eq:sigma_z_even}
\end{equation}

Using $J_k=\frac{J}{2^{(k-1)}}
$, we obtain
\begin{equation}
\left\langle\hat{\sigma}_j^z(t)\right\rangle_{\mathrm{even}}
=
-
\left[
\prod_{k=1}^{N/2-1}
\cos^2
\left(
\frac{2Jt}{2^{k-1}}
\right)
\right]
\cos
\left(
\frac{4Jt}{2^{N/2-1}}
\right).
\label{eq:sigma_z_even_explicit}
\end{equation}

The corresponding stored energy is
\begin{align}
\Delta E_N^{\mathrm{even}}(t)
={}&Nh_z\left\{
1-
\left[
\prod_{k=1}^{N/2-1}
\cos^2\left(\frac{2Jt}{2^{k-1}}\right)
\right]\right. \nonumber\\
&\left.\times
\cos\left(\frac{4Jt}{2^{N/2-1}}\right)
\right\}.
\label{eq:stored_energy_even}
\end{align}
Unlike the odd-$N$ case, the final antipodal contribution is an unsquared cosine and can therefore become negative. Consider the time
\begin{equation}
t_N^*
=
\frac{\pi}{4J_{N/2}}
=
\frac{\pi}{2J}2^{N/2-1}.
\label{eq:t_star_even}
\end{equation}
At this time, $\cos(4J_{N/2}t_N^*)=-1.$ Furthermore, for every $k<N/2$, $2J_k t_N^*
=
\pi\,2^{N/2-k-1},$
which is an integer multiple of $\pi$. Therefore, $\cos^2(2J_k t_N^*)=1.$ Consequently, $\left\langle\hat{\sigma}_j^z(t_N^*)\right\rangle_{\mathrm{even}}
=+1,$ and the battery reaches its fully inverted state, $E_N^{\mathrm{even}}(t_N^*)=+Nh_z.$ Hence, the maximum stored energy is $\Delta E_{\max}^{\mathrm{even}} = 2Nh_z.$ Thus, the maximum dynamically accessible stored energy exhibits a clear odd--even dependence,
\begin{equation}
\Delta E_{\max}(N)
=
\begin{cases}
Nh_z, & N~\mathrm{odd},\\[5pt]
2Nh_z, & N~\mathrm{even}.
\end{cases}
\label{eq:max_stored_odd_even}
\end{equation}
The origin of this parity dependence is the special antipodal interaction present only for even $N$. For odd $N$, every interaction distance occurs in pairs, leading to squared cosine factors and restricting the polarization to non-positive values. For even $N$, the unique antipodal interaction gives rise to an unsquared cosine factor, which can change sign and allow complete inversion of the battery.

\subsection{Charging period of stored energy}
\paragraph{Odd system size.} The oscillatory behavior of the stored energy is governed by the hierarchy of interaction strengths, $J_k=\frac{J}{2^{k-1}}$. For odd $N$, each contribution has the form $\cos^2(2J_k t)$, whose period is $T_k =\frac{\pi}{2J_k}.$ The longest period is associated with the weakest interaction, corresponding to the largest interaction range, $J_{\min}^{\mathrm{odd}}=J_{(N-1)/2}=\frac{J}{2^{(N-3)/2}}.$ Therefore,
\begin{equation}
T_N^{\mathrm{odd}}
=
\frac{\pi}{2J}
2^{(N-3)/2}.
\label{eq:period_odd}
\end{equation}
Consequently, $T_{N+2}^{\mathrm{odd}}
=
2T_N^{\mathrm{odd}},$
or equivalently, $T_{N+2}^{\mathrm{odd}}
=
2T_N^{\mathrm{odd}}.$

\paragraph*{Even system size.} For even $N$, the weakest interaction corresponds to the antipodal distance, $J_{N/2}= \frac{J}{2^{N/2-1}}.$ Because the antipodal interaction is counted twice, its contribution to the polarization is $\cos(4J_{N/2}t).$ The period of this factor is $T_N^{\mathrm{even}} =
\frac{2\pi}{4J_{N/2}} =
\frac{\pi}{2J_{N/2}},$ and hence
\begin{equation}
T_N^{\mathrm{even}}
=
\frac{\pi}{2J}
2^{N/2-1}.
\label{eq:period_even}
\end{equation}
It follows immediately that $T_{N+2}^{\mathrm{even}}
=
2T_N^{\mathrm{even}}.$ Thus, the period doubles whenever the system size is increased by two, independently within the odd-$N$ and even-$N$ sequences.

\subsection{Thermodynamic limit}

As the system size increases, $\mathcal{K}\sim\frac{N}{2},$ and the weakest interaction scale decreases exponentially, $J_{\min} \sim \frac{J}{2^{N/2}}.$ Consequently, the characteristic recurrence time grows as $T_N \sim \frac{2^{N/2}}{J},$ and hence $\lim_{N\rightarrow\infty}T_N=\infty$.

\paragraph*{Odd system size.} For odd $N$, the product in Eq.~\eqref{eq:sigma_z_odd_explicit} can be evaluated using the identity
\begin{equation}
\prod_{r=0}^{K-1}
\cos\left(\frac{x}{2^r}\right)
=
\frac{\sin(2x)}
{2^K\sin\left(x/2^{K-1}\right)}.
\label{eq:product_identity}
\end{equation}
Taking $x=2Jt, \qquad K=\frac{N-1}{2},$ gives
\begin{equation}
\left\langle\hat{\sigma}_j^z(t)\right\rangle_{\mathrm{odd}}
=
-
\left[
\frac{\sin(4Jt)}
{2^{(N-1)/2}
\sin\left(2Jt/2^{(N-3)/2}\right)}
\right]^2.
\label{eq:exact_product_odd}
\end{equation}
Taking the thermodynamic limit along the odd-$N$ sequence yields
\begin{equation}
\lim_{N\rightarrow\infty}
\left\langle\hat{\sigma}_j^z(t)\right\rangle
=
-
\left[
\frac{\sin(4Jt)}
{4Jt}
\right]^2.
\label{eq:thermodynamic_sigma}
\end{equation}
Therefore, the energy density becomes
\begin{equation}
\lim_{N\rightarrow\infty}
\frac{E_N(t)}{N}
=
-h_z
\left[
\frac{\sin(4Jt)}
{4Jt}
\right]^2,
\label{eq:thermodynamic_energy_density}
\end{equation}
and the stored energy per spin is
\begin{equation}
\lim_{N\rightarrow\infty}
\frac{\Delta E_N(t)}{N}
=
h_z
\left[
1-
\left(
\frac{\sin(4Jt)}
{4Jt}
\right)^2
\right].
\label{eq:thermodynamic_stored_energy}
\end{equation}
For long times,
\begin{equation}
\left[
\frac{\sin(4Jt)}
{4Jt}
\right]^2
\sim
\frac{1}{t^2},
\end{equation}
and therefore
\begin{equation}
\frac{\Delta E_N(t)}{N}
\longrightarrow h_z
\qquad
(t\rightarrow\infty).
\end{equation}
Thus, the finite-size periodic charging dynamics crosses over to a non-periodic algebraically relaxing behavior in the thermodynamic limit. The recurrence time diverges exponentially with system size, while the finite-time energy density approaches a smooth thermodynamic-limit form.

\paragraph*{Even system size.} For even $N$, the polarization contains an additional unsquared cosine factor associated with the unique antipodal interaction,
\begin{equation}
\langle \hat{\sigma}_j^z(t)\rangle
=\left[
\prod_{k=1}^{N/2-1}
\cos^2(2J_k t)
\right]
\cos(4J_{N/2}t).
\label{eq:sigma_z_even}
\end{equation}
For the exponentially decaying interaction strength $J_k=\frac{J}{2^{k-1}},$ the antipodal coupling is $J_{N/2} =\frac{J}{2^{N/2-1}},$ which vanishes exponentially with system size. Consequently, for any fixed finite time $t$, $\lim_{N\rightarrow\infty}
\cos\left(4J_{N/2}t\right)
=1$. The remaining product approaches the infinite product
\begin{equation}
\lim_{N\rightarrow\infty}
\prod_{k=1}^{N/2-1}
\cos^2
\left(
\frac{2Jt}{2^{k-1}}
\right)
=\prod_{k=1}^{\infty}
\cos^2
\left(
\frac{2Jt}{2^{k-1}}
\right).
\label{eq:infinite_product_even}
\end{equation}
Using the infinite-product identity
\begin{equation}
\prod_{k=1}^{\infty}
\cos\left(\frac{x}{2^{k-1}}\right)
=\frac{\sin(2x)}{2x},
\label{eq:infinite_product_identity}
\end{equation}
with $x=2Jt$, we obtain
\begin{equation}
\prod_{k=1}^{\infty}
\cos^2
\left(
\frac{2Jt}{2^{k-1}}
\right)
=\left[
\frac{\sin(4Jt)}{4Jt}
\right]^2.
\label{eq:infinite_product_squared}
\end{equation}
Therefore, the thermodynamic-limit polarization for even $N$ becomes
\begin{equation}
\lim_{N\rightarrow\infty}
\langle \hat{\sigma}_j^z(t)\rangle
=\left[
\frac{\sin(4Jt)}{4Jt}
\right]^2.
\label{eq:sigma_z_thermodynamic_even}
\end{equation}
For an initially fully polarized state, the initial battery energy is
\begin{equation}
E_0=-Nh_z.
\end{equation}
The stored energy is therefore
\begin{equation}
\Delta E_N(t)
=\langle \hat{H}_B(t)\rangle-E_0,
\end{equation}
Using translational invariance, the stored energy per spin is
\begin{equation}
\frac{\Delta E_N(t)}{N}
=h_z
\left[
1+\langle\hat{\sigma}_j^z(t)\rangle
\right].
\end{equation}
Consequently, in the thermodynamic limit,
\begin{equation}
\lim_{N\rightarrow\infty}
\frac{\Delta E_N(t)}{N}
=h_z
\left[
1-
\left(
\frac{\sin(4Jt)}{4Jt}
\right)^2
\right].
\label{eq:thermodynamic_stored_energy_even}
\end{equation}
For long times,
\begin{equation}
\left[
\frac{\sin(4Jt)}{4Jt}
\right]^2
\sim
\frac{1}{t^2},
\end{equation}
and hence
\begin{equation}
\lim_{t\rightarrow\infty}
\lim_{N\rightarrow\infty}
\frac{\Delta E_N(t)}{N}
=h_z.
\label{eq:thermodynamic_long_time}
\end{equation}
Thus, at fixed finite time followed by the long-time limit, the thermodynamic-limit stored energy behaves as
\begin{equation}
\Delta E_N(t)
\longrightarrow
Nh_z. 
\end{equation}

\section{System-Size Dependence of the Stored Energy for a Noninteracting Battery with a NN Ising Chargoid}
\label{NN_ana}

When the ATA Ising chargoid is replaced by a nearest-neighbor (NN) interacting Ising chargoid, the odd--even effect observed in the ATA case disappears. Moreover, the charging period no longer exhibits a system-size dependence. These features can be understood analytically from the structure of the NN chargoid Hamiltonian. The NN charger is given by $\hat{H}_{C}^{\mathrm{I,NN}}=J\sum_{j=1}^{N}
\hat{\sigma}{j}^{x}\hat{\sigma}_{j+1}^{x},$ where periodic boundary conditions are imposed. Unlike the ATA charger, the NN Hamiltonian contains only a single interaction range, $\mathcal{K}=1$, hence  $J_1=J,$ which is independent of the system size. Consequently, the interaction structure does not introduce the parity-dependent or exponentially varying energy scales responsible for the odd--even behavior and system-size-dependent recurrence times in the ATA case.
\par
For the NN chargoid, the dynamics of the local magnetization is particularly simple and is given by
\begin{equation}
\left\langle
\hat{\sigma}_j^z(t)
\right\rangle_{\mathrm{NN}}
=\cos^2(2Jt).
\end{equation}
Consequently, the battery energy evolves as
\begin{equation}
E_N^{\mathrm{NN}}(t)
=Nh_z\cos^2(2Jt),
\end{equation}
and the corresponding stored energy is
\begin{equation}
\Delta E_N^{\mathrm{NN}}(t)=Nh_z\sin^2(2Jt).
\label{eq:stored_energy_NN}
\end{equation}
The maximum stored energy is therefore
\begin{equation}
\Delta E_{\max}^{\mathrm{NN}}=Nh_z.
\end{equation}

The periodicity of the charging dynamics follows directly from the $\sin^2(2Jt)$ dependence, yielding
\begin{equation}
T_{\mathrm{NN}}=\frac{\pi}{2J}.
\end{equation}
In contrast to the ATA case, this period is completely independent of the system size,
\begin{equation}
T_{\mathrm{NN}}(N+2)=T_{\mathrm{NN}}(N).
\end{equation}
Thus, increasing the system size affects only the overall magnitude of the stored energy through the extensive factor $N$, while the underlying charging timescale remains unchanged. This absence of a system-size-dependent dynamical timescale is consistent with the lack of an odd--even effect for the NN chargoid.

\begin{figure}
\centering
\includegraphics[width=0.8\linewidth,height=0.6\linewidth]{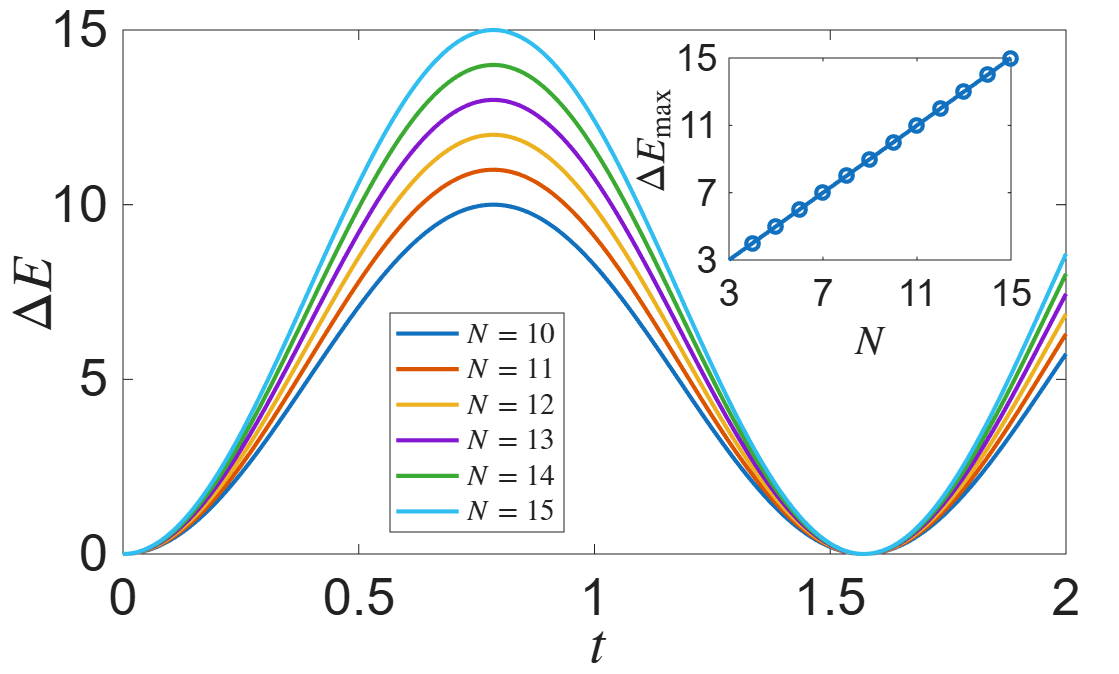}
 \caption{Stored energy $\Delta E$  as functions of time $t$ for a non-interacting battery charged by an NN XX-interacting Ising Hamiltonian for different system sizes $N$. The results are obtained for the countereffect parameter $\lambda=1$. The insets show the corresponding maximum stored energy as a function of system size. The simulations are performed with $J=1$ and $h_z=1$.}

\label{Ising_NN}
\end{figure}

The numerical results for the NN chargoid are shown in Fig.~\ref{Ising_NN} and are consistent with the analytical behavior derived above. In particular, the maximum stored energy increases with the system size, reflecting the extensive growth of the battery energy scale with the number of spins. Meanwhile, the charging dynamics for different system sizes exhibit essentially the same oscillation period, in agreement with the analytical result $T_{\mathrm{NN}}=\pi/(2J)$. Thus, increasing the system size changes the amount of energy stored but does not introduce a new dynamical timescale.

The system-size dependence of the maximum stored energy is further illustrated in the inset of Fig.~\ref{Ising_NN}, where we find an approximately linear scaling,
\begin{equation}
\Delta E_{\max}^{\mathrm{NN}}\approx h_zN.
\end{equation}
This extensive scaling indicates that the stored energy grows proportionally with the number of spins, while the relative fraction of the available battery energy remains approximately unchanged with increasing $N$. Hence, the NN chargoid exhibits a simple extensive charging behavior governed by a characteristic timescale set by the local interaction strength $J$, with no appreciable dependence on the system size.

\begin{figure*}
    \centering
\includegraphics[width=0.30\linewidth,height=0.25\linewidth]{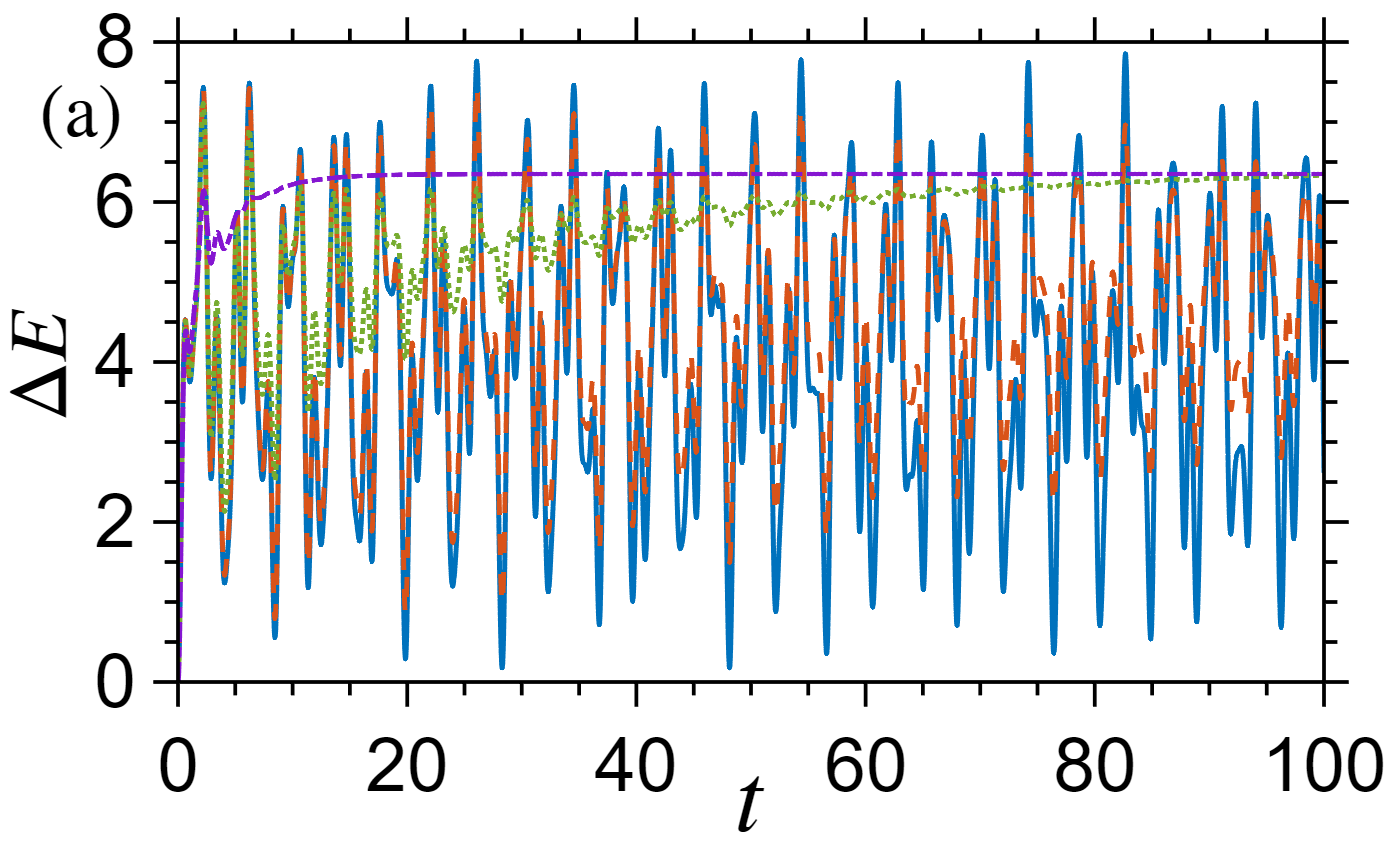}
\includegraphics[width=0.30\linewidth,height=0.25\linewidth]{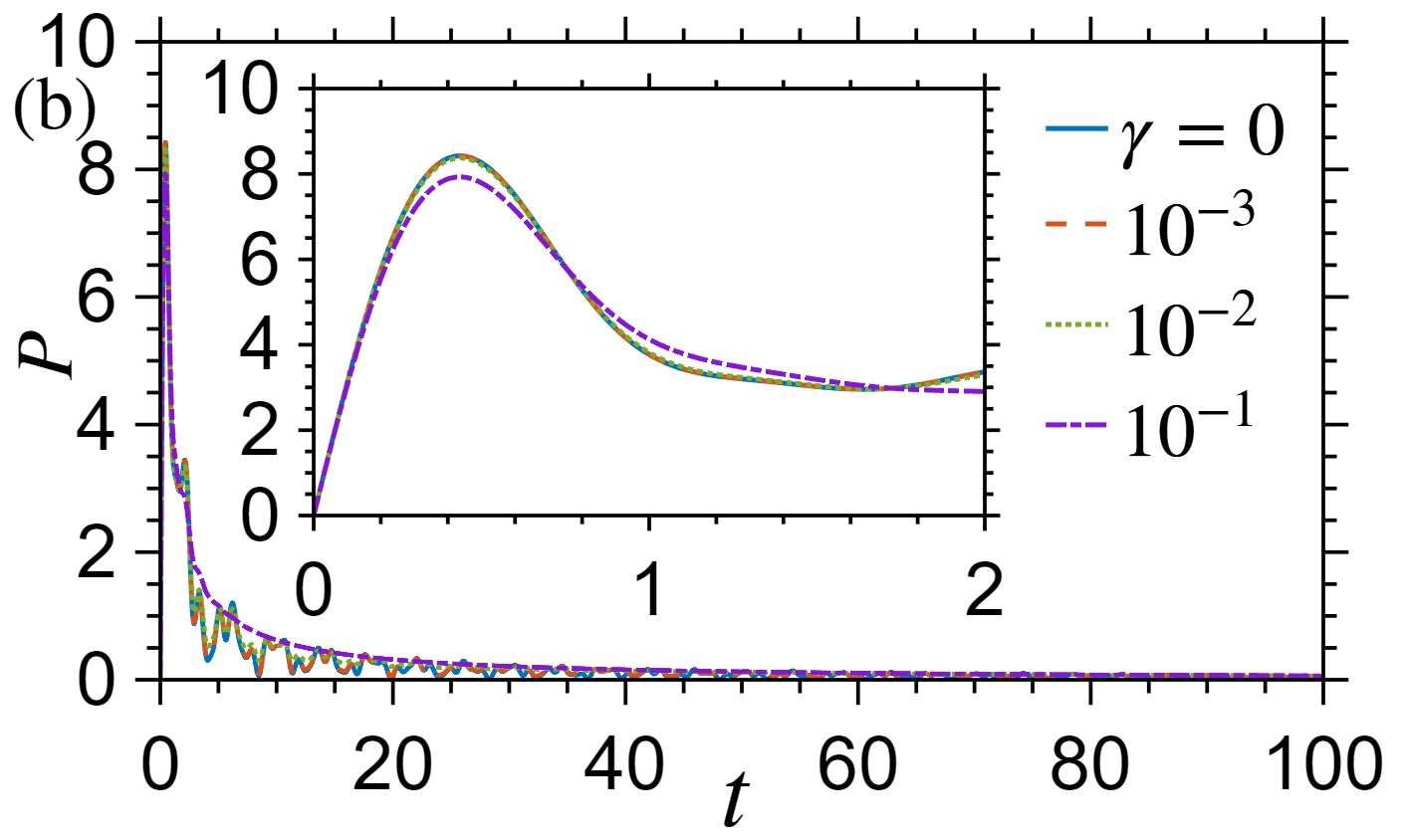}
\includegraphics[width=0.30\linewidth,height=0.25\linewidth]{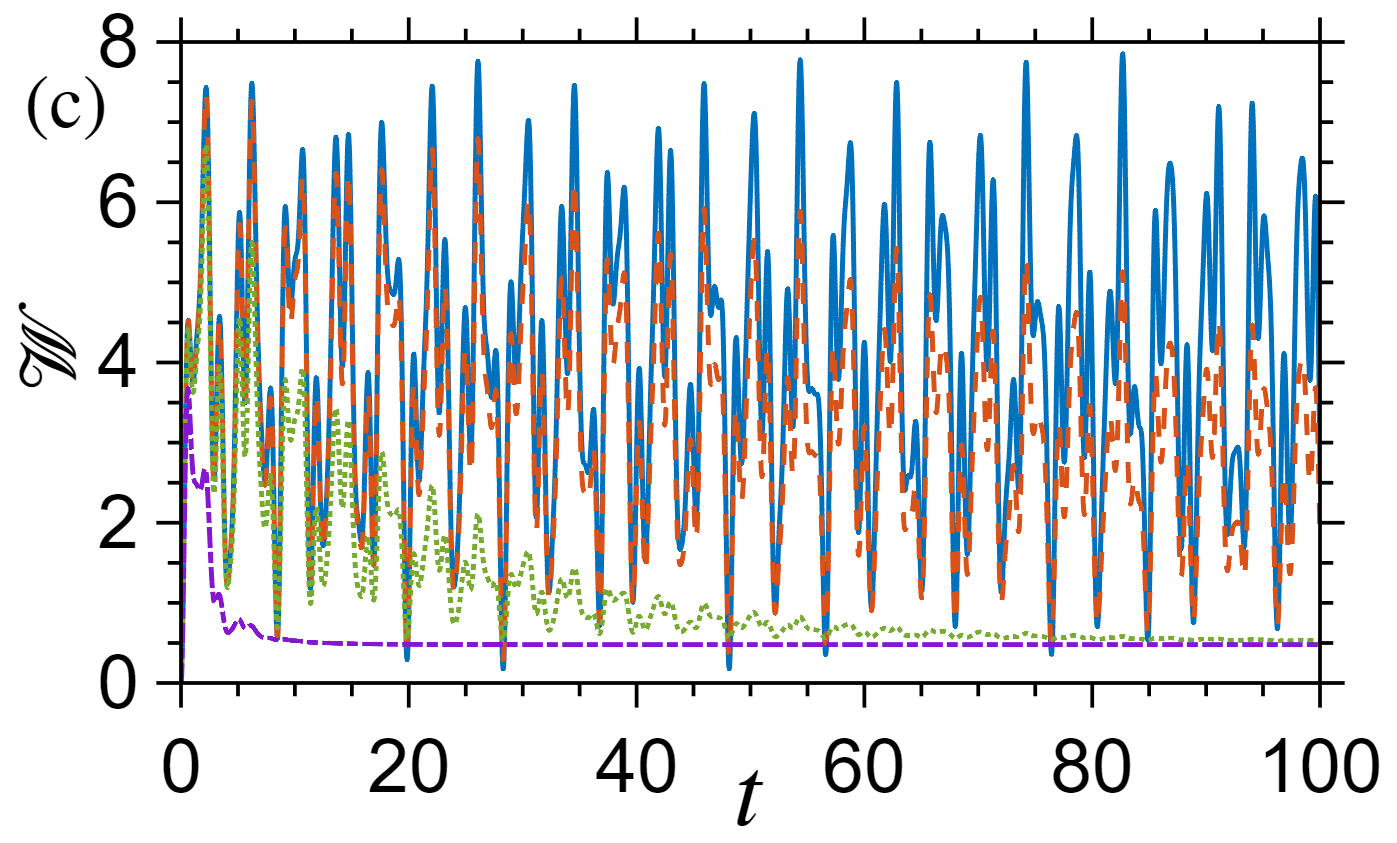}
\includegraphics[width=0.30\linewidth,height=0.25\linewidth]{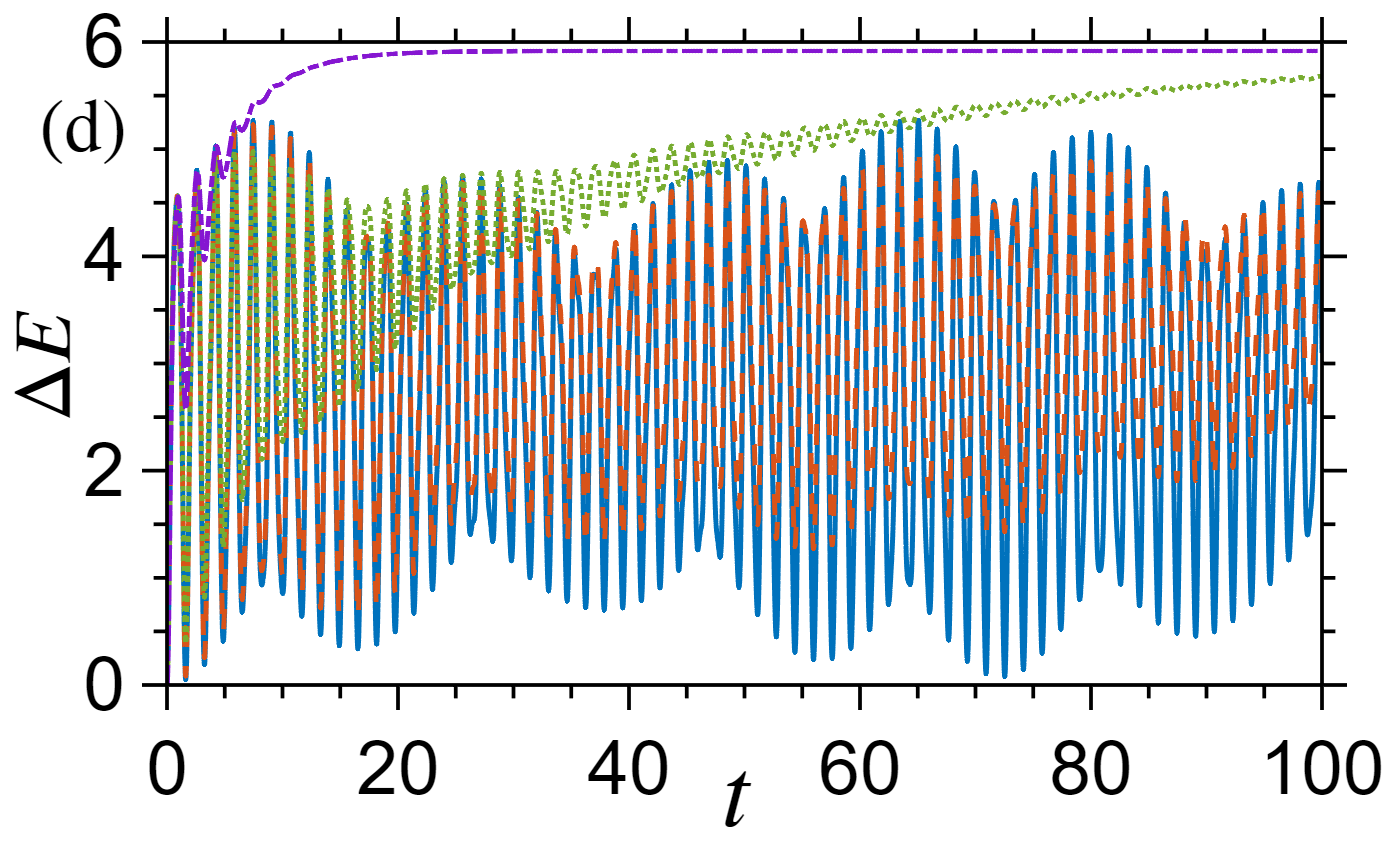}
\includegraphics[width=0.30\linewidth,height=0.25\linewidth]{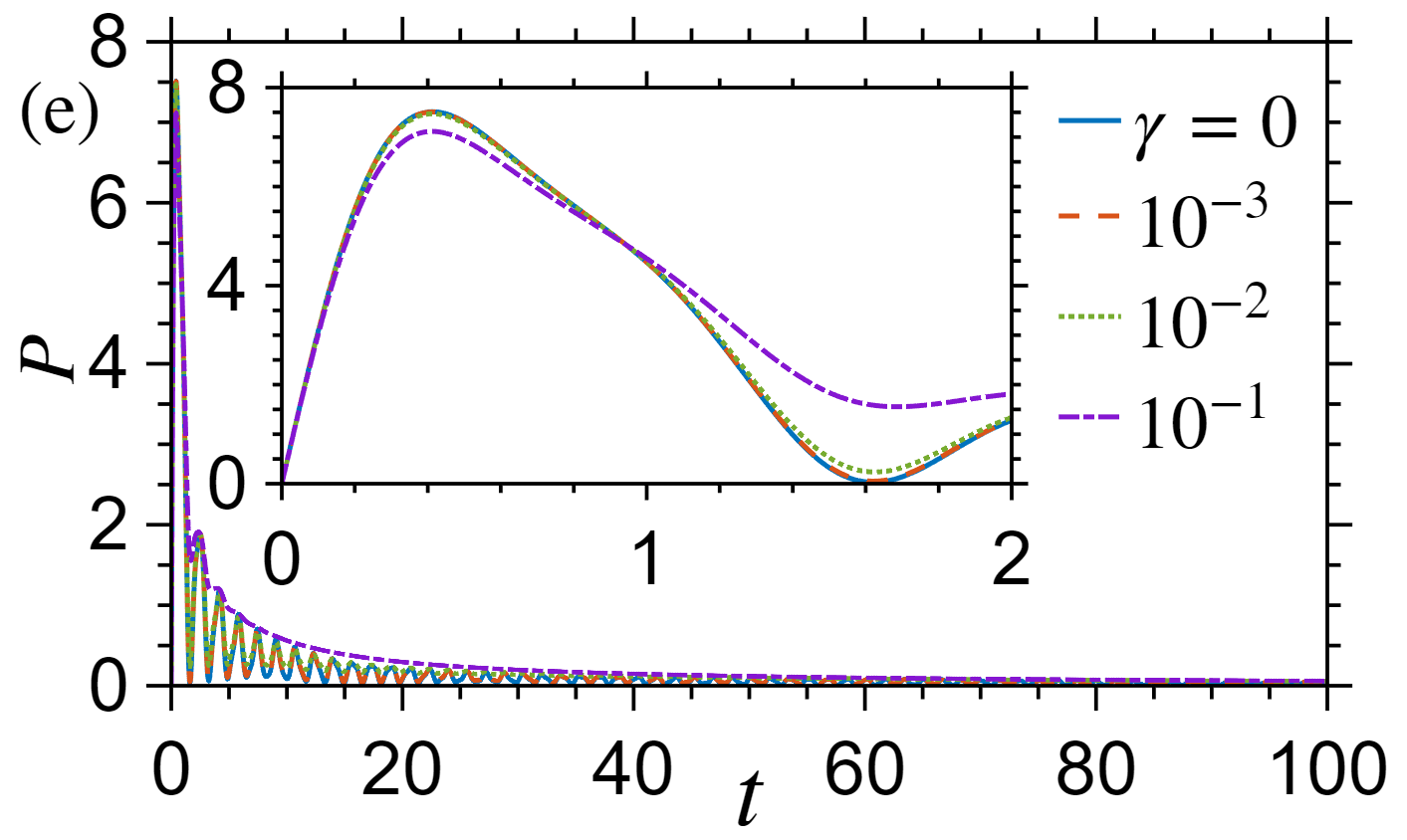}
\includegraphics[width=0.30\linewidth,height=0.25\linewidth]{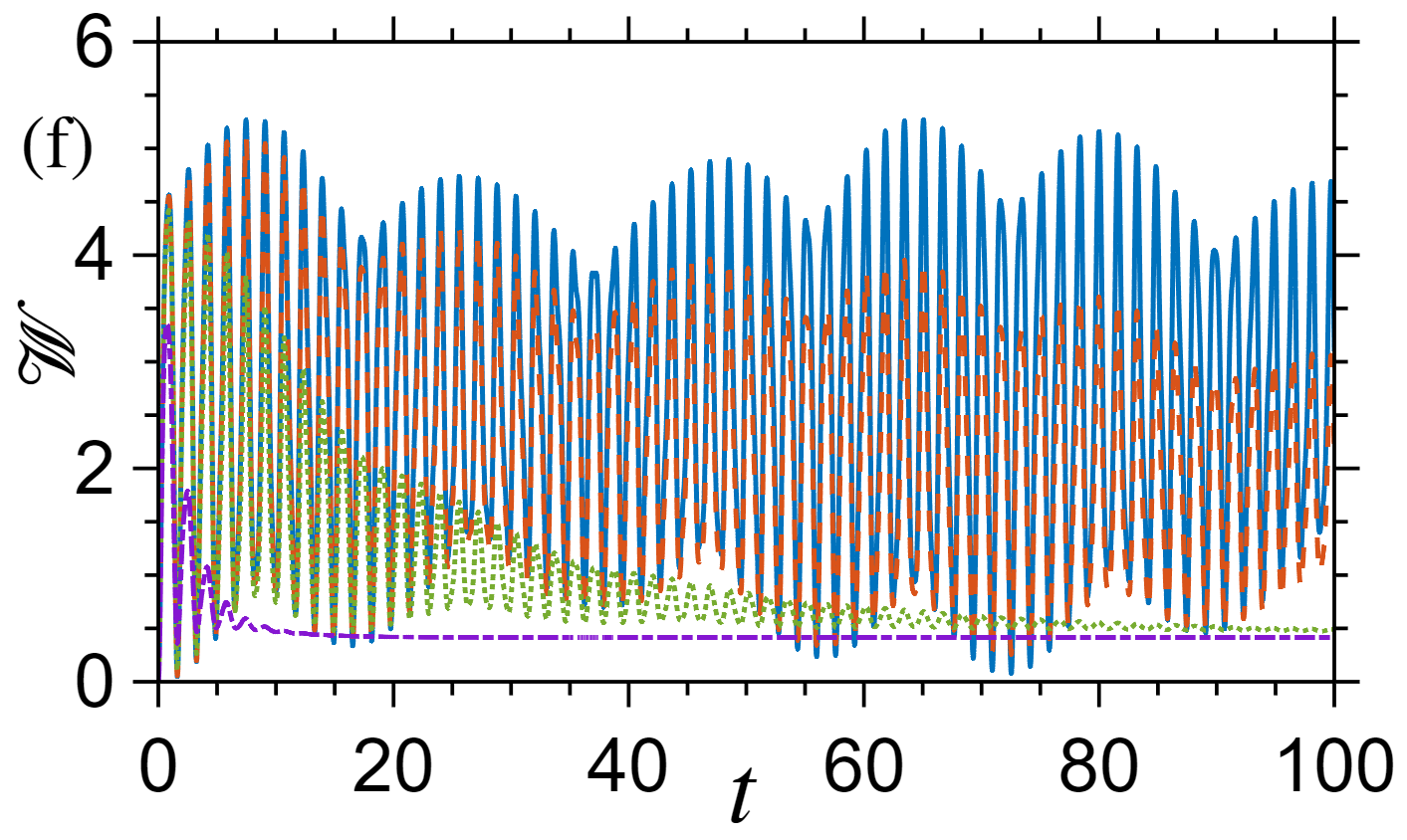}
    \caption{Same as Fig.~\ref{open_nonint_B_int_C}, but for $\lambda=0$.
}
    \label{open_nonint_B_int_C_l0}
\end{figure*}

\begin{figure*}
    \centering
\includegraphics[width=0.30\linewidth,height=0.25\linewidth]{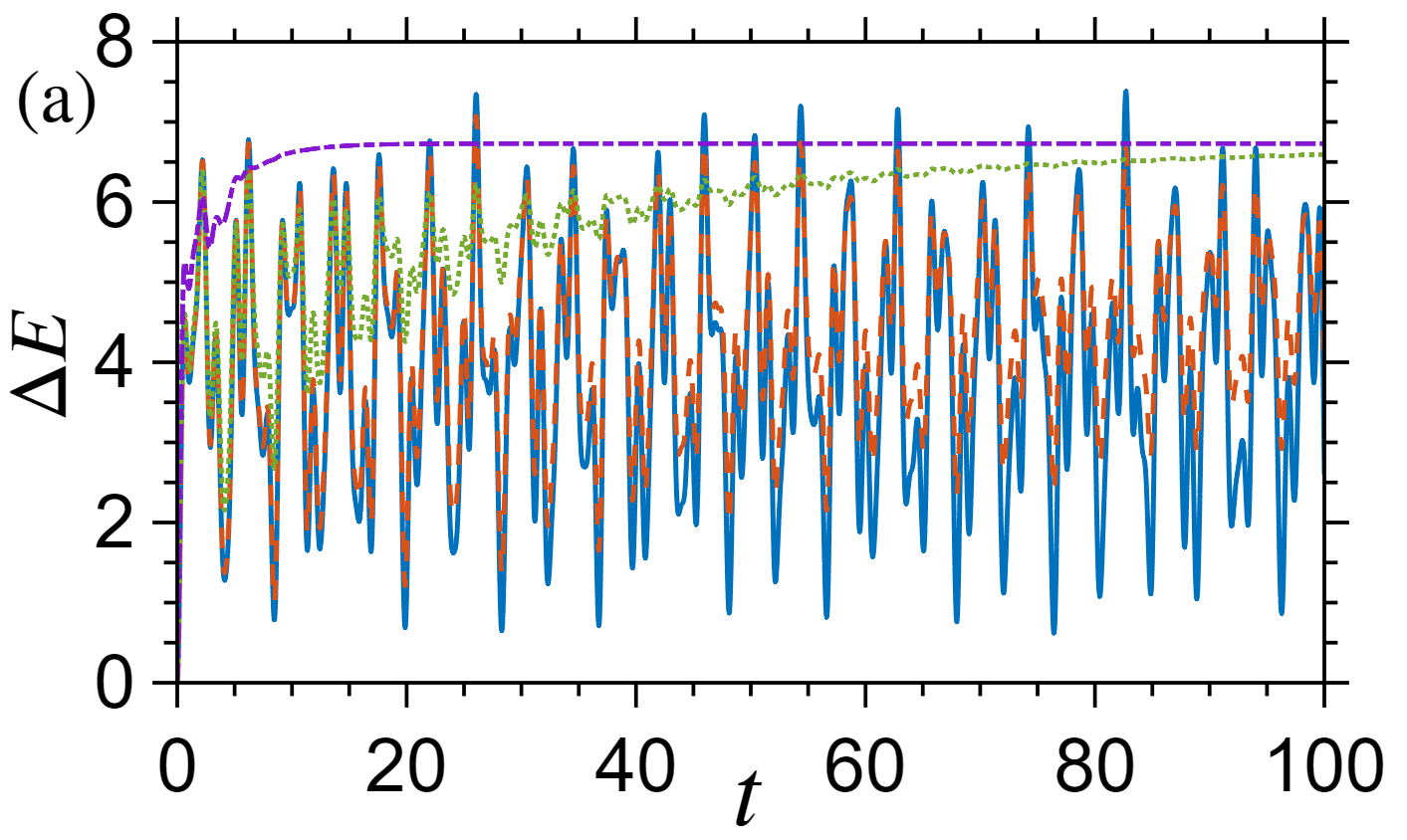}
\includegraphics[width=0.30\linewidth,height=0.25\linewidth]{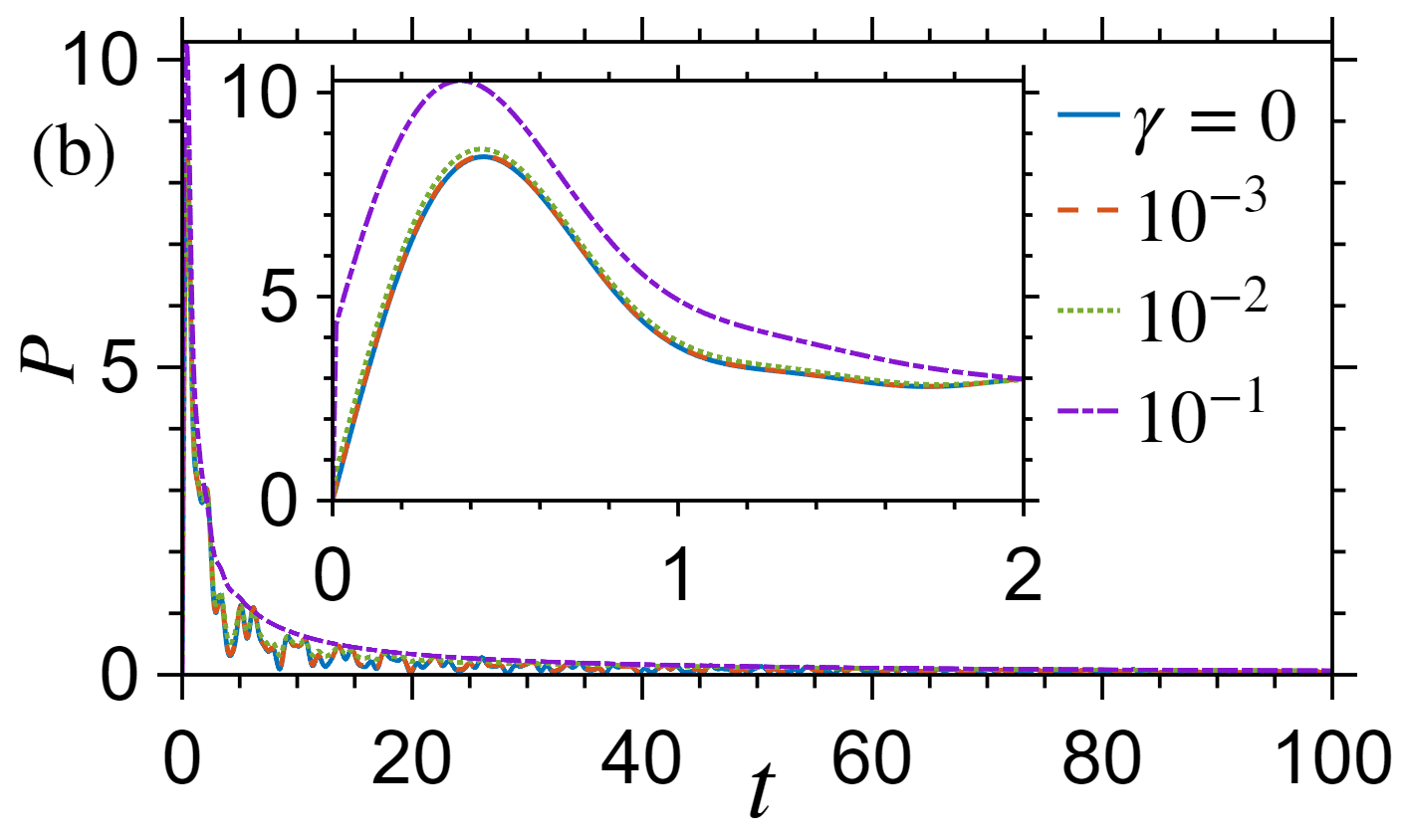}
\includegraphics[width=0.30\linewidth,height=0.25\linewidth]{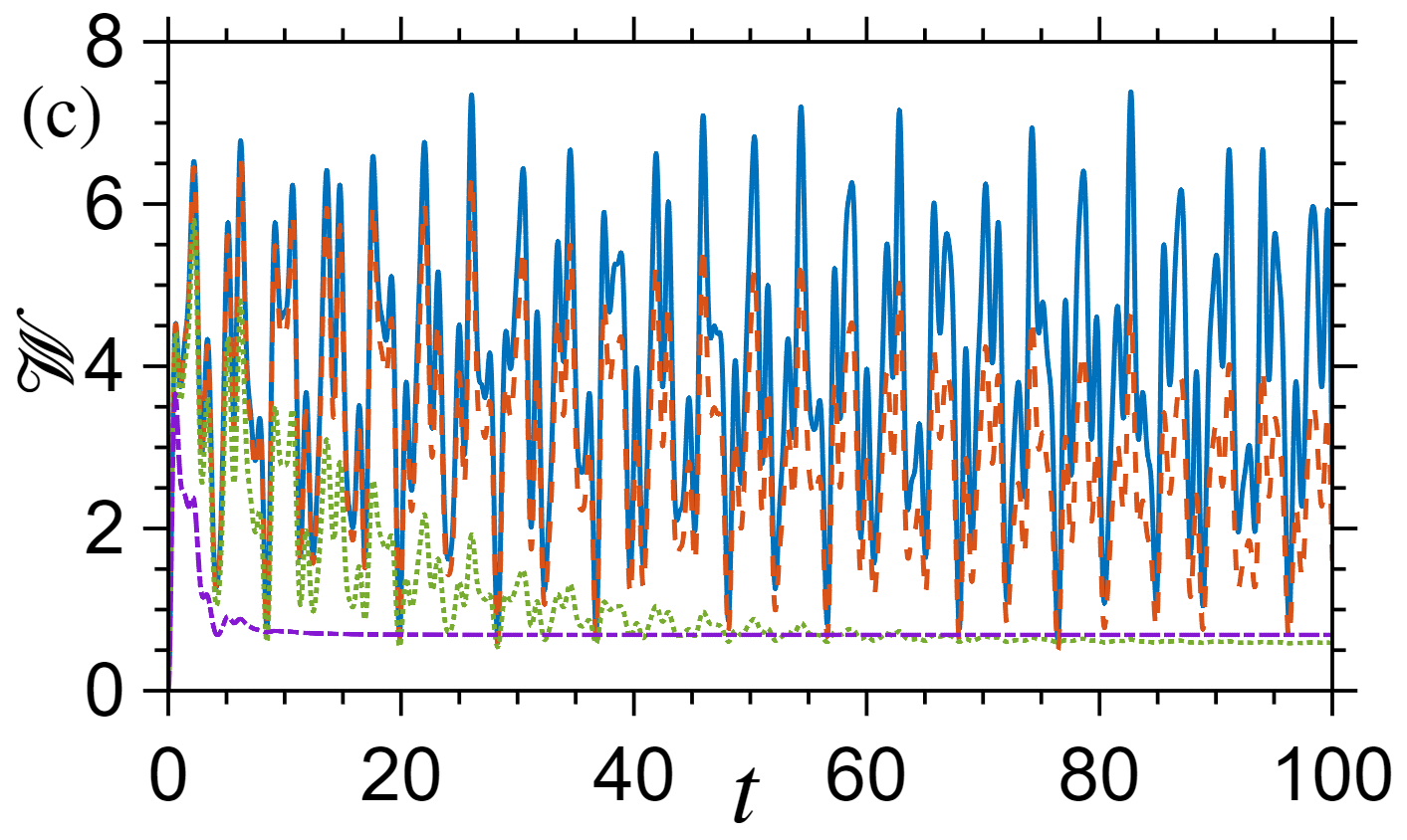}
\includegraphics[width=0.30\linewidth,height=0.25\linewidth]{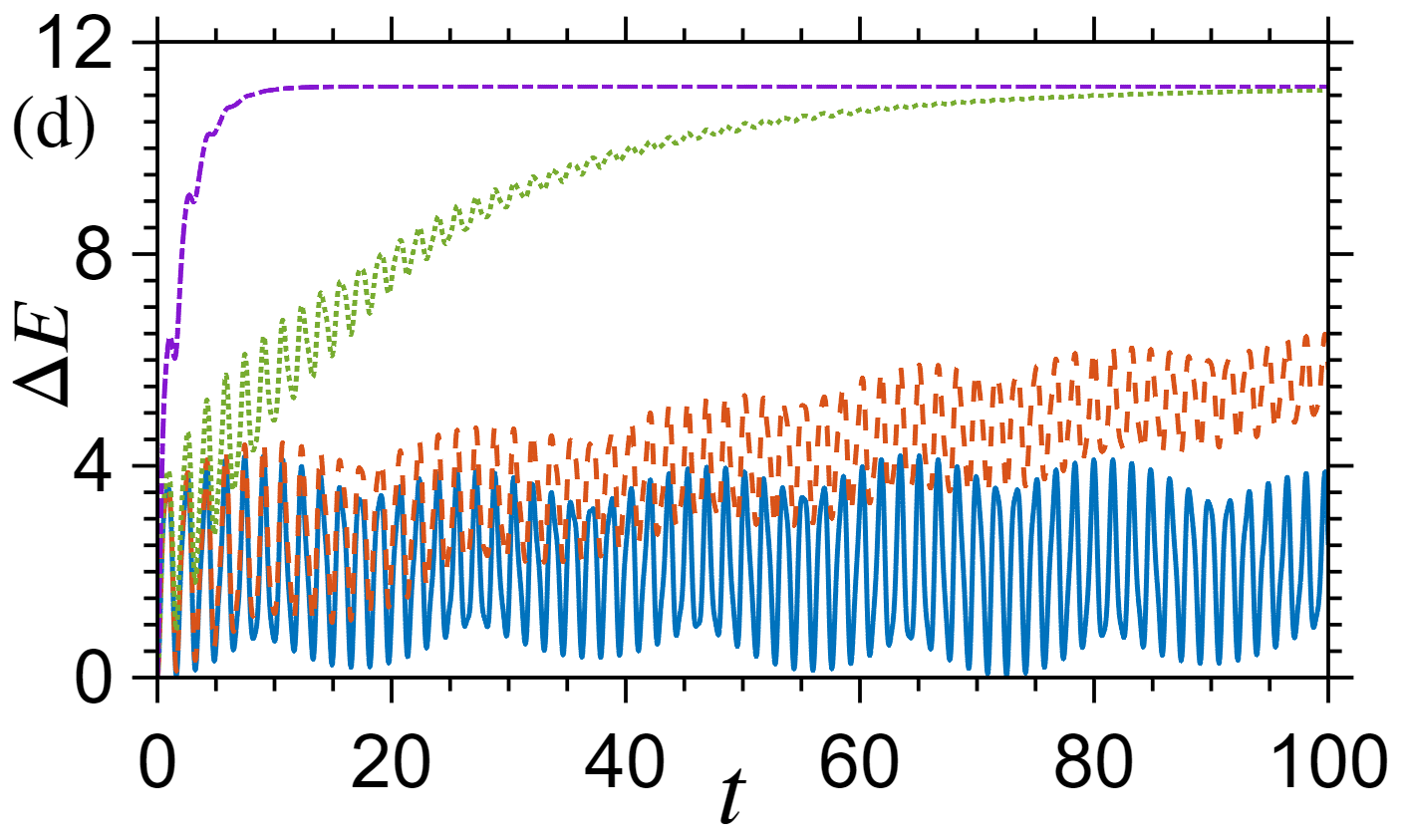}
\includegraphics[width=0.30\linewidth,height=0.25\linewidth]{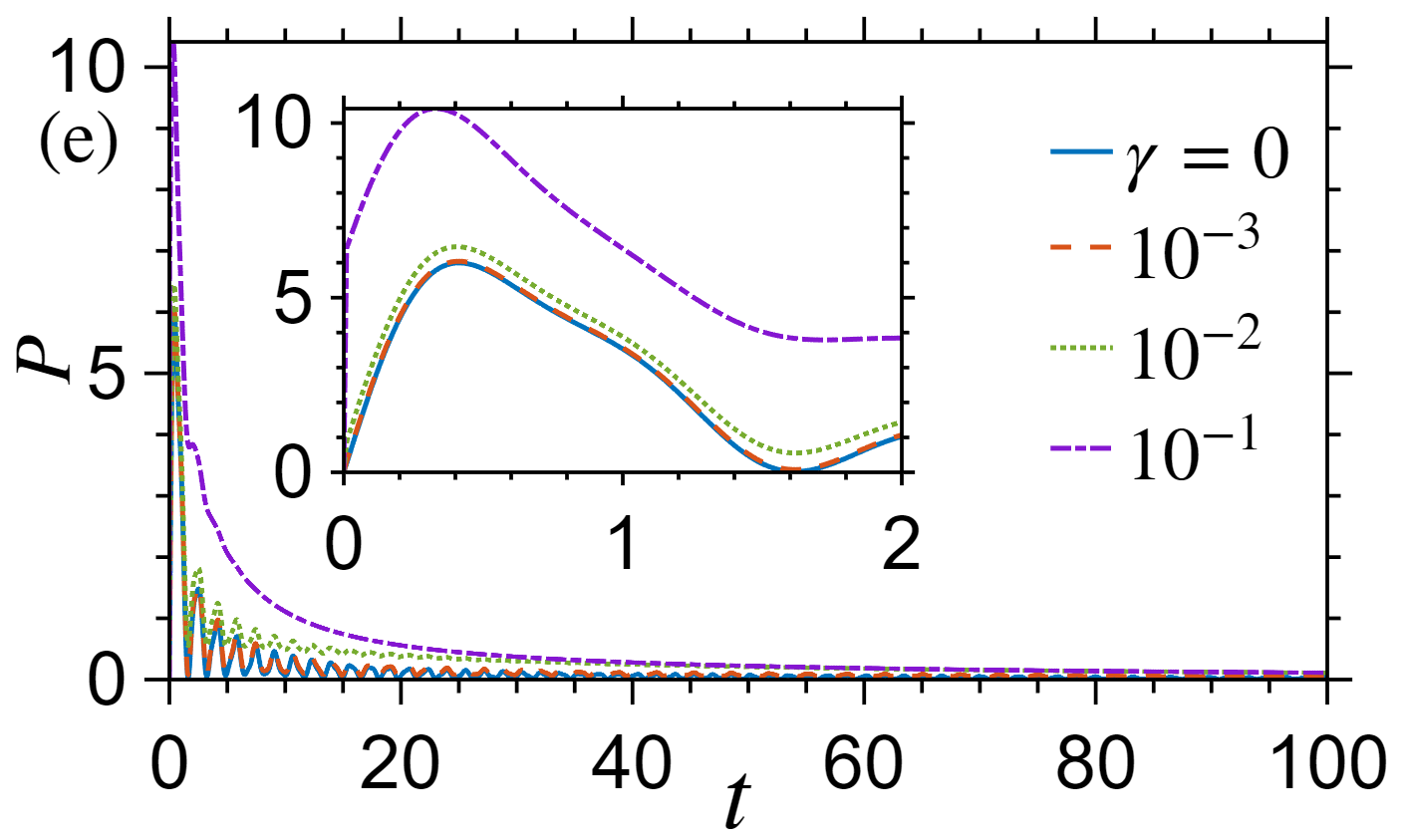}
\includegraphics[width=0.30\linewidth,height=0.25\linewidth]{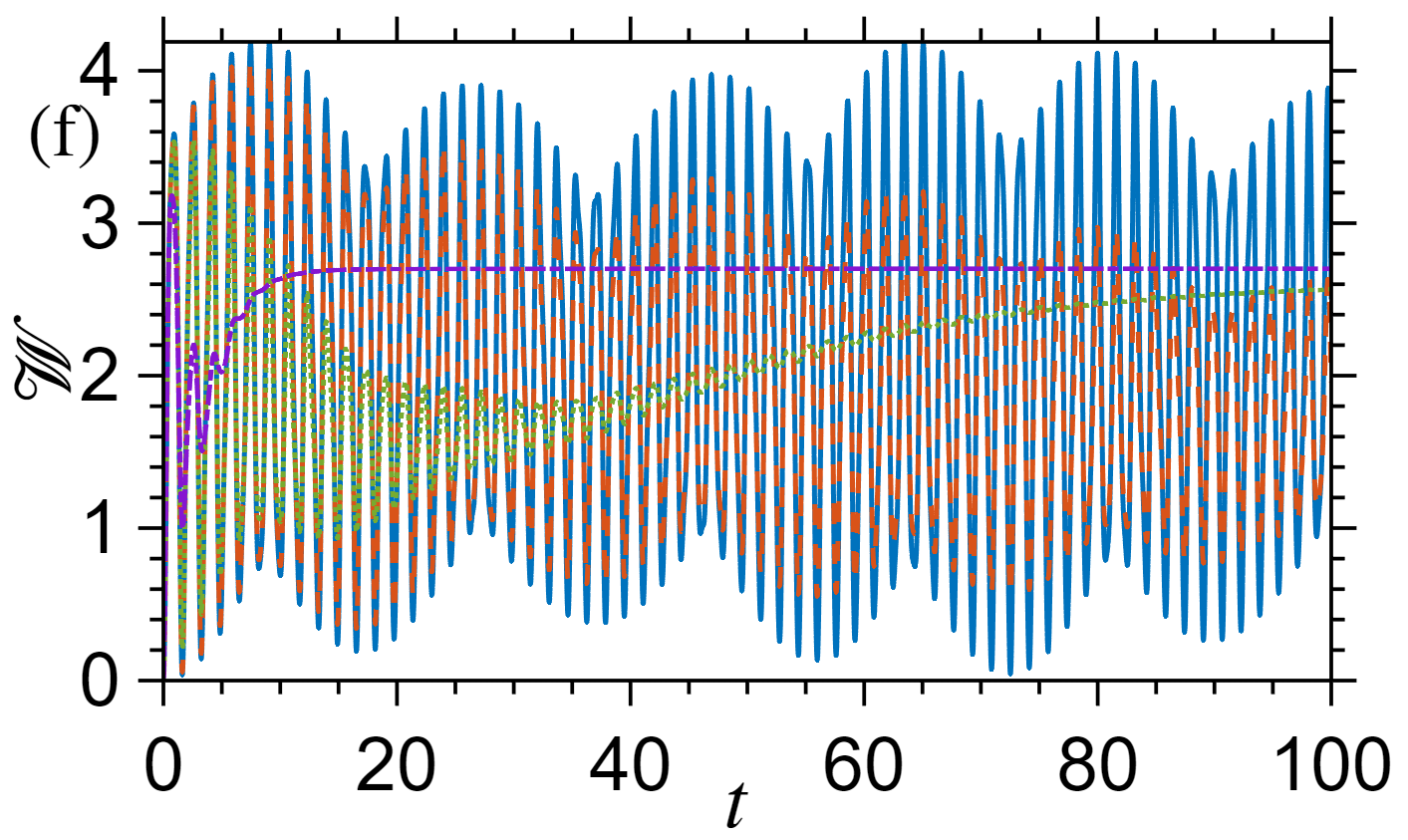}
    \caption{Same as Fig.~\ref{open_int_B_nonint_C}, but for $\lambda=0$.
}
    \label{open_int_B_nonint_C_l0}
\end{figure*}

\begin{figure*}
    \centering
\includegraphics[width=0.30\linewidth,height=0.25\linewidth]{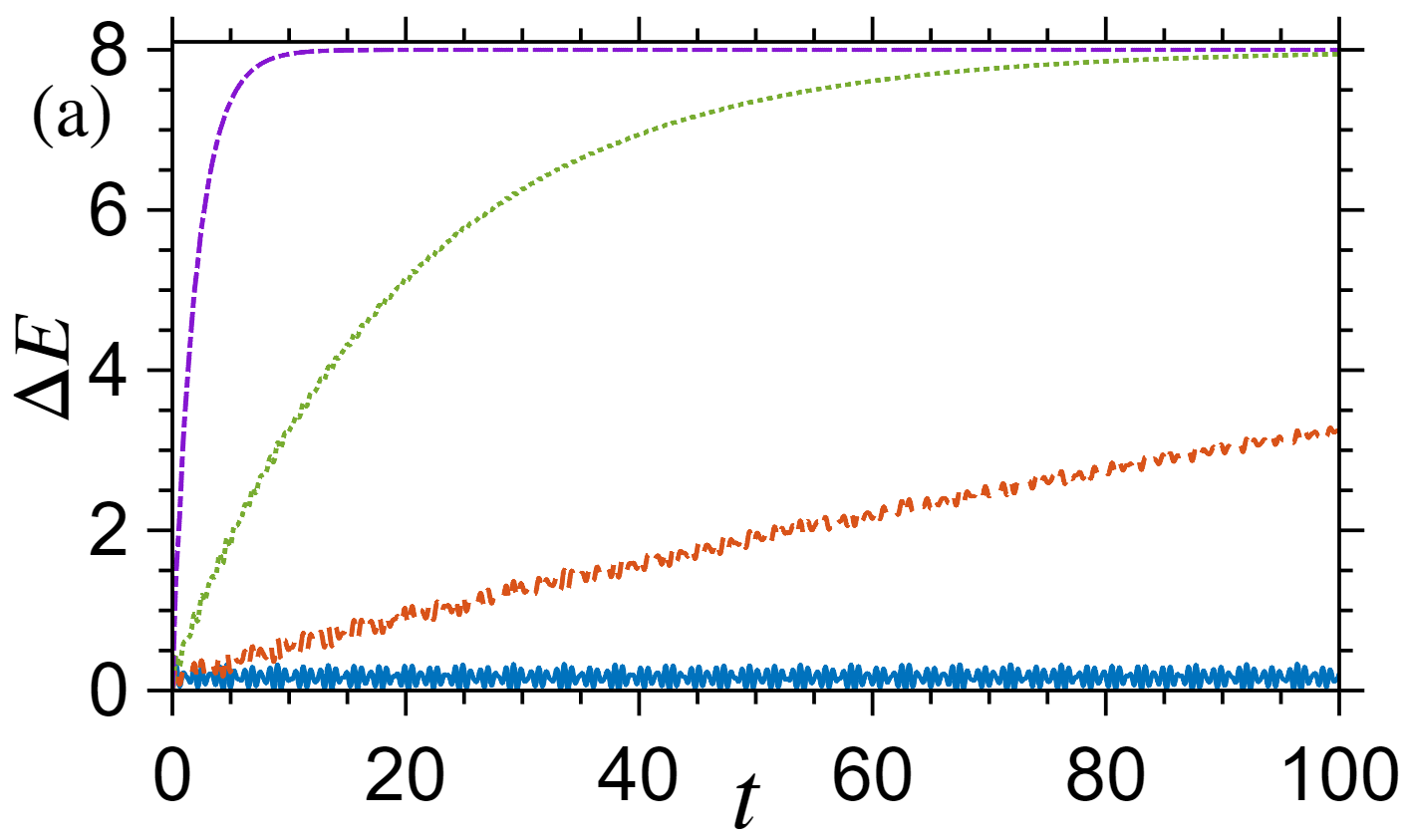}
\includegraphics[width=0.30\linewidth,height=0.25\linewidth]{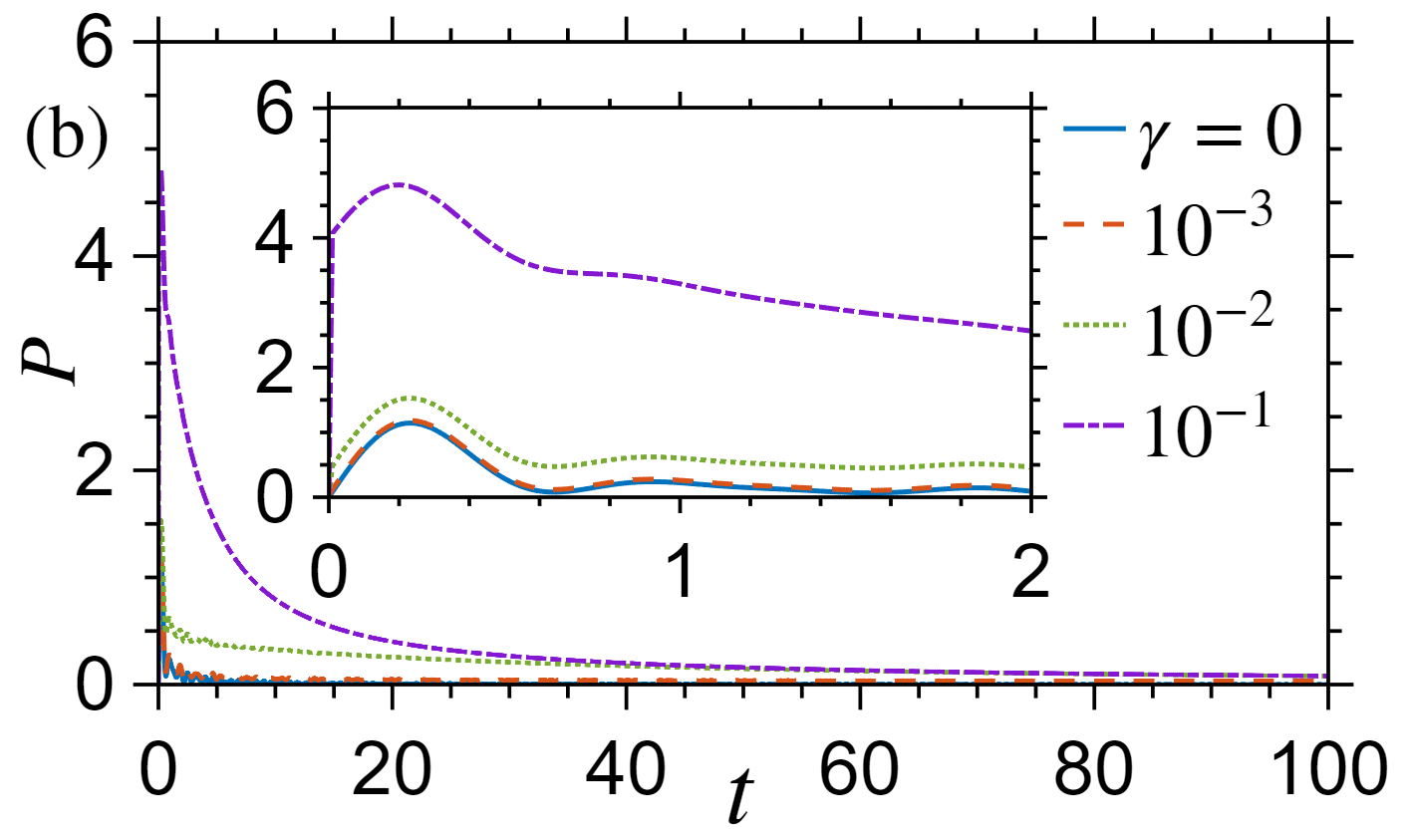}
\includegraphics[width=0.30\linewidth,height=0.25\linewidth]{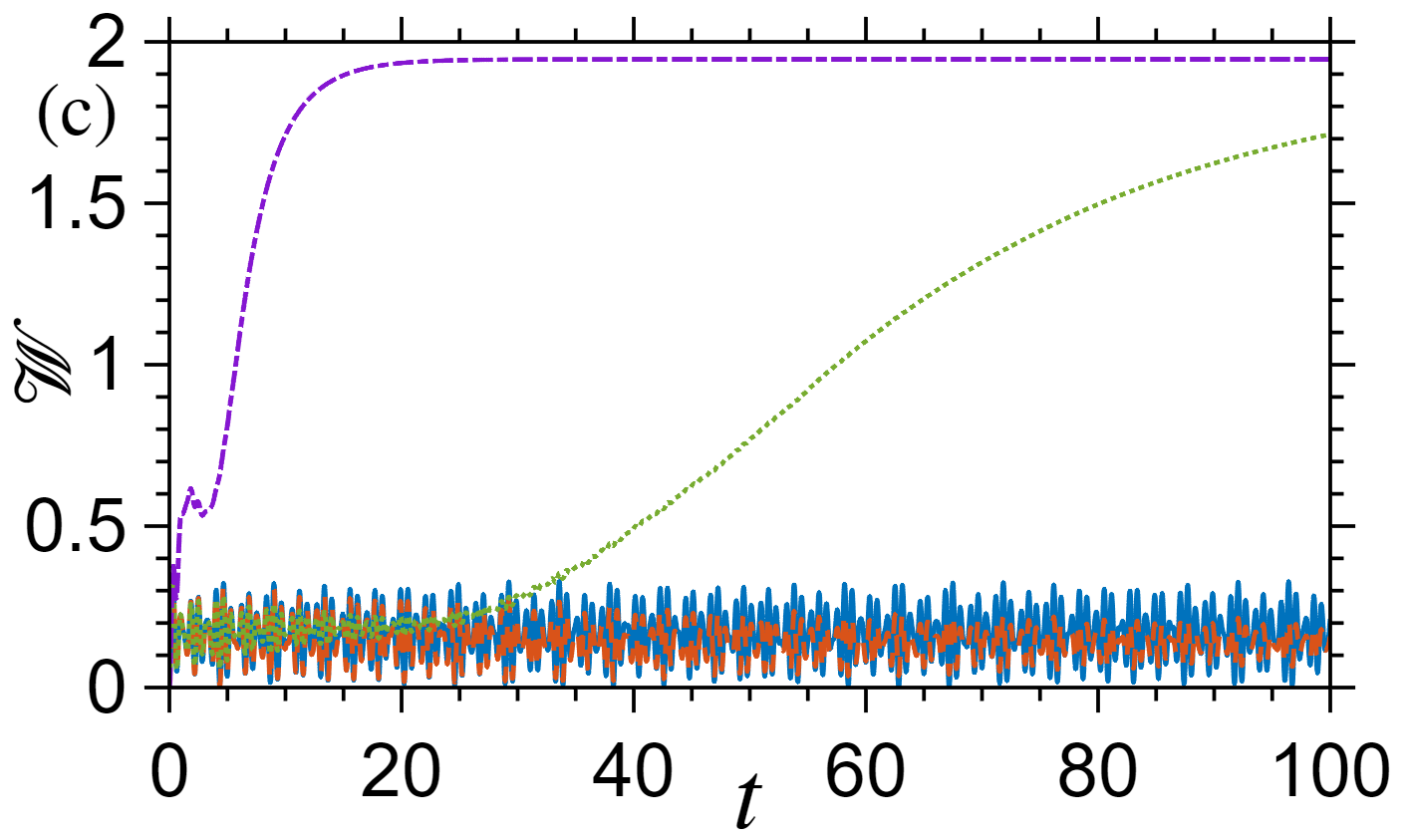}
\includegraphics[width=0.30\linewidth,height=0.25\linewidth]{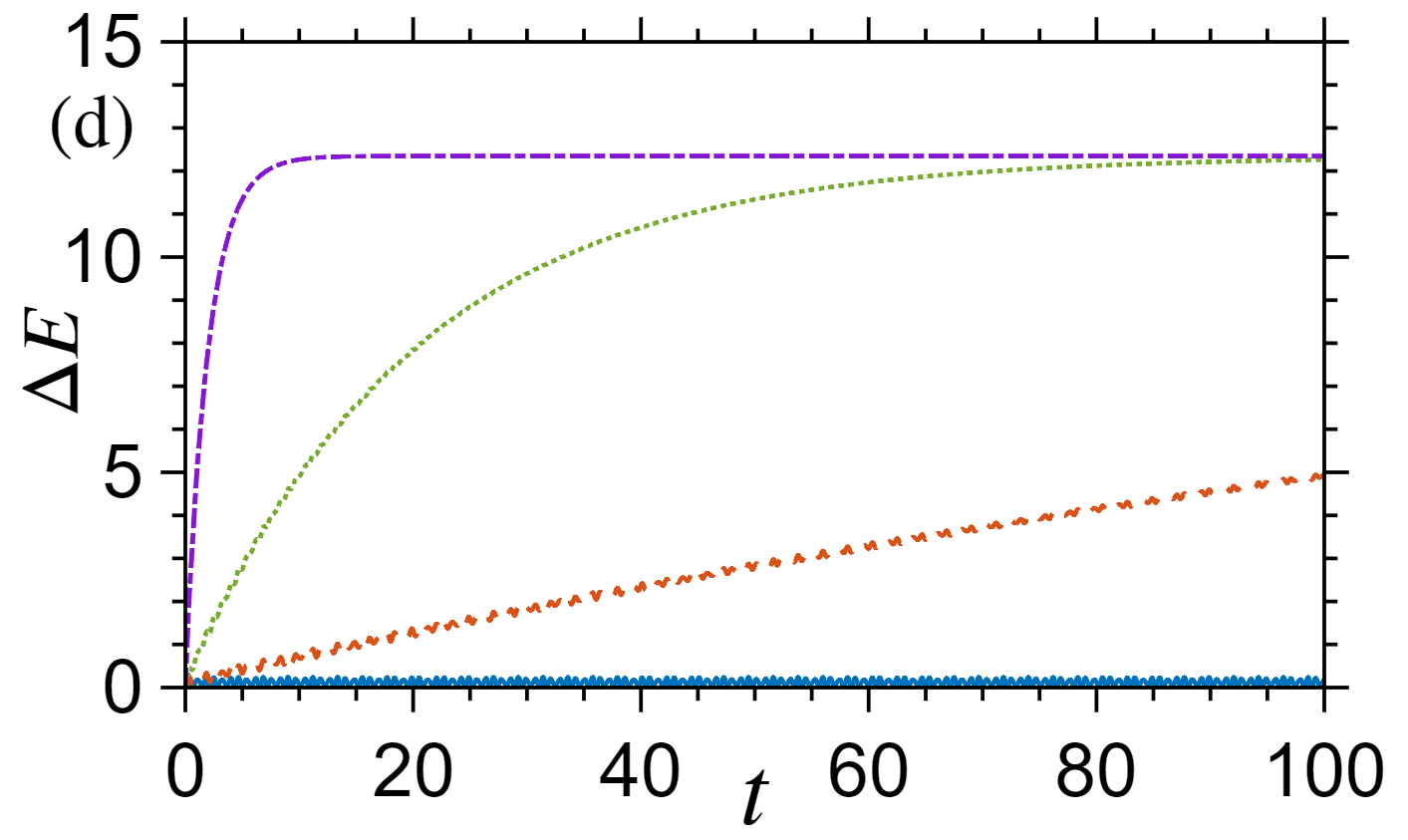}
\includegraphics[width=0.30\linewidth,height=0.25\linewidth]{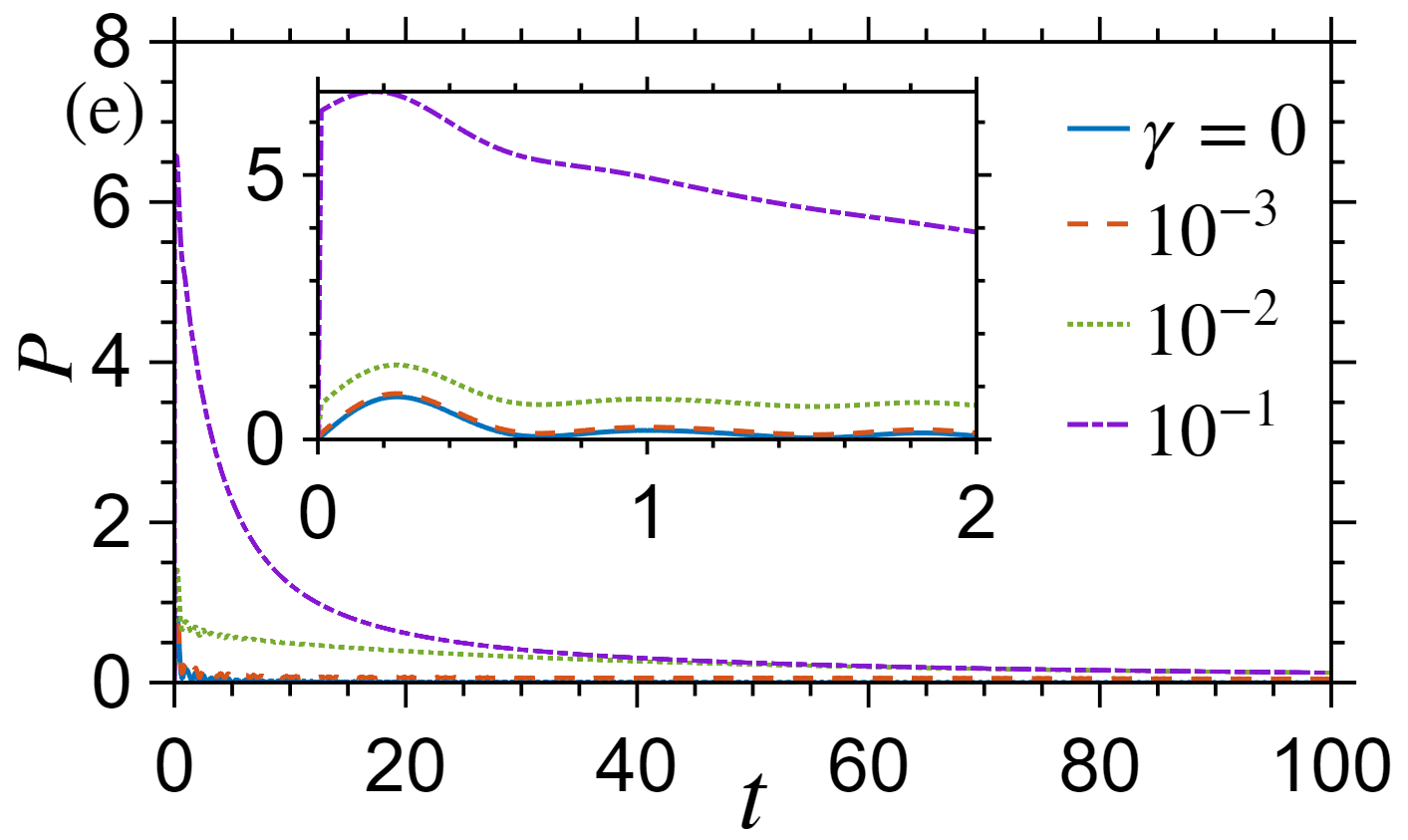}
\includegraphics[width=0.30\linewidth,height=0.25\linewidth]{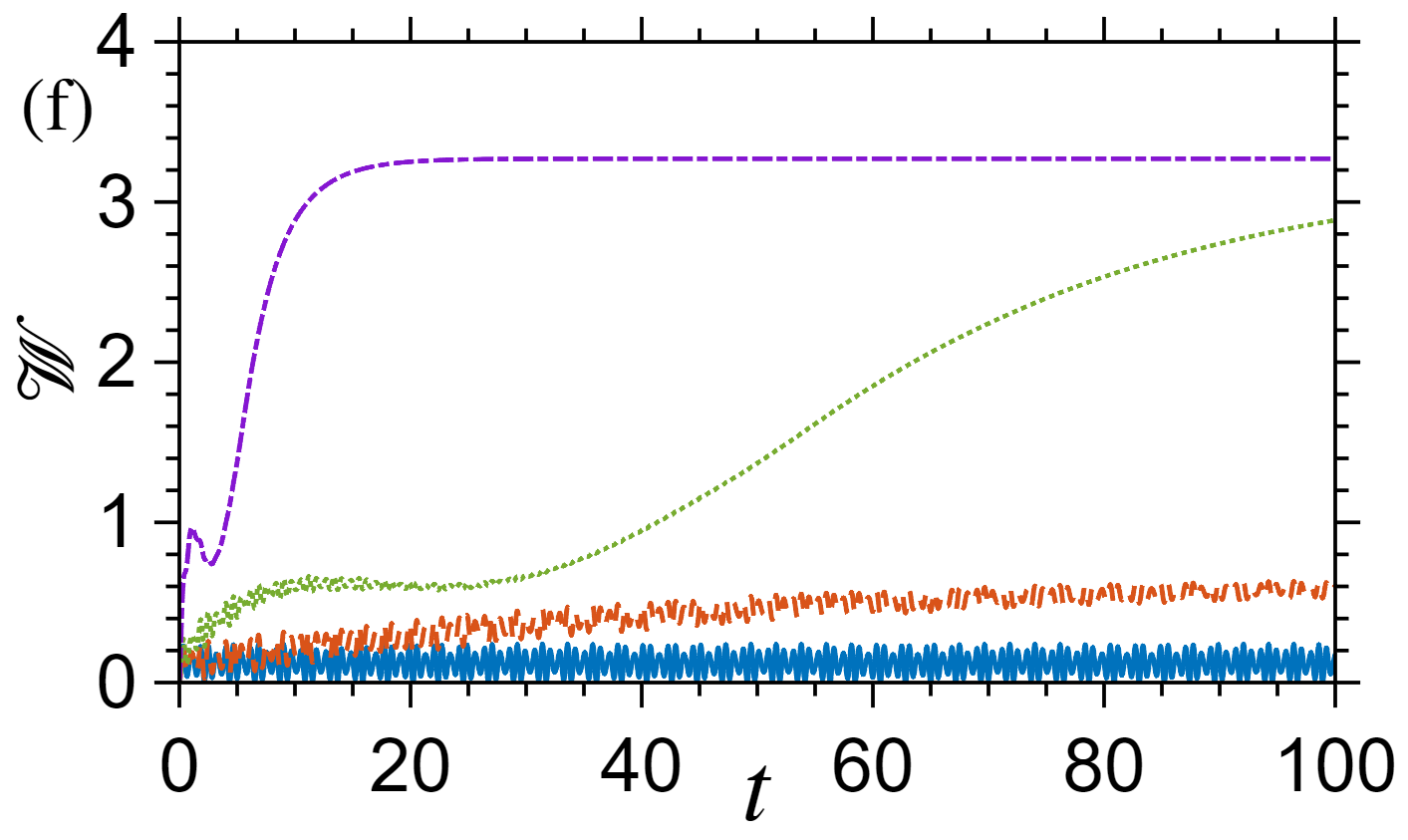}
    \caption{Same as Fig.~\ref{open_int_B_int_C}, but for $\lambda=0$.
}
    \label{open_int_B_int_C_l0}
\end{figure*}

\section{Open-System Quantum Battery Dynamics with $\lambda=0$}
\label{open_System_dynamics_l0}
In the main text, we investigated the effect of environmental decoherence on the charging dynamics by considering $\lambda=1$, for which the battery contribution is completely suppressed during the charging process. Here, we consider the complementary case $\lambda=0$, where the battery Hamiltonian remains fully active throughout the charging dynamics. This comparison allows us to examine how the intrinsic battery contribution modifies the response of the system to environmental relaxation and dephasing. Following the analysis presented in the main text, we systematically investigate the stored energy, charging power, and ergotropy for the different battery--chargoid configurations considered in our study.

\subsection{Configuration I: noninteracting battery and interacting chargoid}We first consider a noninteracting battery for which the charging dynamics is generated by an interacting chargoid Hamiltonian, taken to be either the XX or XY nearest-neighbor interaction. The corresponding results for $\lambda=0$ are shown in Fig.~\ref{open_nonint_B_int_C_l0}. For the XX chargoid, the inclusion of the battery contribution leads to a qualitatively similar suppression of the coherent oscillatory dynamics with increasing decoherence, but with a noticeable quantitative difference from the $\lambda=1$ case. In particular, the stored energy approaches a saturation value that remains below the maximum energy attained in the closed-system limit, $\gamma=0$, as shown in Fig.~\ref{open_nonint_B_int_C_l0}(a). The charging power and ergotropy are likewise reduced with increasing decoherence. Thus, retaining the battery contribution does not remove the detrimental effect of environmental coupling for the XX chargoid, but further limits the energy accumulation compared with the $\lambda=1$ case.

For the XY chargoid, the characteristic response observed for $\lambda=1$ is preserved: increasing the decoherence strength enhances the maximum stored energy while reducing the charging power, as shown in Fig.~\ref{open_nonint_B_int_C_l0}(d,e). Hence, the contrasting responses of the XX and XY chargoids to environmental coupling persist when the battery contribution remains fully active during charging.

A comparison of the two values of $\lambda$ further shows that the maximum stored energy, charging power, and ergotropy obtained for $\lambda=0$ are consistently smaller than or equal to their corresponding values for $\lambda=1$ for all decoherence strengths considered, as illustrated by Figs.~\ref{open_nonint_B_int_C_l0} and \ref{open_nonint_B_int_C}. This indicates that, for the present configurations, retaining the intrinsic battery contribution during the charging dynamics does not enhance the charging performance. Instead, it modifies the quantitative energy-transfer dynamics while preserving the characteristic dependence on the underlying chargoid interaction.

\subsection{ Configuration II: Interacting battery and non-interacting chargoid}
We next consider an interacting battery with a noninteracting chargoid, where the battery is described by either an XX or an XY nearest-neighbor interaction and the charging dynamics is generated by a uniform external magnetic field. The results for $\lambda=0$ are shown in Fig.~\ref{open_int_B_nonint_C_l0}. The qualitative response to environmental coupling remains consistent with that obtained for $\lambda=1$. For the XX-interacting battery, the maximum stored energy remains approximately insensitive to the decoherence strength, while the coherent oscillations are progressively suppressed and the charging power increases with $\gamma$. The ergotropy follows the corresponding behavior of the stored energy, with the oscillatory dynamics being damped as the environmental coupling is increased. For the XY-interacting battery, the enhancement of the stored energy, charging power, and ergotropy with increasing decoherence is also preserved. Thus, the characteristic distinction between the XX and XY interactions remains present when the battery contribution is fully retained during the charging dynamics.

A quantitative comparison with the $\lambda=1$ results shows that the maximum stored energy, charging power, and ergotropy for $\lambda=0$ are smaller than or equal to their corresponding values for $\lambda=1$ over the range of decoherence strengths considered, as illustrated by Figs.~\ref{open_int_B_nonint_C_l0} and \ref{open_int_B_nonint_C}. The difference is particularly relevant for the XY-interacting battery. Although environmental coupling enhances the stored energy and charging power, the corresponding maximum ergotropy does not exceed its closed-system value, as shown in Fig.~\ref{open_int_B_nonint_C_l0}(f), in contrast to the behavior observed for $\lambda=1$. This indicates that the additional energy accumulated through the dissipative dynamics is not necessarily accompanied by a proportional increase in its extractable component. Nevertheless, the persistence of the same qualitative trends for both values of $\lambda$ demonstrates that the dominant effect of environmental coupling is determined by the underlying interaction structure, while the battery contribution primarily affects the quantitative charging performance.

\subsection{Configuration III: Interacting battery and interacting chargoid}
Finally, we consider the case in which both the battery and chargoid are described by interacting Hamiltonians, with the battery contribution remaining fully active during the charging dynamics, $\lambda=0$. The results for the two complementary interaction arrangements are shown in Fig.~\ref{open_int_B_int_C_l0}(a--f). Similar to the $\lambda=1$ case, increasing the decoherence strength enhances the maximum stored energy, charging power, and ergotropy for both arrangements, while progressively suppressing the coherent oscillations. Thus, the constructive influence of environmental coupling on the charging performance persists even when the battery contribution is retained during the charging dynamics.

A comparison with the corresponding $\lambda=1$ results shows that the maximum stored energy, charging power, and ergotropy for $\lambda=0$ remain smaller than or equal to those obtained for $\lambda=1$ over the decoherence strengths considered. Nevertheless, the same qualitative dependence on $\gamma$ is preserved. In particular, the simultaneous enhancement of stored energy and ergotropy indicates that the additional energy accumulated in the presence of environmental coupling retains a significant extractable component. The persistence of this behavior for both interaction arrangements and for both values of $\lambda$ demonstrates that the constructive role of environmental coupling originates primarily from the underlying interacting many-body dynamics, while the battery contribution mainly affects the quantitative magnitude of the charging performance.

\section{ Dynamics beyond the physical regime}
\label{Appendix_lambd_g1}
\vspace{0.25cm}
\begin{figure}
\includegraphics[width=0.48\linewidth,height=0.50\linewidth]{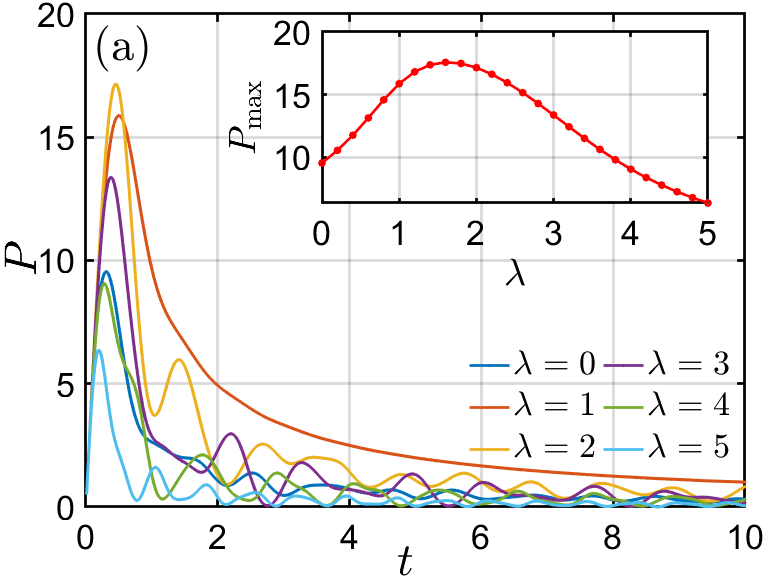}
\includegraphics[width=0.49\linewidth,height=0.50\linewidth]{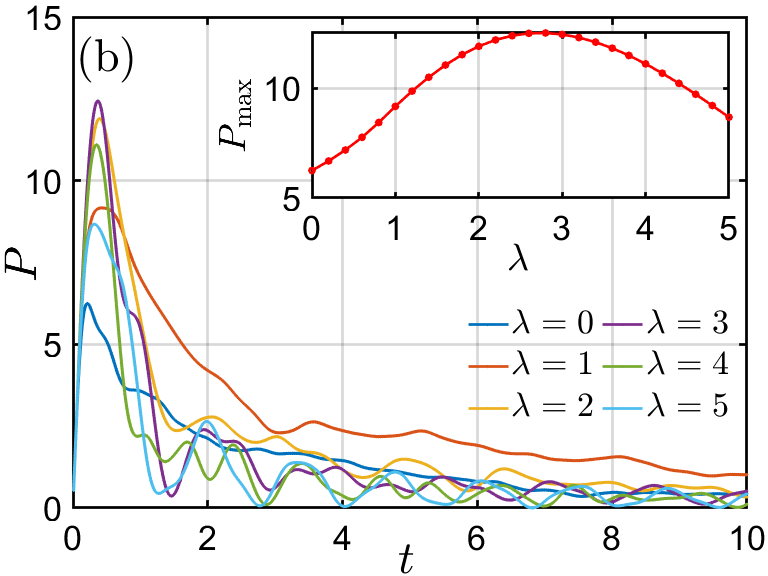}
 \caption{Charging power $P(t)$ as a function of time $t$ for a non-interacting battery charged by an ATA interacting (a) Ising and (b) XY Hamiltonian. Results are shown for several values of the countereffect parameter $\lambda$. The insets display the corresponding maximum charging power as a function of $\lambda$. The simulations use $J = 1$, $h_z = 1$, system size $N = 10$, and anisotropy parameter $\Gamma = 0.5$.}

\label{Ising_long_fig}
\end{figure}

In the main manuscript, we investigate the charging dynamics of quantum batteries within the physically acceptable range of the countereffect parameter, $\lambda \in [0,1]$. Within this regime, both the storage energy and power attain their maximum at $\lambda = 1$. Interestingly, our analysis further reveals that the storage power reaches an even higher value beyond this physical regime ($\lambda > 1$) in the case of ATA interacting Ising and $XY$ chargoid Hamiltonians. However, the storage energy still achieves its maximum at $\lambda = 1$. Although the region $\lambda > 1$ falls outside the standard physical constraints, we emphasize this result to highlight its potential implications and to encourage future investigations into such unconventional regimes.
\par
We analyze the charging performance for Configuration I, in which the battery is noninteracting and described by Eq.~(\ref{eq:HB0}), while the chargoid is described by either the ATA Ising Hamiltonian in Eq.~(\ref{eq:HC_IATA}) or the ATA anisotropic-XY Hamiltonian in Eq.~(\ref{eq:HC_XYATA}). We examine the dependence of the charging power on the countereffect parameter over the range $\lambda\in[0,5]$. In both cases, for all values of \(\lambda\), the storage power exhibits a consistent trend: it initially increases, reaches a maximum value, and subsequently decreases after a certain time [Fig.~\ref{Ising_long_fig}(a,b)]. The maximum storage power, \(P_{\rm max}\), depends on \(\lambda\) and increases with it, attaining its peak at \(\lambda = 1.6\) for the ATA interacting Ising chargoid and \(\lambda = 2.8\) for the ATA interacting XY chargoid. Beyond these points, the storage power decreases with further increases in \(\lambda\) [inset of Fig.~\ref{Ising_long_fig}(a,b)]. Remarkably, the highest \(P_{\rm max}\) values occur outside the physically accepted regime (\(\lambda > 1\)).

\bibliography{QB}

\end{document}